\documentclass[aps,prd,reprint,superscriptaddress,floatfix,showpacs]{revtex4-1}
\usepackage{amsfonts}
\usepackage{amsmath}
\usepackage{amssymb}
\usepackage{subfigure}
\usepackage{mathrsfs}
\usepackage{graphicx}
\usepackage{color}

\newcommand{\CKM}{V}

\newcommand{\V}[1]{\CKM_{#1}^{\,}}
\newcommand{\Vc}[1]{\CKM_{#1}^{\ast}}
\newcommand{\U}[1]{U_{#1}^{\,}}

\newcommand{\La}[2]{\lambda^{#1}_{#2}}
\newcommand{\asld}{A^d_{SL}}
\newcommand{\asls}{A^s_{SL}}
\newcommand{\aslb}{A^b_{SL}}
\newcommand{\aslq}{A^q_{SL}}

\newcommand{\DGd}{\Delta\Gamma_d}
\newcommand{\DGs}{\Delta\Gamma_s}
\newcommand{\DGq}{\Delta\Gamma_q}

\newcommand{\AJPKs}{A_{J/\Psi K_S}}
\newcommand{\AJPP}{A_{J/\Psi \Phi}}
\newcommand{\BBdmix}{$B^0_d$--$\bar B^0_d$}
\newcommand{\BBsmix}{$B^0_s$--$\bar B^0_s$}
\newcommand{\BBqmix}{$B^0_q$--$\bar B^0_q$}
\newcommand{\DDmix}{$D^0$--$\bar D^0$}
\newcommand{\KKmix}{$K^0$--$\bar K^0$}

\newcommand{\Hq}{\mathscr H_{(q)}}
\newcommand{\Mq}{M^{(q)}}
\newcommand{\Gq}{\Gamma^{(q)}}
\newcommand{\Hd}{\mathscr H_{(d)}}
\newcommand{\Mmix}[1]{M_{12}^{({#1})}}
\newcommand{\Mmixq}{\Mmix{q}}
\newcommand{\Mmixd}{\Mmix{d}}
\newcommand{\Mmixs}{\Mmix{s}}
\newcommand{\Gmix}[1]{\Gamma_{12}^{({#1})}}
\newcommand{\Gmixq}{\Gmix{q}}
\newcommand{\Gmixd}{\Gmix{d}}
\newcommand{\Gmixs}{\Gmix{s}}
\newcommand{\DMBd}{\Delta M_{B_d}}
\newcommand{\DMBs}{\Delta M_{B_s}}
\newcommand{\DMBq}{\Delta M_{B_q}}
\newcommand{\DC}{\Delta\chi^2}

\newcommand{\re}[1]{\text{Re}\left(#1\right)}
\newcommand{\im}[1]{\text{Im}\left(#1\right)}

\newcommand{\refeq}[1]{(\ref{#1})}
\newcommand{\eq}[1]{eq.\refeq{#1}}

\graphicspath{{Figs/}}

\begin{document}


\title{Mixing asymmetries in B meson systems, the D0 like-sign dimuon asymmetry and generic New Physics}
\author{F. J. Botella}\email{fbotella@uv.es}
\affiliation{
Departament de F\'isica Te\`orica and IFIC,\\
Universitat de Val\`encia - CSIC, E-46100, Burjassot, Spain
}
\author{G. C. Branco}\email{gbranco@tecnico.ulisboa.pt}
\author{M. Nebot}\email{nebot@cftp.ist.utl.pt}
\affiliation{Centro de F\'isica Te\'orica de Part\'iculas, and Departamento de F\' \i sica\\
Instituto Superior T\'ecnico, Universidade de Lisboa, Av. Rovisco Pais, P-1049-001 Lisboa,
Portugal}
\author{A. S\'anchez}\email{asanchez@ific.uv.es}
\affiliation{
Departament de F\'isica Te\`orica and IFIC,\\
Universitat de Val\`encia - CSIC, E-46100, Burjassot, Spain
}

\preprint{IFIC/14-04}\preprint{CFTP/14-005}
\pacs{12.60.-i,12.15.Ff,11.30.Er,14.65.Jk,14.40.Nd}

\begin{abstract}
{The measurement of a large like-sign dimuon asymmetry $\aslb$ by the D0 experiment at the Tevatron departs noticeably from Standard Model expectations and it may be interpreted as a hint of physics beyond the Standard Model contributing to $\Delta B\neq 0$ transitions. 
In this work we analyse how the natural suppression of $\aslb$ in the SM can be circumvented by New Physics.
 We consider generic Standard Model extensions where the charged current mixing matrix is enlarged with respect to the usual $3\times 3$ unitary Cabibbo-Kobayashi-Maskawa matrix, and show how, within this framework, a significant enhancement over Standard Model expectations for $\aslb$ is easily reachable through enhancements of the semileptonic asymmetries $\asld$ and $\asls$ of both \BBdmix\ and \BBsmix\ systems. Despite being insufficient to reproduce the D0 measurement, such deviations from SM expectations may be probed by the LHCb experiment.}
\end{abstract}

\date{\today}
\maketitle

\section{Introduction\label{SEC:Intro}}
Flavour Physics and CP violation provide a magnificient laboratory to probe our fundamental understanding of Nature and to test, at unprecedented levels, the Standard Model (SM) and any of its extensions. The impact of the B-factories Babar and Belle operating at $e^+e^-$ machines, of the D0 and CDF experiments operating at the Tevatron and lately of the ATLAS, CMS and LHCb experiments at the LHC, is of paramount importance.

Among the plethora of results on CP violating phenomena, the measurement by the D0 collaboration \cite{Abazov:2011yk, *Abazov:2013uma} of the like-sign dimuon asymmetry $\aslb$ has received much attention. Schematically, (i) $b\bar b$ pairs are strongly produced, (ii) they hadronize into $B_d$ or $B_s$ mesons/antimesons and (iii) they decay weakly. Semileptonic decays are flavour specific and ``tag'' the nature of the decaying $B$ depending on the charge of the produced lepton $\ell$: meson for $\ell^+$ or antimeson for $\ell^-$. If it were not for $B_q$ -- $\bar B_q$ oscillations, both decays could not produce leptons\footnote{We directly refer in the following to \emph{muons} since they are the cleanest case from the experimental point of view.} of the same charge. In the presence of $B_q$ -- $\bar B_q$ oscillations, such like-sign muon double decay channels occur, and one defines the asymmetry\footnote{Although central in any experimental analysis, we omit any discussion on issues such as efficiencies or backgrounds.}
\begin{equation}
\aslb=\frac{N^{++}-N^{--}}{N^{++}+N^{--}}\,,
\end{equation}
with $N^{++}$ ($N^{--}$) denoting the number of events with both $B$ mesons decaying to $\mu^+$ ($\mu^-$).
The values reported by the D0 collaboration \cite{Abazov:2011yk, *Abazov:2013uma} are around ``$3\sigma$'' away from Standard Model expectations, and this triggered intense activity to explore the potential of an ample variety of models beyond the SM to produce such values. As $\aslb$ can be expressed in terms of the individual asymmetries $\asld$ and $\asls$ of \BBdmix\ and \BBsmix\ systems \cite{Lees:2013sua, *Aaij:2013gta}, it is customary to discuss in terms of them. In particular, since it is a common thought that the B factories have left little space for New Physics (NP) to contribute new sources of CP violation in the \BBdmix\ system, the focus\footnote{Nevertheless, as we will show, since significant cancellations are at work in the SM case, large NP contributions are not necessary to obtain significant enhancements in $\asld$.} has been on $\asls$. New Physics has been invoked to modify the dispersive mixing amplitude $\Mmixs$ and/or the absorptive one $\Gmixs$ in specific scenarios, such as supersymmetric extensions of the SM \cite{Ko:2010mn,*Parry:2010ce,*Ishimori:2011nv}, extra-dimensions \cite{Datta:2010yq,*Goertz:2011nx}, $Z^\prime$ models \cite{Deshpande:2010hy,*Alok:2010ij,*Kim:2010gx,*Kim:2012rpa}, left-right models \cite{Lee:2011kn}, extended scalar sectors \cite{Jung:2010ik,*Dobrescu:2010rh,*Trott:2010iz,*Bai:2010kf}, axigluon exchange \cite{Chen:2010wv} or additional fermion generations \cite{Hou:2007ps,*Soni:2010xh,*Chen:2010aq,*Botella:2008qm,*Botella:2012ju,*Alok:2012xm,*Alok:2014yua}. New Physics in $\aslb$ has also been explored through model independent analyses \cite{Ligeti:2010ia,*Bauer:2010dga,*Bobeth:2011st,*Bobeth:2014rda} or through NP modifying highly suppressed (within the SM) additional contributions \cite{DescotesGenon:2012kr}.
This article is organised as follows. In section \ref{SEC:mix} we review the well known SM prediction for the semileptonic asymmetries $\asld$ and $\asls$, and the dimuon asymmetry $\aslb$. In section \ref{SEC:3x3NP} we revisit a model independent analysis where New Physics is allowed to modify the mixing amplitudes $\Mmixq$, and show how the previous asymmetries can be significantly larger than SM expectations. In section \ref{SEC:no3x3NP} we consider NP scenarios where the mixing matrix is not the usual $3\times 3$ unitary Cabibbo-Kobayashi-Maskawa, but an enlarged one, and thus study for the first time how the values that the asymmetries of interest can span, differ from the SM ones. We also analyse the prospects to, eventually, distinguish if the mixing matrix is $3\times 3$ unitary or not. In the last section we present our conclusions.
\section{Mixing in $\mathbf{B_d}$ and $\mathbf{B_s}$ meson systems\label{SEC:mix}}
Under general conditions, the time evolution of the \BBqmix\ systems, $q=d,s$, is described by an effective weak hamiltonian $\Hq$ according to the Schr\"odinger equation \cite{Branco:1999fs}
\begin{equation}
i\frac{d}{dt}\begin{pmatrix}B_q(t)\\ \bar B_q(t)\end{pmatrix}\,=\,\Hq\,\begin{pmatrix}B_q(t)\\ \bar B_q(t)\end{pmatrix}\ .\label{eq:schrodinger:01}
\end{equation}
$\Hq$ has hermitian and anti-hermitian parts $\Mq$ and $-i\Gq/2$:
\begin{equation}
\Hq=\Mq-\frac{i}{2}\,\Gq\,,\ \Mq={\Mq}^{\dag}\,,\ \Gq={\Gq}^{\dag}\,.\label{eq:effH:01}
\end{equation}
In the Standard Model, the dispersive part $\Mmixq$ of the transition amplitude $B_q\to\bar B_q$ is dominated by one loop box diagrams with virtual $t$ quarks and $W$ bosons\footnote{Equation (\ref{eq:M12q:01}) includes perturbative QCD corrections $\eta_B$, and non-perturbative information, i.e. the decay constant $f_{B_q}$ and the bag parameter $B_q$. Subleading contributions from virtual $u$ or $c$ quarks are neglected.}
\begin{equation}
\left[\Mmixq\right]_{\text{SM}}=\frac{G_F^2M_W^2}{12\pi^2}\,M_{B_q}\,f_{B_q}^2\,B_{B_q}\,\eta_B\,(\V{tb}\Vc{tq})^2\,S_0(x_t)\ .\label{eq:M12q:01}
\end{equation}
On the other hand, the absorptive part, $\Gmixq$, is dominated by intermediate real (on-shell) $u$ and $c$ quarks. The corresponding SM short-distance prediction is more involved \cite{Beneke:1998sy,*Beneke:2003az,*Ciuchini:2003ww,*Lenz:2012mb,*Hagelin:1981zk,Lenz:2006hd}: a Heavy Quark Expansion is carried out, yielding $\Gmixq$ as an expansion in $\alpha_s(m_b)$ and $\Lambda/m_b$. Focusing on the flavour structure, it has in general the following form
\begin{multline}
\frac{\Gmixq}{\Mmixq}= -\Bigg[\frac{\Gamma_{12}^{cc}}{\Mmixq}(\V{cb}\Vc{cq})^2\\
+2\frac{\Gamma_{12}^{uc}}{\Mmixq}(\V{ub}\Vc{uq}\V{cb}\Vc{cq}) +\frac{\Gamma_{12}^{uu}}{\Mmixq}(\V{ub}\Vc{uq})^2\Bigg]\,,\label{eq:G12q:01}
\end{multline}
and in particular in the SM the flavour structure is
\begin{multline}
\left[\frac{\Gmixq}{\Mmixq}\right]_{\textrm{SM}}\propto\\ 
\Gamma_{12}^{cc}\frac{(\V{cb}\Vc{cq})^2}{(\V{tb}\Vc{tq})^2}+2\Gamma_{12}^{uc}\frac{\V{ub}\Vc{uq}\V{cb}\Vc{cq}}{(\V{tb}\Vc{tq})^2}+\Gamma_{12}^{uu}\frac{(\V{ub}\Vc{uq})^2}{(\V{tb}\Vc{tq})^2}
\ .\label{eq:G12q:01b}
\end{multline}
It is important to notice that, in terms of the weak interactions, the coefficients $\Gamma_{12}^{uu}$, $\Gamma_{12}^{uc}$ and $\Gamma_{12}^{cc}$ are dominated by tree level contributions. We can then rely on \eq{eq:G12q:01} without qualms about New Physics contributions invalidating it: only if a given scenario beyond the Standard Model can give competing contributions to tree level SM predictions, should we worry and consider a specific analysis. 
The coefficients $\Gamma_{12}^{ab}$ are in turn
\begin{equation}
-\Gamma_{12}^{cc}=c\,,\quad -2\Gamma_{12}^{uc}=2c-a\,,\quad -\Gamma_{12}^{uu}=b+c-a\,,\label{eq:G12d:01:coeff1}
\end{equation}
where \cite{Lenz:2006hd}
\begin{eqnarray}
a&=& (10.5\pm 1.8)\times 10^{-4}\,,\nonumber\\ 
b&=& (0.2\pm 0.1)\times 10^{-4}\,,\nonumber\\
c&=& (-53.3\pm 12.0)\times 10^{-4}\,.\label{eq:G12d:01:coeff2}
\end{eqnarray}
It is important to stress that in an expansion in powers of $(m_c/m_b)^2$, only $c$ is present at zero-th order.
Then, unitarity of the CKM mixing matrix, implying the orthogonality condition $\V{ub}\Vc{uq}+\V{cb}\Vc{cq}+\V{tb}\Vc{tq}=0$, can be used to write 
\begin{equation}
\left[\frac{\Gmixq}{\Mmixq}\right]_{\textrm{SM}}= 
 K_{(q)}\left[c+a\,\frac{\V{ub}\Vc{uq}}{\V{tb}\Vc{tq}}+b\,\left(\frac{\V{ub}\Vc{uq}}{\V{tb}\Vc{tq}}\right)^2\right]\ ,\label{eq:G12q:03}
\end{equation}
where
\[
K_{(q)}=\frac{12\pi^2}{M_{B_q}\,f_{B_q}^2\,B_{B_q}\,G_F^2M_W^2\,\eta_B\,S_0(x_t)}\ .
\]
$\Gmixq/\Mmixq$ is accessed through two observables; at leading order in $\Gmixq/\Mmixq$ powers, one has
\begin{equation}
-\frac{\DGq}{\DMBq}=\re{\frac{\Gmixq}{\Mmixq}}\,,\quad \aslq=\im{\frac{\Gmixq}{\Mmixq}} \,.\label{eq:G12qObs:01}
\end{equation}
The real part $\re{{\Gmixq}/{\Mmixq}}$ controls the width difference between the eigenstates of $\Hq$. The imaginary part $\im{{\Gmixq}/{\Mmixq}}$ is genuinely CP violating and only involves mixing amplitudes; as anticipated, it is accessed through asymmetries in flavour specific, semileptonic decays.
The SM expectations for those observables, with the inputs in table \ref{AP:tab:data} (appendix \ref{APP:Input}), are
\begin{eqnarray}
\left[\asld\right]_{\textrm{SM}} & = (-4.2\pm 0.7)\cdot 10^{-4}\,,\nonumber\\
\left[\DGd\right]_{\textrm{SM}}  &= (2.60\pm 0.25)\cdot 10^{-3}\,\text{ps}^{-1}\,,\nonumber\label{eq:SMpred:d:01}\\
\left[\asls\right]_{\textrm{SM}} & = (2.0\pm 0.3)\cdot 10^{-5}\,,\nonumber\\
\left[\DGs\right]_{\textrm{SM}}  &= (0.090\pm 0.008)\,\text{ps}^{-1} \,.\hfill \label{eq:SMpred:s:01}
\end{eqnarray}
The following comments are in order:
\begin{itemize}
\item \emph{both} $\asld$ and $\asls$ are small, $\mathcal O(10^{-4})$ and $\mathcal O(10^{-5})$ respectively, with room for variation at the $\pm 20$\% level.
This smallness can be traced back to the $(m_c/m_b)^2$ suppression in \eq{eq:G12q:03}: the leading contribution, proportional to $c$, is real and does not contribute to the semileptonic asymmetries; furthermore, since the hierarchy of the CKM matrix gives
\begin{align}
&\left|\frac{\Vc{ud}\V{ub}}{\Vc{td}\V{tb}}\right|\simeq 0.40\,,\ &&\arg\left(\frac{\Vc{ud}\V{ub}}{\Vc{td}\V{tb}}\right)\simeq -1.57\,,\label{eq:SMmod:d:01}\\
&\left|\frac{\Vc{us}\V{ub}}{\Vc{ts}\V{tb}}\right|\simeq 0.02\,,\ &&\arg\left(\frac{\Vc{us}\V{ub}}{\Vc{ts}\V{tb}}\right)\simeq \pi-1.18\,,\label{eq:SMmod:s:01}
\end{align}
one could expect $|\asld|\gg |\asls|$.
\item $\DGd$ is $\mathcal O(10^{-3})$ ps$^{-1}$ and $\DGs$ is $\mathcal O(10^{-1})$ ps$^{-1}$: while $\Gamma_d\simeq\Gamma_s\simeq (1.5\text{ ps})^{-1}$, the hierarchy in $\DGd$ and $\DGs$ can be anticipated with the leading term in \eq{eq:G12q:03}, giving $\DGs/\DMBs\sim \DGd/\DMBd$.
\end{itemize}
\noindent Underlying this simple SM analysis are two important assumptions:
\begin{itemize}
\item[(1)] $\Mmixq$ is dominated by a single weak amplitude (the one with virtual top quarks),
\item[(2)] the CKM matrix is $3\times 3$ unitary.
\end{itemize}
Another interesting possibility is to rewrite the flavour structure of \eq{eq:G12q:01}, as done in \cite{Botella:2006va}, in terms of a priori measurable quantities, as we now illustrate for $q=d$. 
Since the mass difference\footnote{In both \BBdmix\ and \BBsmix\ systems, $\DMBq=2|\Mmixq|$ since $|\Gmixq|\ll |\Mmixq|$ \cite{Branco:1999fs}.} between the eigenstates of $\Hd$ is $\DMBd=2|\Mmixd|$, and the ``golden'' time-dependent CP asymmetry $\AJPKs$ in $B_d\to J/\Psi K_S$ is controlled by $\arg{(\Mmixd)}$, one can use
\begin{equation}
\Mmixd=\frac{\DMBd}{2}\,e^{i\,2\bar\beta}\,,\label{eq:M12d:01}
\end{equation}
with the effective phase $\bar\beta$ given by $\AJPKs\equiv\sin(2\bar\beta)$, to rewrite
\begin{multline}
\frac{\Gmixd}{\Mmixd}= \frac{2}{\DMBd}
\Big[
c\,e^{-i\,2\bar\beta}\left(|\V{cb}\Vc{cd}|-|\V{ub}\Vc{ud}|e^{-i\gamma}\right)^2\\
+a\,|\V{ub}\Vc{ud}|e^{-i(2\bar\beta+\gamma)}\left(|\V{cb}\Vc{cd}|-|\V{ub}\Vc{ud}|e^{-i\gamma}\right)\\
+b\,|\V{ub}\Vc{ud}|^2e^{-i2(\bar\beta+\gamma)}
\Big]\ .\label{eq:G12d:01}
\end{multline}
We use the physical rephasing invariant phases \cite{Branco:1999fs} $\gamma\equiv\arg(-\V{ud}\Vc{ub}\V{cb}\Vc{cd})$ and $\beta\equiv\arg(-\V{cd}\Vc{cb}\V{tb}\Vc{td})$; even though in the SM $\bar\beta=\beta$, we introduce $\bar\beta$ for later use since it is directly related to an observable. 
Equation \refeq{eq:G12d:01} provides a particularly interesting expression\footnote{Notice that \eq{eq:G12d:01} is written, as it should, in terms of quantities invariant under rephasings of the CKM elements and of the $B_d^0$ and $\bar B_d^0$ states, even if, for the sake of brevity, intermediate expressions such as \eq{eq:M12d:01} are not.} for $\Gmixd/\Mmixd$. It involves: (i) tree level CKM moduli $|\V{ub}|$, $|\V{ud}|$, $|\V{cb}|$ and $|\V{cd}|$, (ii) the mass difference $\DMBd$, and (iii) the phases $2\bar\beta$, $2\bar\beta+\gamma$ and $2(\bar\beta+\gamma)$. All of them are, in principle, directly measurable\footnote{Besides $2\bar\beta$ from the golden channel $B_d\to J/\Psi K_s$, $\gamma$ is accessed through \emph{tree} level decays such as $B_d\to DK$, while the combination $2(\bar\beta+\gamma)$ is obtained in decay channels $B_d\to\pi\pi, \rho\pi,\rho\rho$.} and furthermore, if New Physics contributes to $\Delta B=2$ transitions, it can manifest through non-standard values of the mass difference $\DMBd$ or the mixing phase $2\bar\beta$, which are automatically incorporated into \eq{eq:G12d:01}; the remaining quantities are, in terms of weak interactions, \emph{tree level}, hence a priori safe from potential contributions from New Physics.
Analogous expressions for the \BBsmix\ case can be readily obtained:
\begin{multline}
\frac{\Gmixs}{\Mmixs}= \frac{2}{\DMBs}
\Big[c\,e^{i2\bar\beta_s}\left(|\V{cb}\Vc{cs}|+|\V{ub}\Vc{us}|e^{-i\gamma}\right)^2\\
-a\,|\V{ub}\Vc{us}|e^{i(2\bar\beta_s-\gamma)}\left(|\V{cb}\Vc{cs}|+|\V{ub}\Vc{us}|e^{-i\gamma}\right)\\
+b\,|\V{ub}\Vc{us}|^2e^{i\,2(\bar\beta_s-\gamma)}
\Big]\ .\label{eq:G12s:01}
\end{multline}
For the \BBsmix\ system, the ``golden'' decay channel is $B_s\to J/\Psi \Phi$ and the corresponding time-dependent CP asymmetry is $\AJPP\equiv \sin \bar\beta_s$.\\ 
The $(m_c/m_b)^2$ suppression of $\asld$ and $\asls$ within the SM manifests itself in \eq{eq:G12d:01} and in \eq{eq:G12s:01} through the unitarity relations
\begin{multline}
|\V{cb}\Vc{cd}|-|\V{ub}\Vc{ud}|e^{-i\gamma}=|\V{tb}\Vc{td}|e^{-i\beta}\quad \text{and}\\
 |\V{cb}\Vc{cs}|+|\V{ub}\Vc{us}|e^{-i\gamma}=-|\V{tb}\Vc{ts}|e^{-i\beta_s}\,.
\end{multline}
Following the previous discussion of the semileptonic asymmetries $\asld$ and $\asls$, it is easy to grasp how dramatically the dimuon asymmetry observed by D0 \cite{Abazov:2011yk, *Abazov:2013uma} \emph{cannot} be obtained within the SM. The dimuon asymmetry $\aslb$ is essentially a weighted combination of $\asld$ and $\asls$,
\begin{equation}
\aslb=\frac{\asld+g\asls}{1+g}\label{eq:SM:AbSL:01}
\end{equation}
where 
\begin{multline}
g=f\,\frac{\Gamma_d}{\Gamma_s}\frac{(1-y_s^2)^{-1}-(1+x_s^2)^{-1}}{(1-y_d^2)^{-1}-(1+x_d^2)^{-1}}\,,\\
y_q=\frac{\DGq}{2\Gamma_q},\,x_q=\frac{\DMBq}{\Gamma_q}\,.\label{eq:SM:AbSL:02}
\end{multline}
The $B_s$--$B_d$ fragmentation fraction ratio in the $B$ sample is $f=0.269\pm 0.015$. Numerically $g\sim 1$ and thus $\asld$ and $\asls$ have similar weights in \eq{eq:SM:AbSL:01}.\\
With $\asls\sim 2\cdot 10^{-5}$, $\aslb$ is dominated by $\asld\sim -5\cdot 10^{-4}$ and the SM expectation turns out to be
\begin{equation}
\aslb= (-2.40\pm 0.45)\cdot 10^{-4}\,.\label{eq:SM:AbSL:03}
\end{equation}
The values quoted by the D0 collaboration are $\aslb=(-7.87\pm 1.72\pm 0.93)\cdot 10^{-3}$ and $\aslb=(-4.96\pm 1.53\pm 0.72)\cdot 10^{-3}$ in \cite{Abazov:2011yk,*Abazov:2013uma}, so the disagreement with SM expectations, as anticipated, is around the $3\sigma$ level! It is important to stress again that the almost $\mathcal O(10^{-2})$ scale of that measured value is \emph{twenty} times larger than SM expectations.\\
On the other hand, the LHCb collaboration has recently started to measure the semileptonic asymmetry $\asls$ \cite{Aaij:2013gta}; additional results concerning $\asls\pm\asld$ are also expected \cite{Bird:Discrete2012}. If the D0 measurement is to be interpreted as a clear signal of New Physics, LHCb results, in particular $\asls\pm\asld$, should necessarily depart from the SM expectations
\begin{eqnarray}
\asls+\asld & = &(-4.0\pm 0.7)\cdot 10^{-4}\,,\nonumber\\
\asls-\asld & = &(4.1\pm 0.7)\cdot 10^{-4}\,.\label{eq:SM:ASL:01}
\end{eqnarray}
In this section we have analysed the details of the SM expectations for observables genuinely related to mixing in neutral $B_d$ and $B_s$ systems, including the ``problematic'' asymmetry measured at Tevatron. Those are, in any case, well known results, but analysing them in detail, in particular how the use of $3\times 3$ unitarity and the dominance of the top quark contribution in $\Mmixq$ are central in the SM suppressed expectation, paves the way to understand the changes to the picture which one encounters when moving beyond the SM.
\newpage
\section{New Physics within $\mathbf{3\times 3}$ unitarity \label{SEC:3x3NP}}
\begin{figure*}[htb!]
\begin{center}
\subfigure[ $\DC$ vs. $\asld$.\label{fig:UnNP:01a}]{\includegraphics[width=0.38\textwidth]{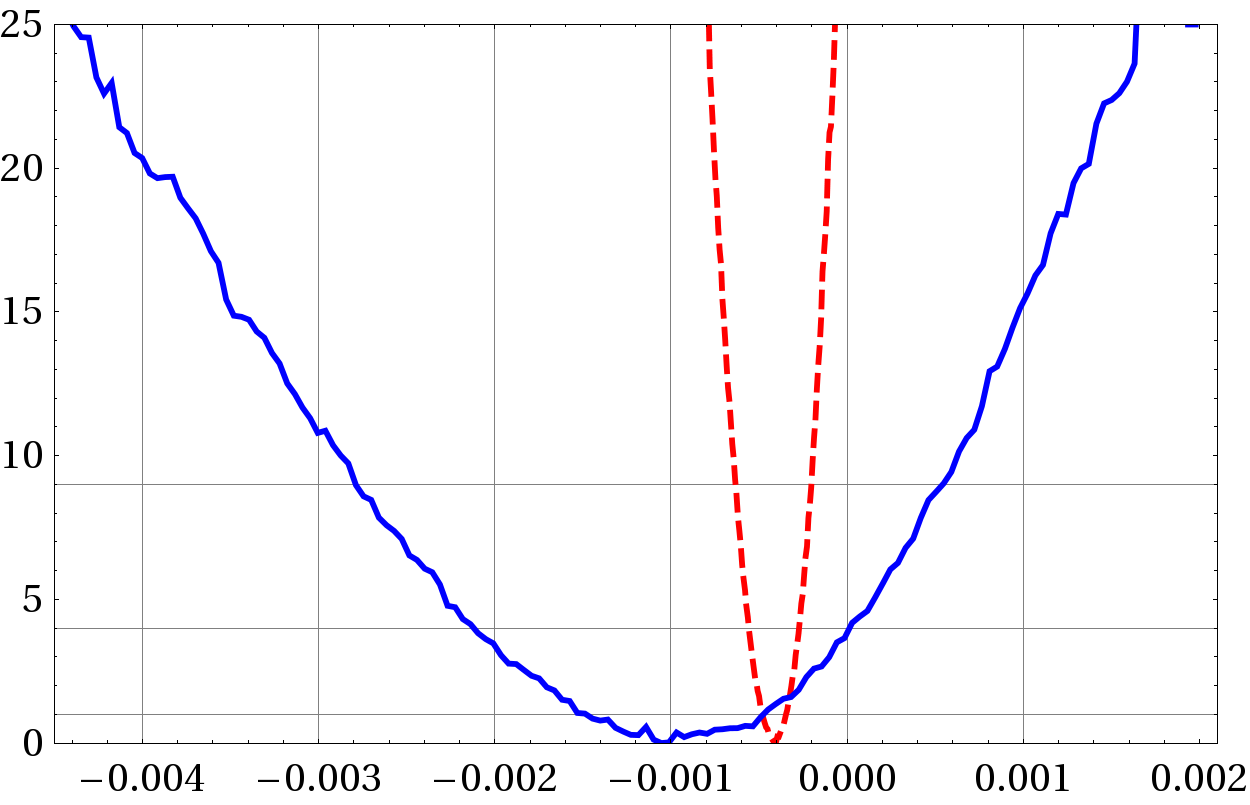}}\qquad
\subfigure[ $\DC$ vs. $\asls$.\label{fig:UnNP:01b}]{\includegraphics[width=0.365\textwidth]{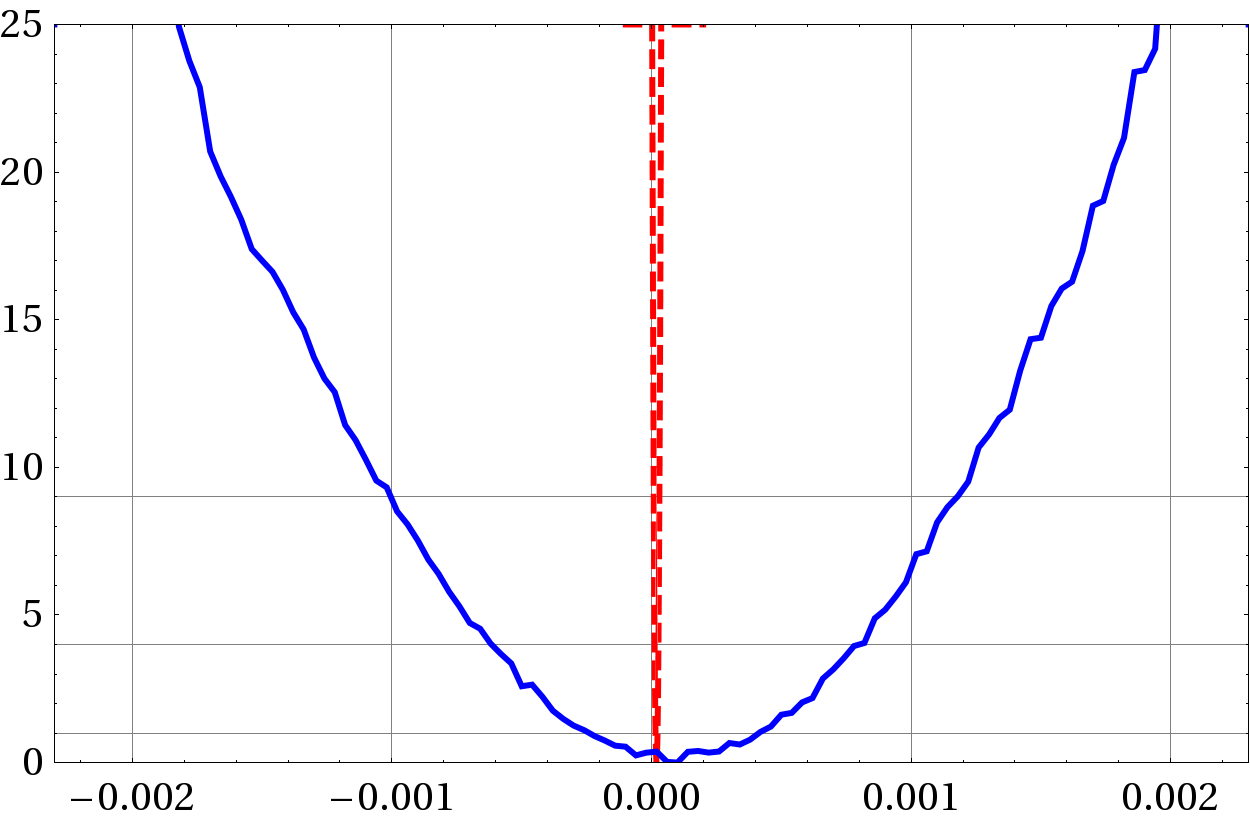}}
\caption{$\DC$ profile of the semileptonic asymmetries $\asld$ and $\asls$; the {\color{blue} blue} line corresponds to the NP scenario -- \eq{eq:33NP:Mix01} --, the {\color{red} red dashed} line corresponds to the SM case. Notice that for $\asls$ the SM range is too narrow to be resolved on this scale.\label{fig:UnNP:01}}
\end{center}
\end{figure*}

\begin{figure*}[th!]
\begin{center}
\subfigure[$\asld$ vs. $2\phi_d$.\label{fig:UnNP:03a}]{\includegraphics[width=0.39\textwidth]{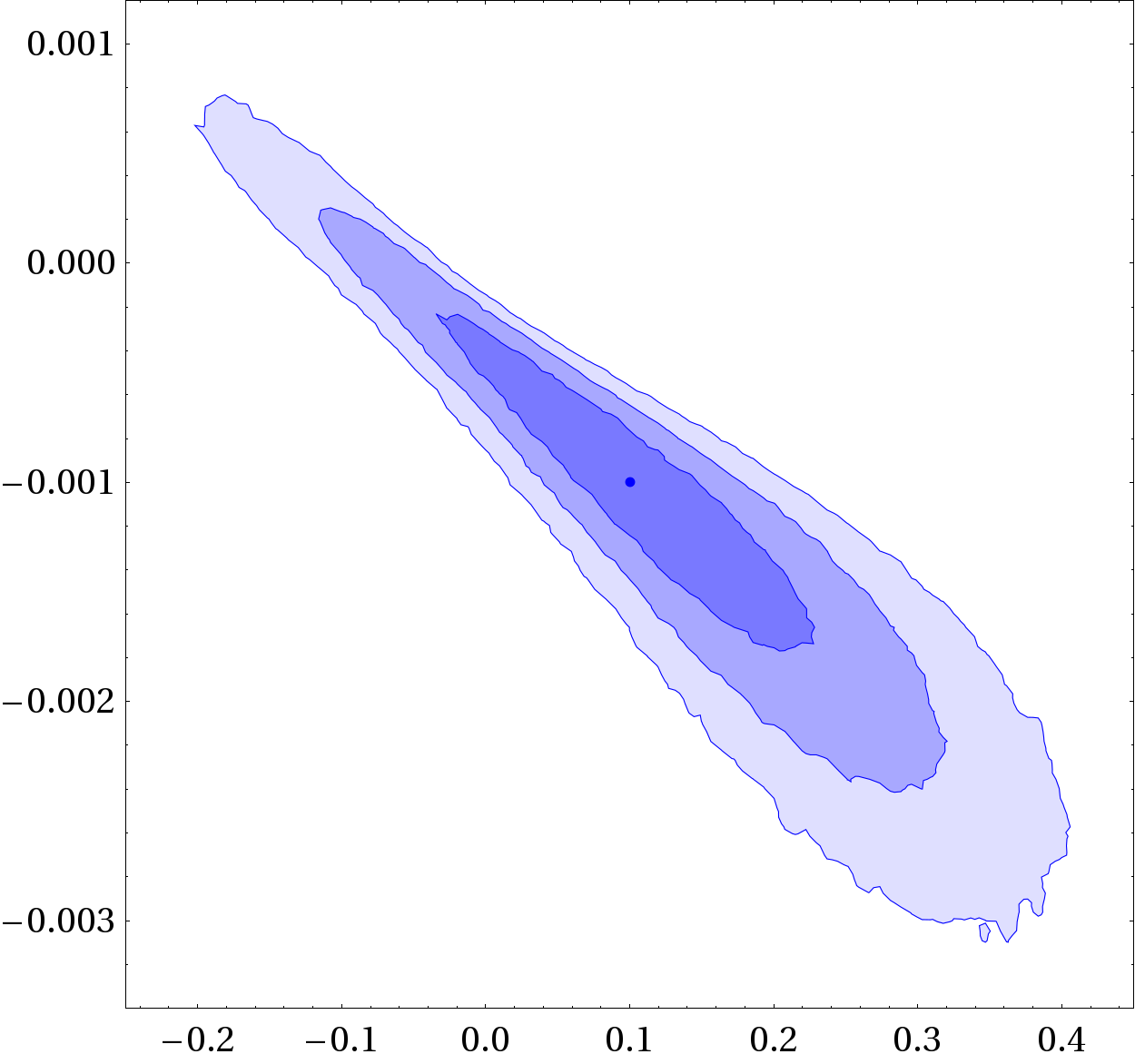}}\qquad
\subfigure[$\asls$ vs. $2\phi_s$.\label{fig:UnNP:03d}]{\includegraphics[width=0.39\textwidth]{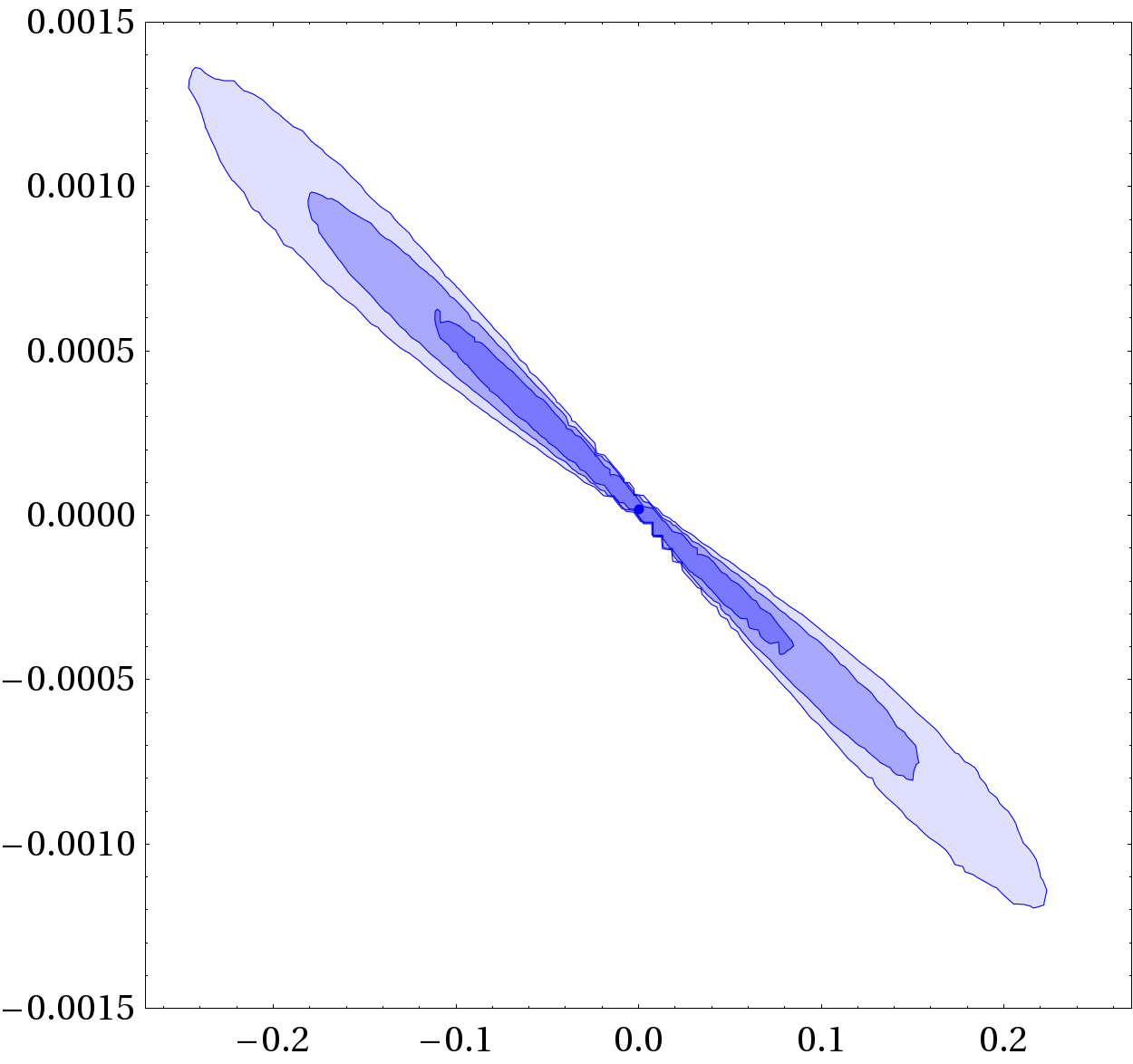}}
\caption{$\DC$ profiles of $\aslq$ vs. $2\phi_q$; 68\%, 95\% and 99\% CL regions are shown.\label{fig:UnNP:03}}
\end{center}
\end{figure*}

In order to move beyond the SM, a general model independent analysis of NP in the flavour sector could start by considering the effective Hamiltonians describing a set of relevant weak transitions. Model independence would be achieved by allowing all independent Wilson coefficients to depart from SM values. 
This task would not only be daunting, but would also be very difficult to extract meaningful information from it, since the NP parameters controlling the generalized Wilson coefficients would typically show a high degree of degeneracy (not to mention the experimental accuracy required to single out any interesting feature in such a scenario). 
One can consider simpler, yet interesting alternatives, by focusing on a few relevant operators that enter multiple observables. In addition, since the SM flavour picture is essentially correct, it is legitimate to circumscribe NP deviations to the Wilson coefficients of operators that do not arise at tree level in the SM. An approach that has been rather popular in recent times \cite{Botella:2006va,Laplace:2002ik,*Ligeti:2006pm,*Ball:2006xx,*Grossman:2006ce,*Bona:2006sa,*Botella:2005fc,*Bona:2007vi,*Lenz:2012az} focuses on meson mixings.
Since the dispersive amplitudes $\Mmixq$ arise at the loop level in the SM, they are appropriate candidates to be polluted by New Physics; the effect of having new contributions may be parameterised in the following manner (with $r_q$, $\phi_q$, new, independent parameters):
\begin{equation}
\Mmixq = \left[\Mmixq\right]_{\text{SM}}\,r_q^2\,e^{-i\,2\phi_q}\ .\label{eq:33NP:Mix01}
\end{equation}
Deviations from $(r_q,\phi_d)=(1,0)$ describe New Physics in the mixing of $B_q$ mesons. Equation (\ref{eq:33NP:Mix01}) has several advantages. Since $2|\Mmixq|=\DMBq$, one concludes that:
\begin{itemize}
\item the prediction for $\DMBq$ is directly modified, with respect to the SM, by $r_q$, but unaffected by $\phi_q$,
\item $\phi_q$ directly modifies, in a common way, the SM prediction for observables where the phase of the (dispersive) mixing enters: this is the case, for example, of the ``golden'' time-dependent CP asymmetries in $B_d^0\to J/\Psi K_S$ and $B_s^0\to J/\Psi\Phi$; these are insensitive to $r_d$ and $r_s$.
\end{itemize}
This ``factorization'' of the effects of New Physics is quite convenient\footnote{Another popular alternative uses the NP parameters $h_q$ and $\sigma_q$ with $r_q^2e^{-i2\phi_q}\equiv 1+h_qe^{i2\sigma_q}$, where this separation of NP is less straightforward. Our results, in any case, do not depend on adopting one parametrization or the other.}.\\ 
In the SM, $\Mmixq$ is controlled by a single dimension six $\Delta B=2$ effective operator; in \eq{eq:33NP:Mix01}, $r_q^2\,e^{-i\,2\phi_q}$ can be interpreted as the factor modifying the corresponding Wilson coefficient in the presence of NP. However, notice that for specific models where $r_d$, $r_s$, $\phi_d$ and $\phi_s$ are not independent, the situation may be more involved.\\
The very first question one should address when considering \eq{eq:33NP:Mix01} is whether this kind of modification could bring something \emph{really} new to the SM picture presented in section \ref{SEC:mix}. Naively, one would expect an affirmative answer. Nevertheless, since (i) $\Gmixd$ and $\Gmixs$ are tree level dominated and (ii) $\Mmixd$ and $\Mmixs$ are constrained by experimental information\footnote{In fact, \emph{tightly} constrained, except for the argument of $\Mmixs$, accessed through $B_s\to J/\Psi\Phi$, where, despite the excellent performance of LHCb, the smallness of the SM expectation still allows for deviations.}, modifying the picture may not be so straightforward. First of all, through \eq{eq:33NP:Mix01}, the predictions for the most important observables are modified according to
\begin{align}
&\DMBd = r_d^2\,|\V{tb}\Vc{td}|^2\,\frac{G_F^2M_W^2}{6\pi^2}\,M_{B_d}\,f_{B_d}^2\,B_{B_d}\,\eta_B\,S_0(x_t)\,,\label{eq:DMBd:01}\\
&\AJPKs = \sin(2\bar\beta)=\sin(2(\beta-\phi_d))\,,\label{eq:AJPKs:01}\\
&\DMBs = r_s^2\,|\V{tb}\Vc{td}|^2\,\frac{G_F^2M_W^2}{6\pi^2}\,M_{B_s}\,f_{B_s}^2\,B_{B_s}\,\eta_B\,S_0(x_t)\,,\label{eq:DMBs:01}\\
&\AJPP =\sin(2\bar\beta_s)=\sin(2(\beta_s+\phi_s))\,,\label{eq:AJPP:01}
\end{align}
where we have introduced for convenience the effective phases $\bar\beta\equiv \beta-\phi_d$ and $\bar\beta_s=\beta_s+\phi_s$. Considering equations \refeq{eq:DMBd:01} to \refeq{eq:AJPP:01}, our skepticism takes a precise form: if the functional form of $\Gmixq$ does not change, while $\Mmixq$ obey the same experimental constraints, how could $\Gmixq/\Mmixq$ differ from SM expectations?\\
\noindent Figure \ref{fig:UnNP:01} displays the individual $\DC$ profiles of $\asld$ and $\asls$ corresponding to this NP scenario together with the SM ones (obtained using the same experimental contraints, the ones in table \ref{AP:tab:data}): there is little doubt that the simple modification in \eq{eq:33NP:Mix01} allows for ample deviations from SM expectations.
Consider for example the \BBdmix\ system. With the following values,
\begin{align}
&2\phi_d\simeq 0.20\,,\ \beta\simeq 0.47\,,\ \gamma\simeq 1.22\,,\ 2(\beta+\gamma)\simeq 3.36\,,\label{eq:uNPpha:d:01}\\
&|\Vc{ud}\V{ub}|\simeq 4.22\cdot 10^{-3}\,,\ |\Vc{cd}\V{cb}|\simeq 9.22\cdot 10^{-3}\,,\label{eq:uNPmod:d:01}
\end{align}
one obtains
\begin{equation}
\DGd= 3.25\cdot 10^{-3}\,\text{ps}^{-1}\ \ \text{and}\quad \asld=-1.92\cdot 10^{-3}\,.\label{eq:uNPpred:d:01}
\end{equation}
While $\DGd$ is rather unchanged, the departure of $\asld$ from the value in \eq{eq:SMpred:s:01} is quite significant: it is larger by a factor of \emph{five}. How such a large enhancement could be achieved? The main differences between the values in eqs.\refeq{eq:uNPmod:d:01},\refeq{eq:uNPpha:d:01} and the ones in the SM case, $\beta\simeq 0.38$, $\gamma\simeq 1.18$, $|\Vc{ud}\V{ub}|\simeq 3.46\cdot 10^{-3}$ and $|\Vc{cd}\V{cb}|\simeq 9.23\cdot 10^{-3}$, are in $|\V{ub}|$ and $\beta$. 
In the SM, $3\times 3$ unitarity of the CKM matrix, nicely illustrated by the usual $bd$ unitarity triangle, forces $|\V{ub}|$ to be tightly related to $\beta$. This is indeed the cornerstone of the so called \emph{tensions} in the $bd$ sector \cite{Bona:2009cj,*Lunghi:2010gv}. In this extended scenario, the situation is changed. While $|\V{ub}|$ is still directly obtained, and it may call for values larger than in the SM fit, the measurement of $\AJPKs$ fixes $\bar \beta$ instead of $\beta$. The new parameter $\phi_d$ breaks the SM tight relation between $\AJPKs$ and $|\V{ub}|$ imposed by $3\times 3$ unitarity and the dominance of the top quark contribution in $\Mmixd$. This is sufficient to remove the $(m_c/m_b)^2$ suppression present in the imaginary part of ${\Gmixd}/{\Mmixd}$. We can read in figure \ref{fig:UnNP:01a} how far from SM expectations could $\asld$ be pushed: at 95\% CL (that is, up to $\DC=4$), $\asld\in[-3.3;-0.8]\cdot 10^{-3}$, to be compared with the SM 95\% CL range $[-0.58;-0.29]\cdot 10^{-3}$. It is important to stress that although the presence of NP induces a departure in the phase of $\Mmixd$ and in $|\V{ub}|$ with respect to SM values at the 20-30\% level, $\asld$ is enhanced by a factor 4-5: it is a priori highly sensitive to the presence of NP in $\Mmixd$ precisely because of the natural suppression present in the SM. 
The increase in $|\V{ub}|$ allowed by $\phi_d\neq 0$ may also enhance $\asls$ marginally -- through the second, $a$-dependent, term in \eq{eq:G12q:03} --, however, the main source of potential deviation from SM expectations is simply $\phi_s\neq 0$: as illustrated in Fig. \ref{fig:UnNP:01b}, $\asls$ can be lifted from values $\mathcal O(10^{-5})$ to values $\mathcal O(10^{-3})$.\\ 
One can indeed express \cite{Laplace:2002ik} the asymmetry $\aslq$ in this scenario as 
\begin{multline}
\frac{\Gmixq}{\Mmixq}=\frac{e^{i2\phi_q}}{r_q^2}\left[\frac{\Gmixq}{\Mmixq}\right]_{\textrm{SM}}\Rightarrow\\
 \aslq=\frac{\cos(2\phi_q)}{r_q^2}[\aslq]_{\textrm{SM}}-\frac{\sin(2\phi_q)}{r_q^2}\left[\frac{\DGq}{\Delta M_{q}}\right]_{\textrm{SM}}\ ,\label{eq:aslq_with_phiq}
\end{multline}
from which we can expect enhancements up to the $10^{-3}$ level in both \BBdmix\ and \BBsmix\ systems for phases $\phi_q\sim 0.1$.\\ 
\noindent It is also interesting to represent the two dimensional $\DC$ profiles of $\aslq$ vs. $\phi_q$: in figures \ref{fig:UnNP:03a} and \ref{fig:UnNP:03d} we plot\footnote{For these and successive two dimensional $\DC$ profiles, we display, for clarity, 68\%, 95\% and 99\% CL regions.} $\asld$ vs. $2\phi_d$ and $\asls$ vs. $2\phi_s$, respectively. 

\begin{figure*}[h]
\begin{center}
\includegraphics[width=0.4\textwidth]{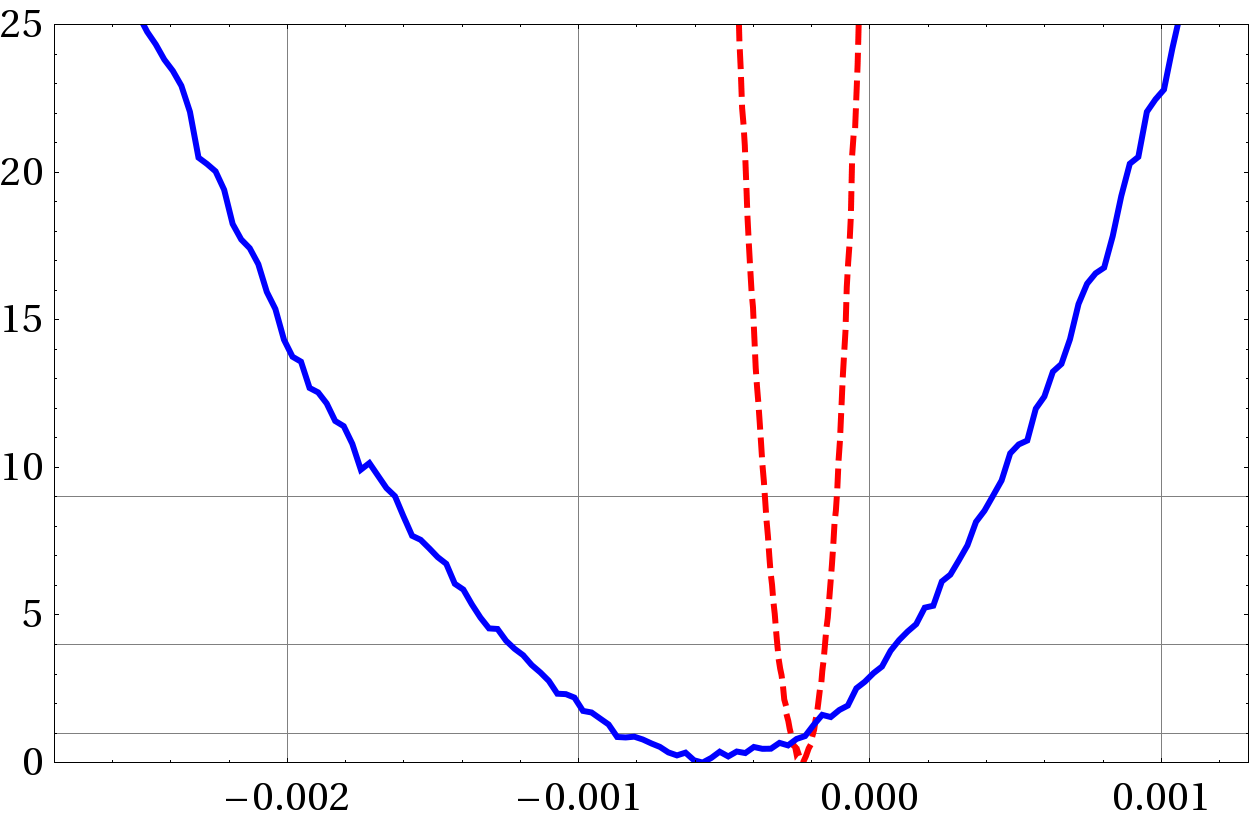}
\caption{$\DC$ profile of $\aslb$; the {\color{blue} blue} line corresponds to the NP scenario, the {\color{red} red dashed} line corresponds to the SM case. The last D0 measurements gives $\aslb=(-4.96\pm 1.69)\cdot 10^{-3}$ \cite{Abazov:2013uma}.\label{fig:UnNP:04}}
\end{center}
\end{figure*}

\begin{figure*}[h]
\begin{center}
\subfigure[$\DC$ vs. $\asls+\asld$.\label{fig:UnNP:05a}]{\includegraphics[width=0.4\textwidth]{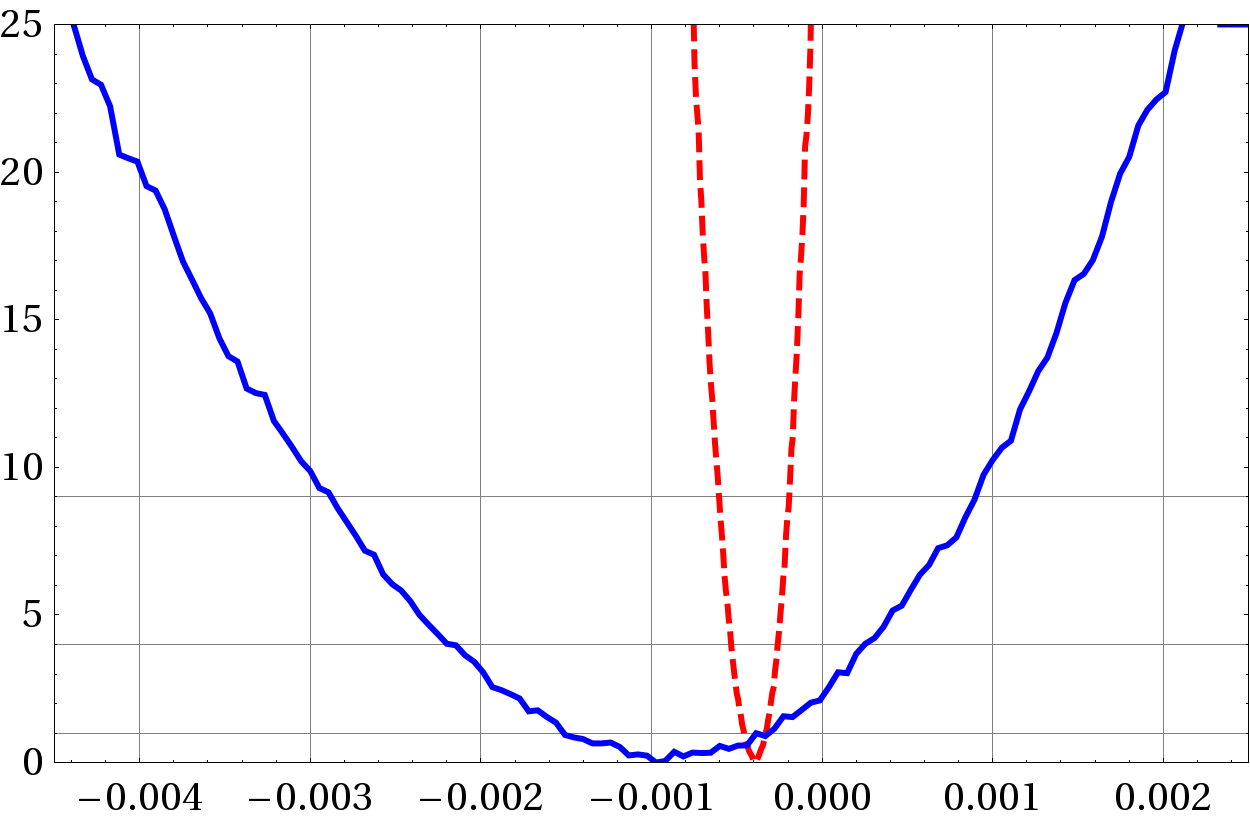}}\qquad
\subfigure[$\DC$ vs. $\asls-\asld$.\label{fig:UnNP:05b}]{\includegraphics[width=0.41\textwidth]{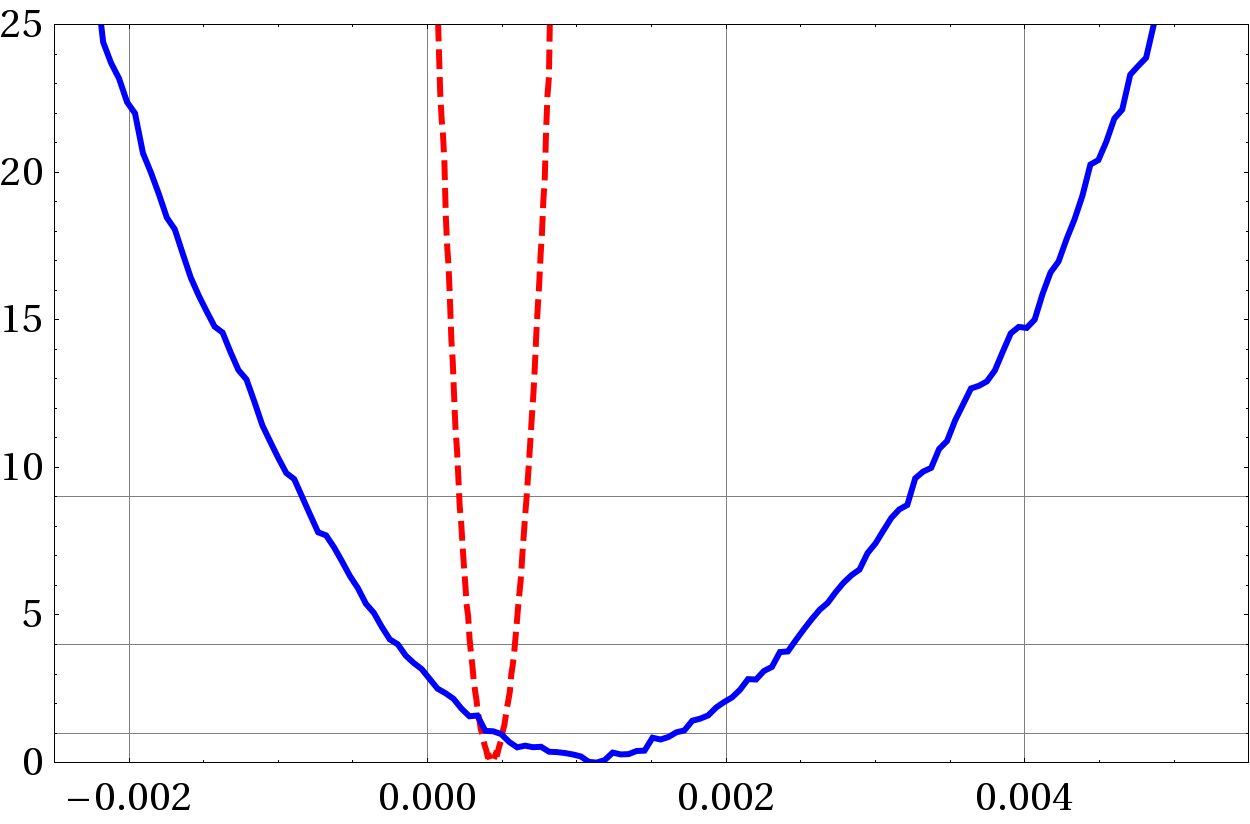}}
\caption{$\DC$ profile of the combinations of semileptonic asymmetries $\asls\pm\asld$; the {\color{blue} blue} lines correspond to the NP scenario -- \eq{eq:33NP:Mix01} --, the {\color{red} red dashed} lines correspond to the SM case.\label{fig:UnNP:05}}
\end{center}
\end{figure*}

\noindent Figure \ref{fig:UnNP:03a} shows how the $\asld$ departure from SM expectations relies on $2\phi_d\neq 0$. How $2\phi_s\neq 0$ can produce $\mathcal O(10^{-3})$ values for $\asls$ is clearly reflected in fig. \ref{fig:UnNP:03d}.\\
The previous analysis provides a clear picture of the deviations from SM expectations for the individual asymmetries $\asld$ and $\asls$. Turning to the D0 asymmetry $\aslb$, figure \ref{fig:UnNP:04} shows the corresponding $\DC$ profile. Within $3\sigma$ it may reach values of $-2.5\cdot 10^{-3}$; this means an enhancement of almost an order magnitude with respect to the SM expectation in \eq{eq:SM:AbSL:01}. However, even if this enhancement softens the disagreement with the experimental value of $\aslb$, that central value is out of the ranges that this New Physics scenario can accommodate. 
\noindent For completeness we display in figure \ref{fig:UnNP:05} the $\DC$ profiles for the combinations $\asls\pm\asld$, of interest for the LHCb experiment. 
\noindent The potential enhancement with respect to SM expectations in \eq{eq:SM:ASL:01} is, once again, noticeable:
\begin{eqnarray*}
 \asls + \asld &=& (-1.0\pm 0.6)\cdot 10^{-3}\,,\\
 \asls - \asld &=& (1.0\pm 0.7)\cdot 10^{-3}\, .
\end{eqnarray*}

\begin{figure*}[h!]
\begin{center}
\subfigure[$\asld$ vs. $|\V{ub}|$.\label{fig:UnNP:06a}]{\includegraphics[width=0.38\textwidth]{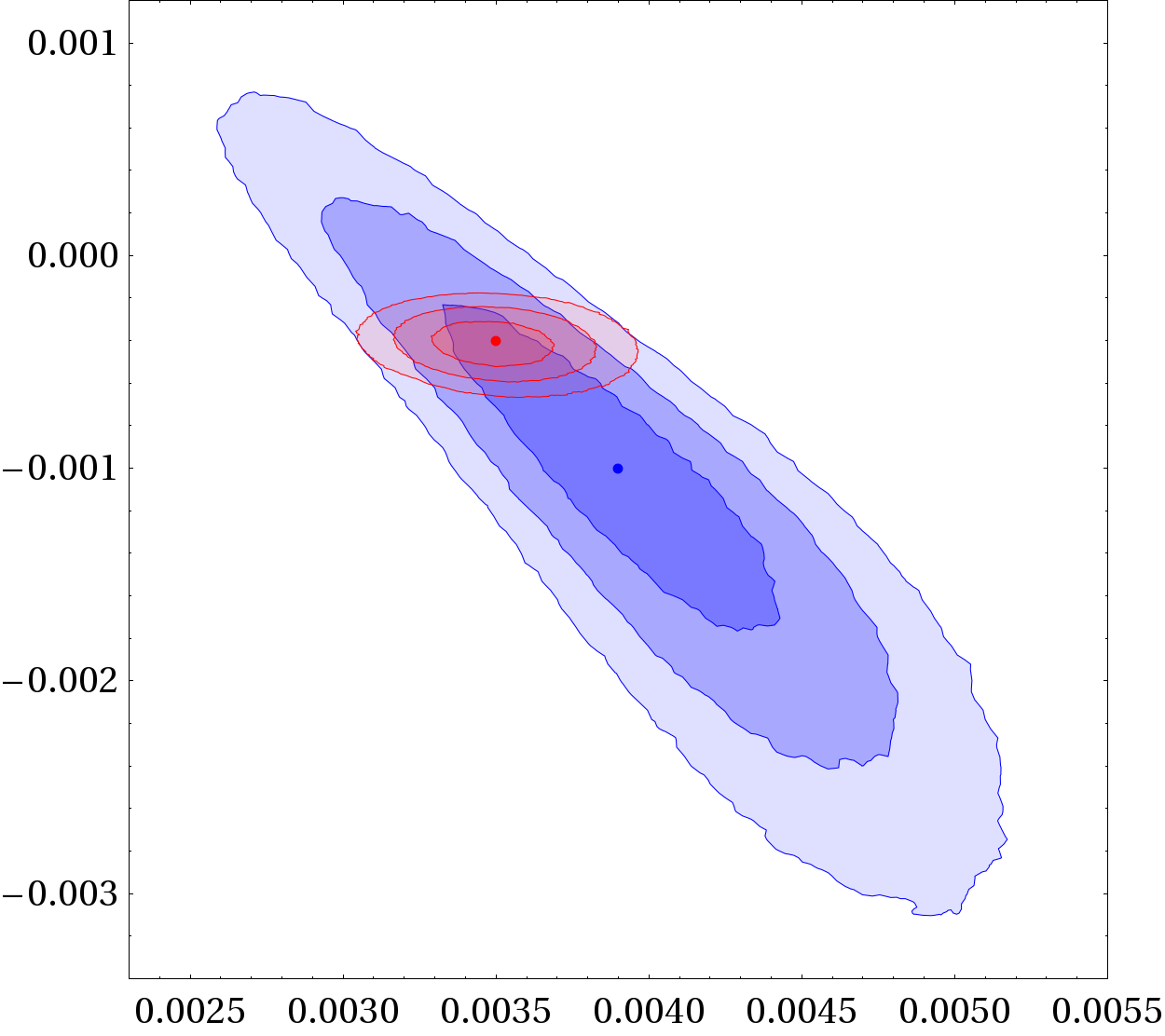}}\qquad
\subfigure[$\asls$ vs. $\AJPP$.\label{fig:UnNP:06b}]{\includegraphics[width=0.385\textwidth]{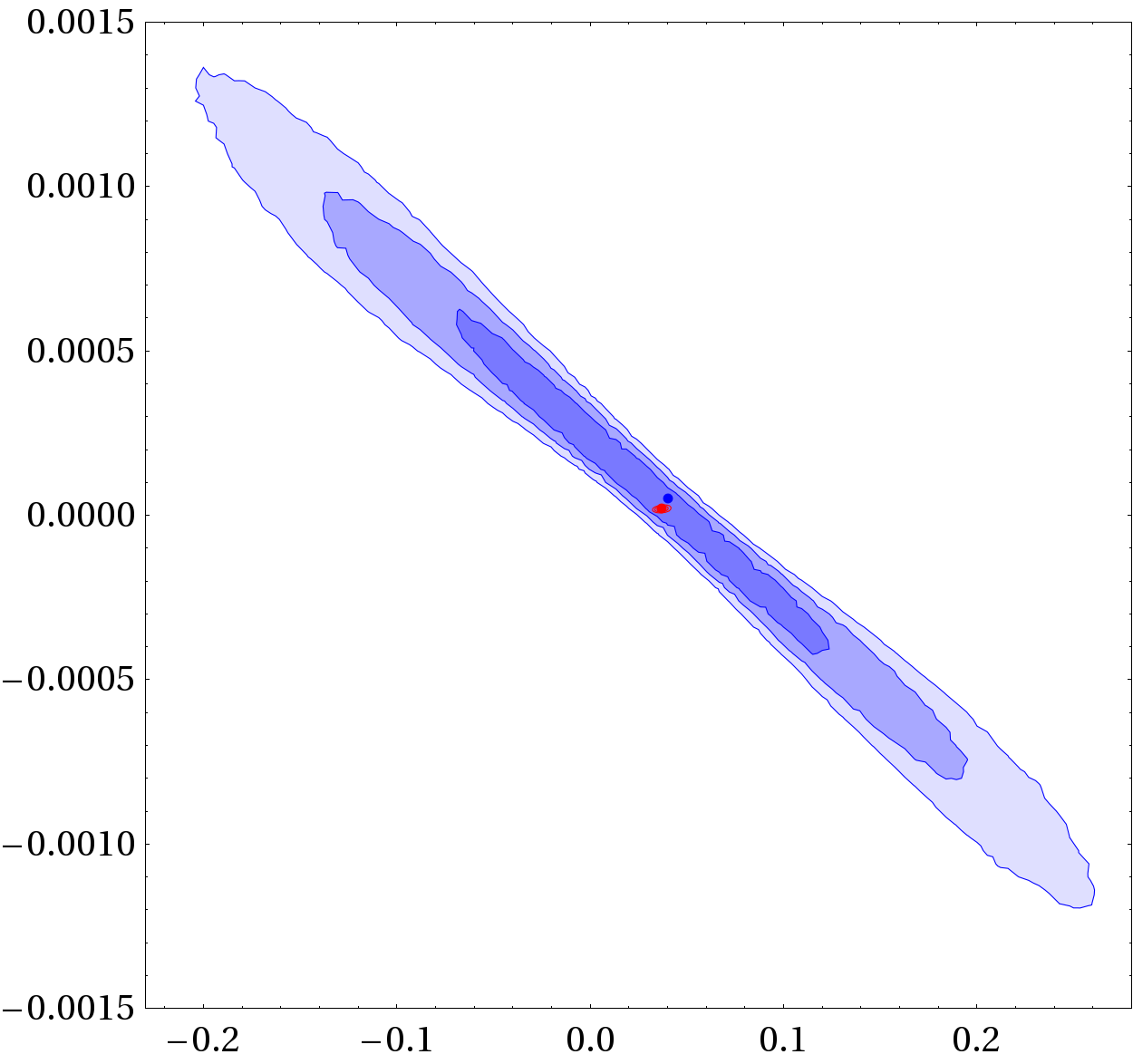}}\\
\subfigure[$\aslb$ vs. $|\V{ub}|$.\label{fig:UnNP:07a}]{\includegraphics[width=0.38\textwidth]{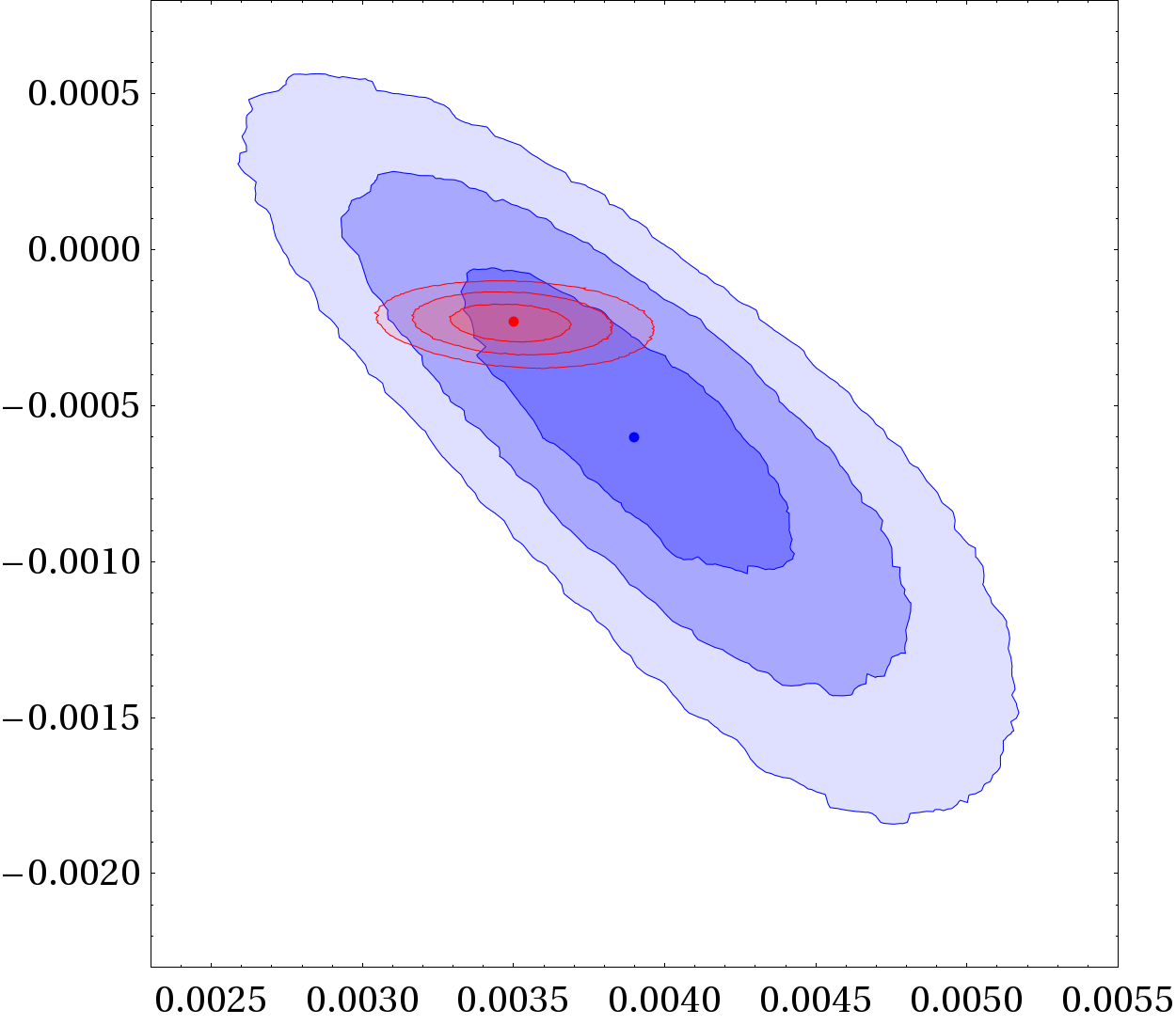}}\qquad
\subfigure[$\aslb$ vs. $\AJPP$.\label{fig:UnNP:07b}]{\includegraphics[width=0.38\textwidth]{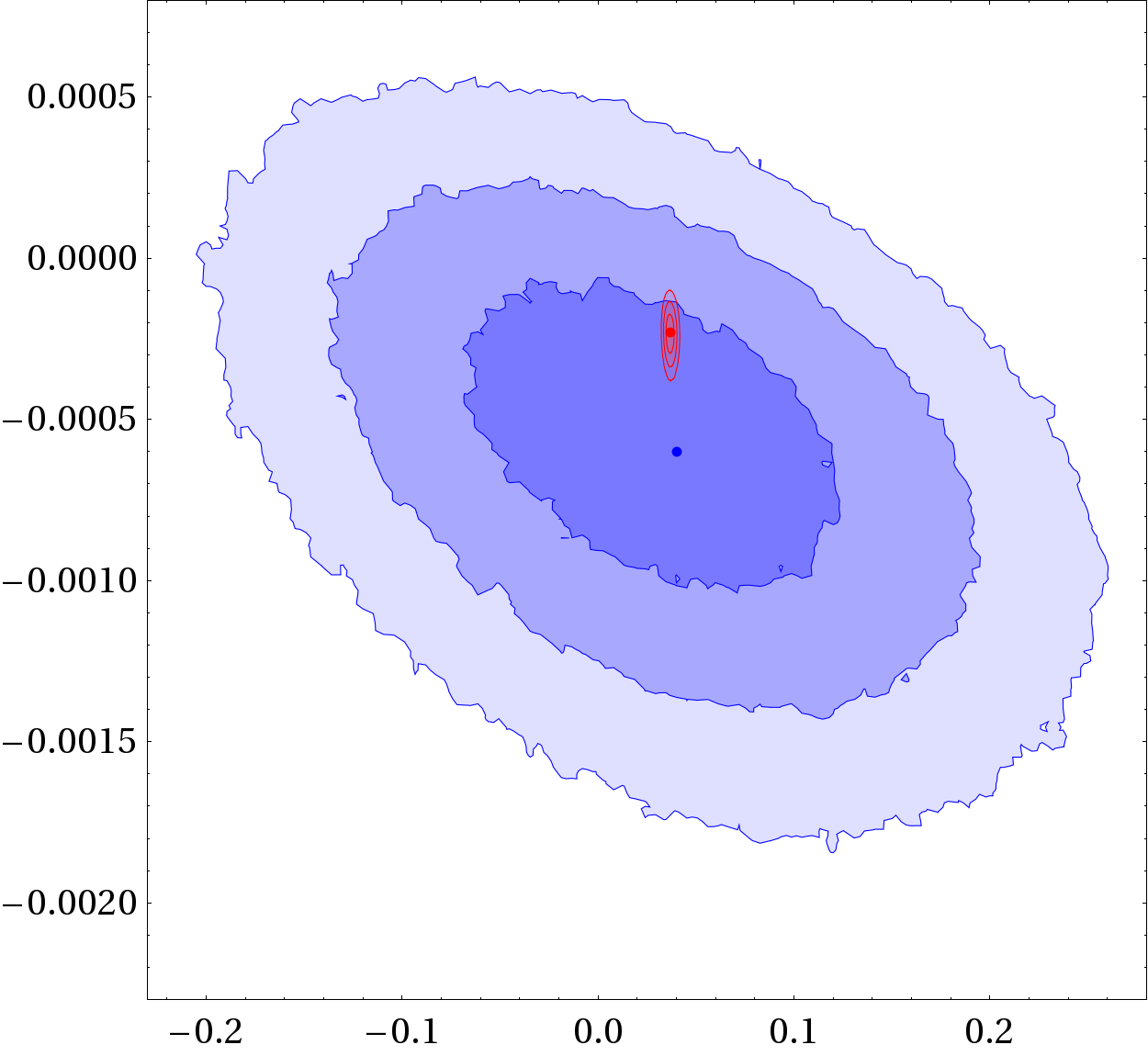}}\\
\caption{$\DC$ 68\%, 95\% and 99\% CL regions. {\color{blue} Blue} regions correspond to the NP scenario, {\color{red} red} regions correspond to the SM case. Notice that with the scales in fig. \ref{fig:UnNP:06b}, the SM region is barely a point.\label{fig:UnNP:06}}
\end{center}
\end{figure*}

\noindent It should be stressed that deviating from SM expectations in $\asld$ and in $\asls$ is intimately related to NP effects in other observables. For $\asld$, large values are associated to ``tensions'' in $bd$ that manifest, for example, through larger than standard values of $|\V{ub}|$. This is illustrated through the correlated $\DC$ profiles of $\asld$ vs. $|\V{ub}|$ shown in figure \ref{fig:UnNP:06a}. On the other hand, for $\asls$, large values of $\asls$ are associated to large values of the CP asymmetry $\AJPP$, as figure \ref{fig:UnNP:06b} confirms (and could be anticipated from fig. \ref{fig:UnNP:03d}). The dimuon asymmetry $\aslb$ is sensitive to both correlations, as figures \ref{fig:UnNP:07a} and \ref{fig:UnNP:07b} illustrate.

Along this section we have analysed in detail how the introduction of New Physics in the mixings $\Mmixq$ allows for significant deviations from SM expectations in the semileptonic asymmetries $\aslq$. The key point at the origin of those deviations is the effect of the phases $\phi_q$:
(1) within the SM, $3\times 3$ unitarity and the top quark dominance of $\Mmixq$ together, enforce a natural suppression of $\asld$ and $\asls$;
(2) the presence of $\phi_q\neq 0$ misaligns the phases of the would-be leading contribution to $\Gmixq$ (the one not suppressed by $m_c/m_b$) and $\Mmixq$; 
(3a) in the \BBsmix\ system, since $[\AJPP]_{\textrm{SM}}\sim\mathcal O(10^{-2})$, but the experimental sensitivity has just started to explore that ground, there is still ample room for $\phi_s\neq 0$, and thus $\asls\sim \mathcal O(10^{-3})$ can be achieved; 
(3b) in the \BBdmix\ system, $\AJPKs\simeq 0.68$ has been measured to a few percent precision; in addition, unitarity imposes a close relation between $|\V{ub}|$ and $\beta$ that is transmitted, within the SM, to $\AJPKs$. Having $\phi_d\neq 0$ requires, necessarily, that both $|\V{ub}|$ and $\beta$ deviate from their SM values while $\AJPKs$ remains unchanged. The presence of ``tensions'' between the $|\V{ub}|$ and the $\AJPKs$ measurements favors, in this simple NP scenario,  $\phi_d\neq 0$, thus evading the SM suppression and obtaining a significant enhancement of $\asld$.
New Physics at the 20-30\% level in $\Mmixd$ does not give a 20-30\% modification in $\asld$, it gives a much larger effect, contrary to what one can naively expect \cite{Hou:2007ps}.
It should be stressed that, despite the significant increase with respect to the SM, the values that can be reached for $\asld$ and $\asls$ are too small to reproduce the D0 value of the $\aslb$ asymmetry.
\clearpage
\section{New Physics beyond $\mathbf{3\times 3}$ unitarity \label{SEC:no3x3NP}}
\begin{figure*}[hb!]
\begin{center}
\subfigure[$\DC$ vs. $\asld$.\label{fig:NoUnNP:01a}]{\includegraphics[width=0.4\textwidth]{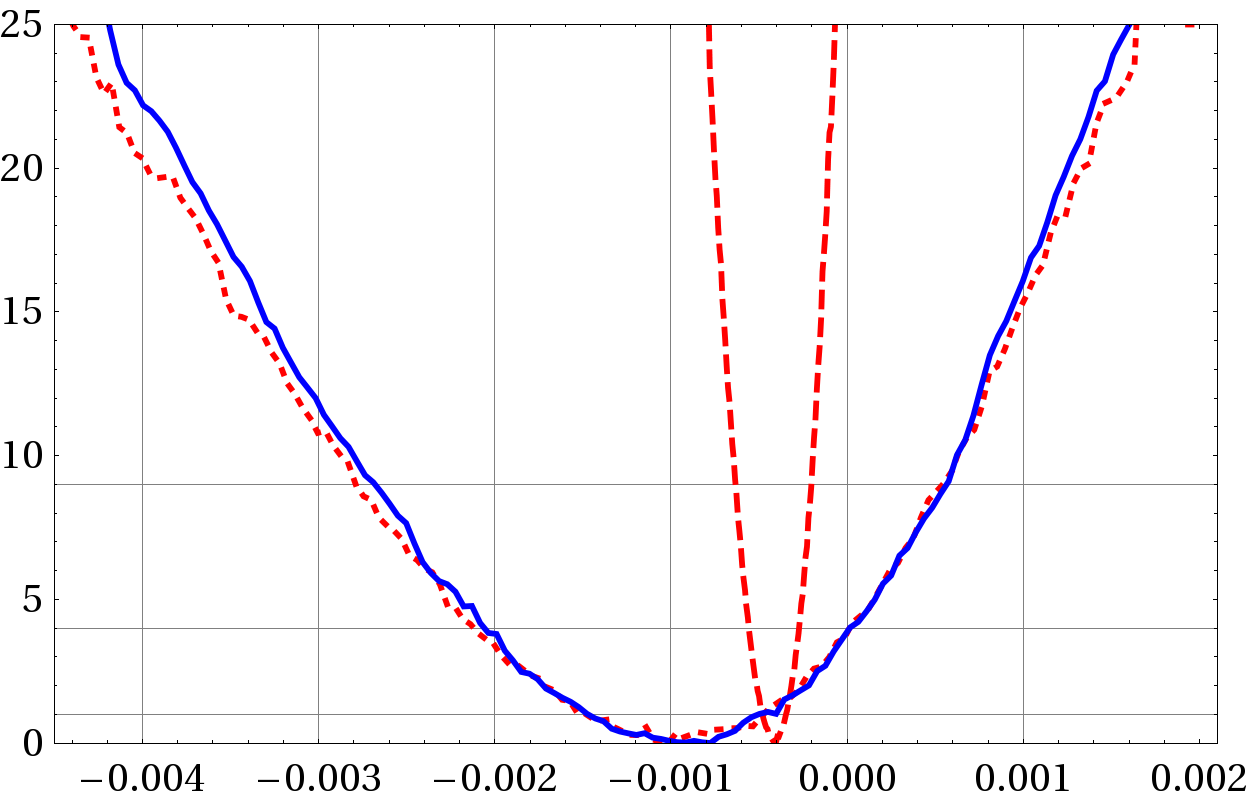}}\qquad
\subfigure[$\DC$ vs. $\asls$.\label{fig:NoUnNP:01b}]{\includegraphics[width=0.39\textwidth]{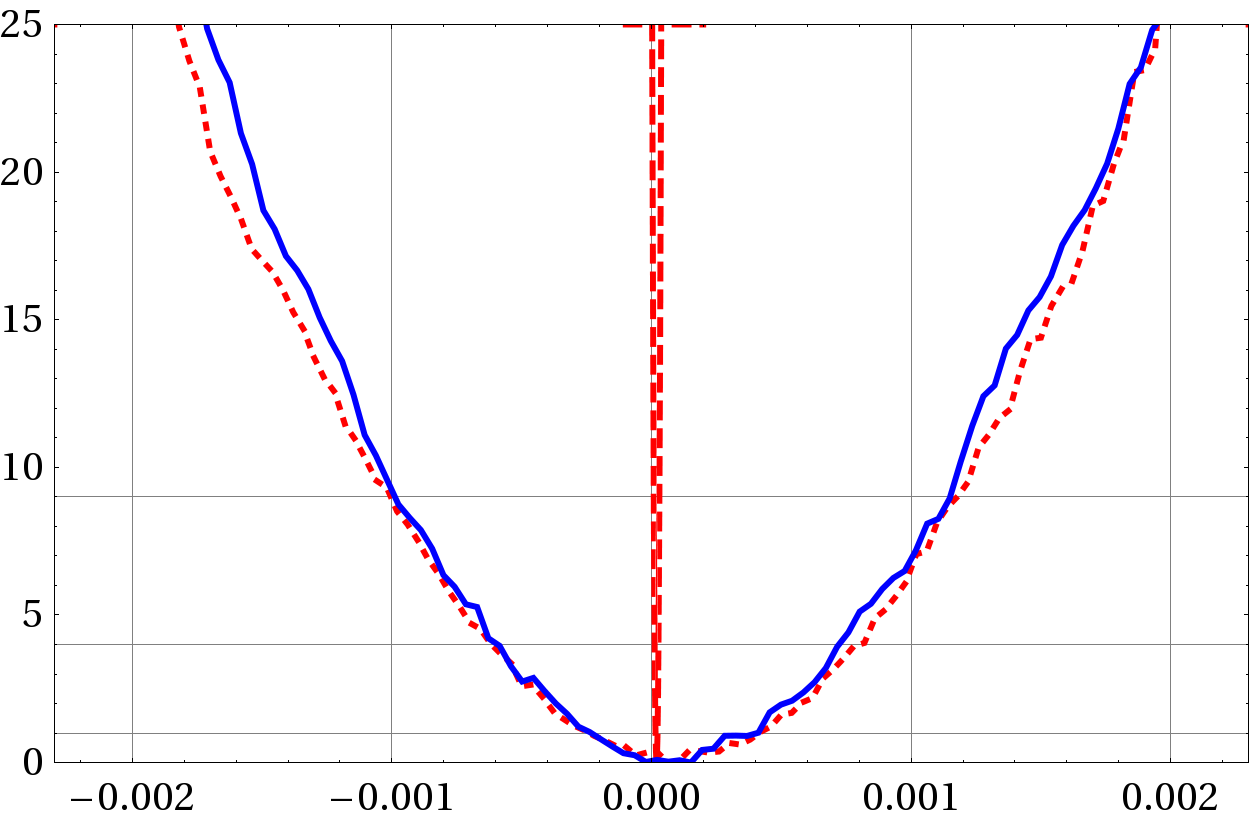}}\\
\subfigure[$\DC$ vs. $\aslb$.\label{fig:NoUnNP:03}]{\includegraphics[width=0.45\textwidth]{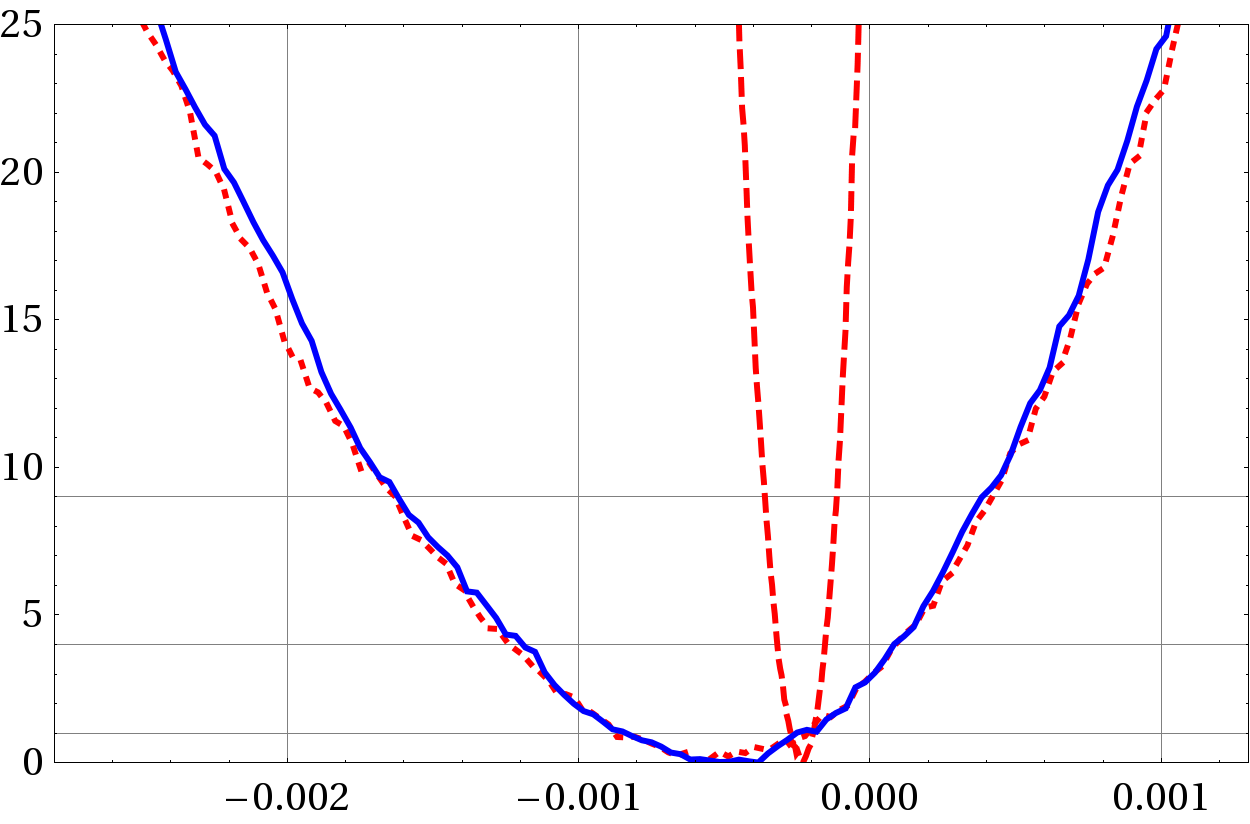}}
\caption{$\DC$ profiles of semileptonic asymmetries $\aslq$; the {\color{blue} blue} lines correspond to the $4\times 4$ unitary NP scenario -- eqs. \refeq{eq:Un44Mixq:01} and \refeq{eq:Un44Mixq:02} --, the {\color{red} red dotted} lines correspond to the $3\times 3$ unitary NP scenario of section \ref{SEC:3x3NP}, the {\color{red} red dashed} lines correspond to the SM case. The last D0 measurement gives $\aslb=(-4.96\pm 1.69)\cdot 10^{-3}$ \cite{Abazov:2013uma}.\label{fig:NoUnNP:01}}
\end{center}
\end{figure*}

\begin{figure}[thb]
\subfigure[$\DC$ vs. $|\V{tb}|$.\label{fig:Undev:Vtb}]{\includegraphics[width=0.32\textwidth]{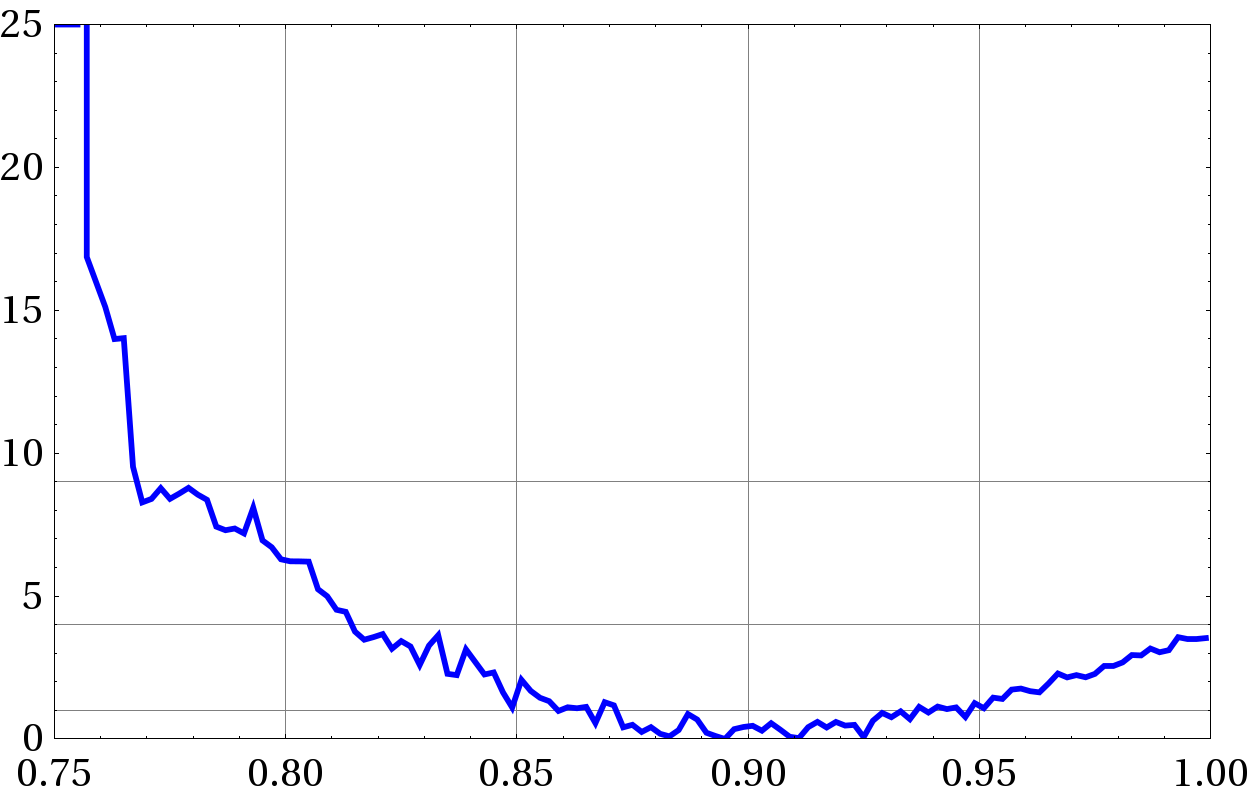}}\\
\subfigure[$\DC$ vs. ${1-|\V{ud}|^2-|\V{us}|^2-|\V{ub}|^2}$.\label{fig:Undev:uRow}]{\includegraphics[width=0.325\textwidth]{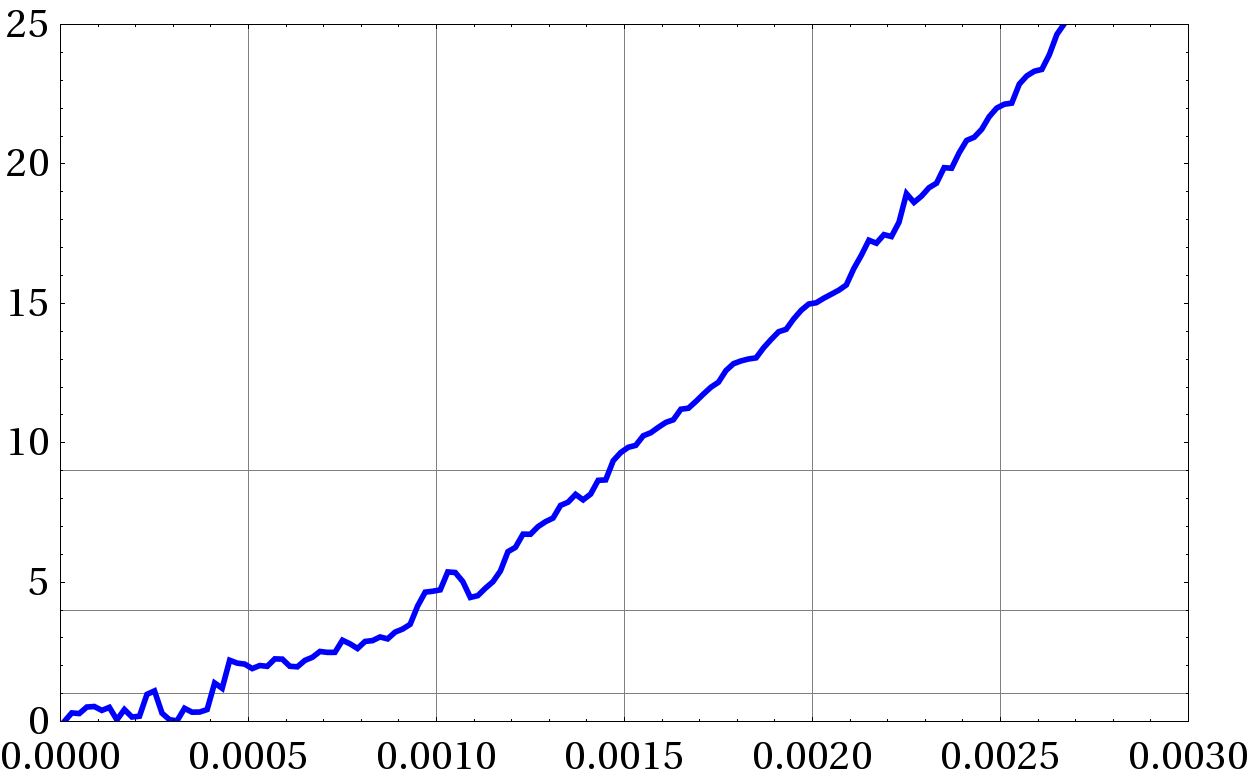}}\\
\subfigure[$\DC$ vs. ${1-|\V{cd}|^2-|\V{cs}|^2-|\V{cb}|^2}$.\label{fig:Undev:cRow}]{\includegraphics[width=0.32\textwidth]{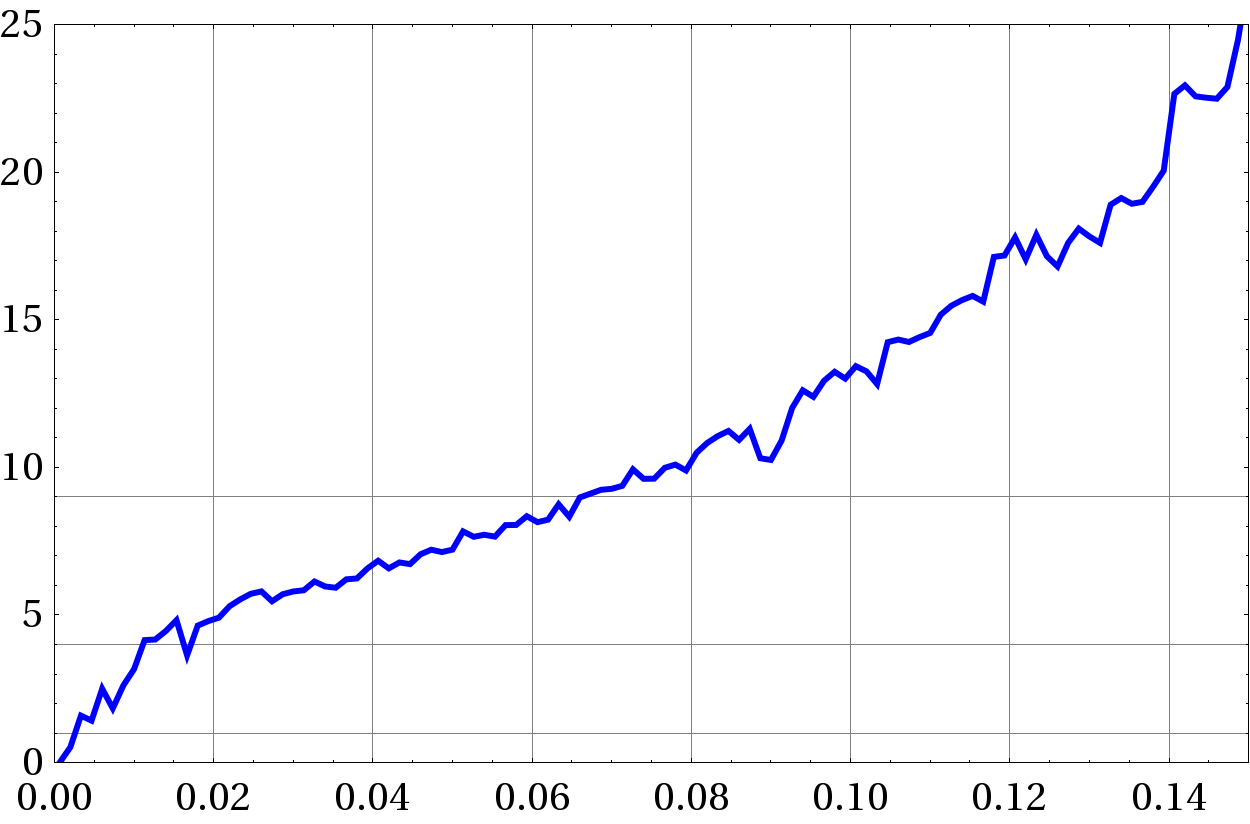}}
\caption{$\DC$ profiles of the deviations from $3\times 3$ unitarity in $|\V{tb}|$ and in the first and second rows of the mixing matrix.\label{fig:Undev:Rows}}
\end{figure}

The model independent parameterizations in \eq{eq:33NP:Mix01} do not exhaust the NP scenarios that could give rise to an enhancement of the mixing  asymmetries $\asld$ and $\asls$. One can consider scenarios in which the CKM matrix is no longer $3\times 3$ unitary and it is, on the contrary, part of a larger unitary matrix. 
If the CKM matrix is part of a larger unitary matrix, there are, necessarily, additional fields beyond the standard three chiral ones; since they may couple to known quarks and weak bosons, they can give new contributions to $\Mmixq$, controlled by the matrix elements beyond the $3\times 3$ usual CKM matrix. If, for instance,
\begin{equation}
\V{ub}\Vc{uq}+\V{cb}\Vc{cq}+\V{tb}\Vc{tq}\equiv-N_{bq}\neq 0\,,\label{eq:NoUn:01}
\end{equation}
one should consider modified $\Mmixq$ expressions with the following structure \cite{Barenboim:1997pf,*Barenboim:1997qx,*Barenboim:2000zz,*Barenboim:2001fd,*Eyal:1999ii}:
\begin{multline}
\Mmixq = \frac{G_F^2M_W^2}{12\pi^2}\,M_{B_q}f_{B_q}^2B_{B_q}\eta_B\\
\left((\V{tb}\Vc{tq})^2S_0(x_t)+(\V{tb}\Vc{tq})\,N_{bq}\,C_1+N_{bq}^2\,C_2\right)\,.\label{eq:NoUnMixq:01}
\end{multline}
$C_1$ and $C_2$, both real, are the model dependent parameters that control the terms linear and quadratic (respectively) in the deviation $N_{bq}$ of the mixing matrix with respect to $3\times 3$ unitarity. We consider $C_1$ and $C_2$ common to both $\Mmixd$ and $\Mmixs$, and real, confining all the new flavour dependence and CP violation to the mixings $N_{bq}$. Examples of such scenarios are models where the fermion content is extended through additional chiral or vectorlike quarks \cite{Barenboim:1997pf,*Barenboim:1997qx,*Barenboim:2000zz,*Barenboim:2001fd,*Eyal:1999ii,Frampton:1999xi}. 
Equations \refeq{eq:NoUn:01} and \refeq{eq:NoUnMixq:01} provide indeed the ingredients, analysed in the previous section, that could induce deviations from SM expectations both in $\asld$ and in $\asls$ \cite{Botella:2008qm,Botella:2012ju}. Notice that we include new terms in $\Mmixq$, not in $\Gmixq$. To include new terms in $\Gmixq$ one should expand the present analysis since those eventual new contributions would be model dependent and constrained by additional information concerning $\Delta B\neq 0$ processes, as done, e.g. in \cite{Branco:1992uy}. Since the simplest realization of this extended scenario is to consider the CKM matrix to be embedded in a $4\times 4$ unitary matrix, we restrict our analyses of the next subsections to such a case. 
Then, \eq{eq:NoUn:01} gives
\begin{equation}
\La{u}{bq}+\La{c}{bq}+\La{t}{bq}=-\La{4}{bq}\,,\quad q=d,s\,,\label{eq:Un44Mixq:01}
\end{equation}
where $U$ is $4\times 4$ unitary, $\U{ij}=\V{ij}$ for $i,j\leq 3$ and $\La{a}{bq}\equiv\V{ab}\Vc{aq}$. The \BBqmix\ mixing amplitude $\Mmixq$ is
\begin{multline}
\Mmixq = \frac{G_F^2M_W^2}{12\pi^2}\,m_{B_q}f_{B_q}^2B_{B_q}\eta_B\\
  \times\left((\La{t}{bq})^2S_0(x_t)+2(\La{t}{bq}\La{4}{bq}) C_{1}+(\La{4}{bq})^2 C_{2}\right)\,.\label{eq:Un44Mixq:02}
\end{multline}
Then, instead of \eq{eq:G12q:03}, we have
\begin{multline}
\frac{\Gmixq}{\Mmixq}= 
 K_{(q)}S_0(x_t)\\
\left[
\frac{c\,(\La{u}{bq}+\La{c}{bq})^2-a\,\La{u}{bq}(\La{u}{bq}+\La{c}{bq})+b\,(\La{u}{bq})^2}{(\La{t}{bq})^2S_0(x_t)+2(\La{t}{bq}\La{4}{bq}) C_{1}+(\La{4}{bq})^2 C_{2}}
\right]\ ,\label{eq:Un44Mixq:03}
\end{multline}
and unitarity -- \eq{eq:Un44Mixq:01} -- allows to write the first term as
\begin{equation}
c\,\frac{(\La{t}{bq}+\La{4}{bq})^2}{(\La{t}{bq})^2S_0(x_t)+2(\La{t}{bq}\La{4}{bq}) C_{1}+(\La{4}{bq})^2 C_{2}}\,,
\end{equation}
which is not, in general, real. In this kind of New Physics scenario, the SM suppression of $\aslq$ is naturally removed: the two ingredients which align the phase of this would-be-leading term with that of $\Mmixq$, namely $3\times 3$ unitarity and $\Mmixq$ dominated by the top quark contribution, are absent. In order to illustrate how deviations of $3\times 3$ unitarity provide the ingredients that may enhance the semileptonic asymmetries, figures \ref{fig:UTMixbd} and \ref{fig:UTMixbs} in appendix \ref{APP:NonUnitarity} show the modifications brought by this scenario with respect to the previous $3\times 3$ unitary one and with respect to the SM.
The analysis of section \ref{SEC:3x3NP} is rather simple and general, because of the complete parametric freedom and independence accorded to $r_d$, $r_s$, $\phi_d$ and $\phi_s$. The present scenario with a $4\times 4$ unitary mixing is somehow different: beside $C_1$ and $C_2$, all the available freedom is the freedom that $4\times 4$ unitarity provides to have $bd$ and $bs$ quadrangles instead of triangles. 
For specific models, $C_1$ and $C_2$ will have well defined functional forms (involving new fermion masses, for example): to maintain full generality, $C_1$ and $C_2$ are allowed to vary freely within reasonable ranges, in particular we consider\footnote{Those are sufficiently generous ranges: for example, for an additional up quark $T$, we will have $C_{1}\to S_0(x_t,x_T)$ and $C_{2}\to S_0(x_T)$, ($x_T=m_T^2/M_W^2$); with a mass $m_T$ ranging up to $5$ TeV, $C_{1}\leq 9.3$ and $C_{2}\leq 980$.} $C_1\in [-10;10]$ and $C_2\in [-10^3;10^3]$.\\
\noindent The previous prospects translate into results for the relevant observables: figure \ref{fig:NoUnNP:01} shows the $\DC$ profiles of $\asld$ and $\asls$, together with the ones corresponding to the NP scenario of section \ref{SEC:3x3NP} and the SM ones for easy comparison. The values that the semileptonic asymmetries may reach are similar to the ones that can be obtained in the $3\times 3$ unitary scenario with NP in $\Mmixq$. With the results for the single $\asld$ and $\asls$ asymmetries, one can expect the dimuon asymmetry $\aslb$ to span a range similar to the one in fig. \ref{fig:UnNP:04}: figure \ref{fig:NoUnNP:03} shows the $\DC$ profile of $\aslb$ for the $4\times 4$ unitary case, together with the ones in fig. \ref{fig:UnNP:04} for comparison. As in the $3\times 3$ unitary case with NP in $\Mmixq$, the value of $\aslb$ is enhanced, thus reducing the discrepancy with the D0 result, but the enhancement is insufficient to reproduce the measurement.\\
\begin{figure*}[htb!]
\subfigure[$\asld$ vs. $|\V{ub}|$.\label{fig:NoUnNP:02a}]{\includegraphics[width=0.35\textwidth]{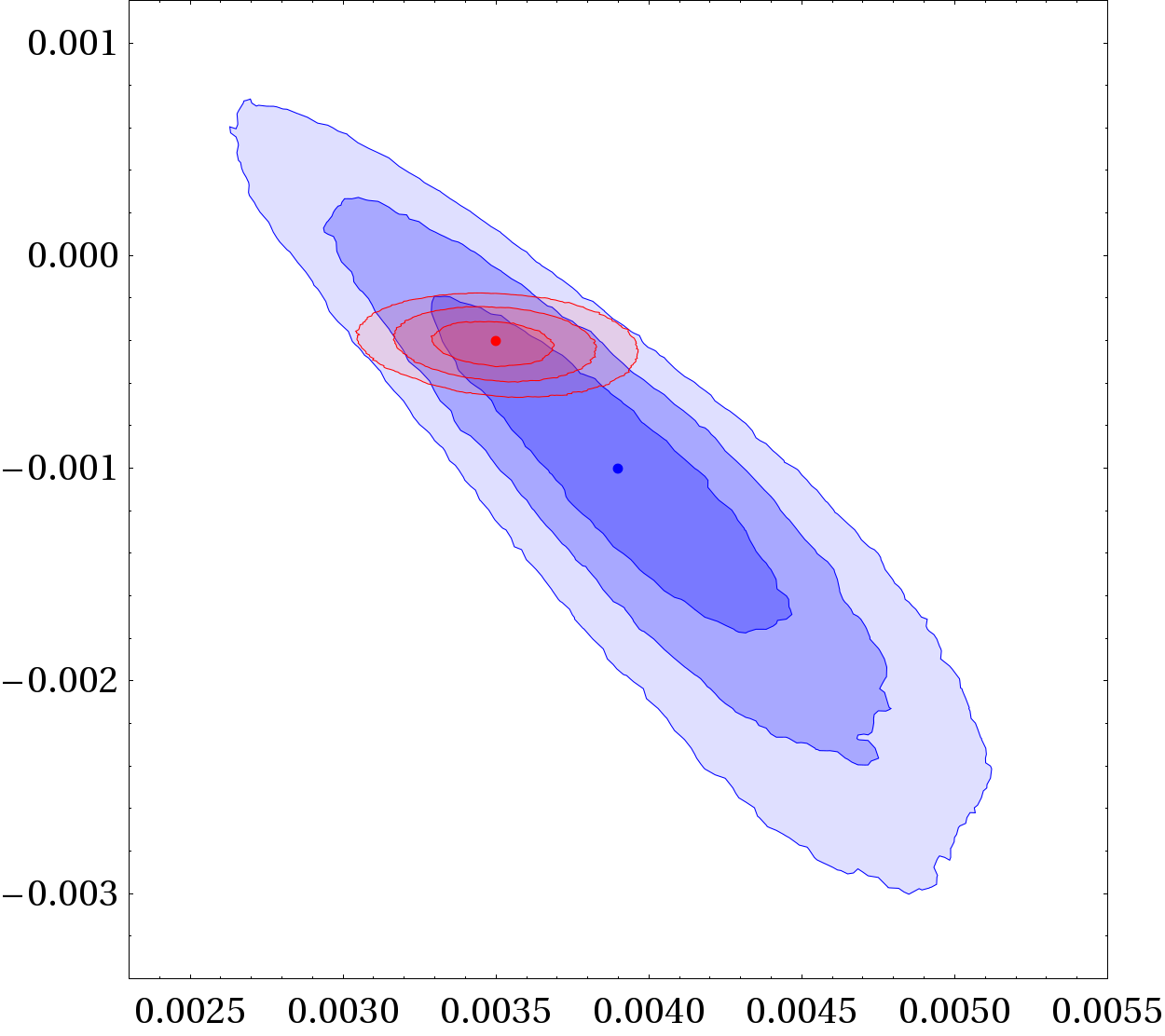}}\qquad
\subfigure[$\asls$ vs. $\AJPP$.\label{fig:NoUnNP:02b}]{\includegraphics[width=0.352\textwidth]{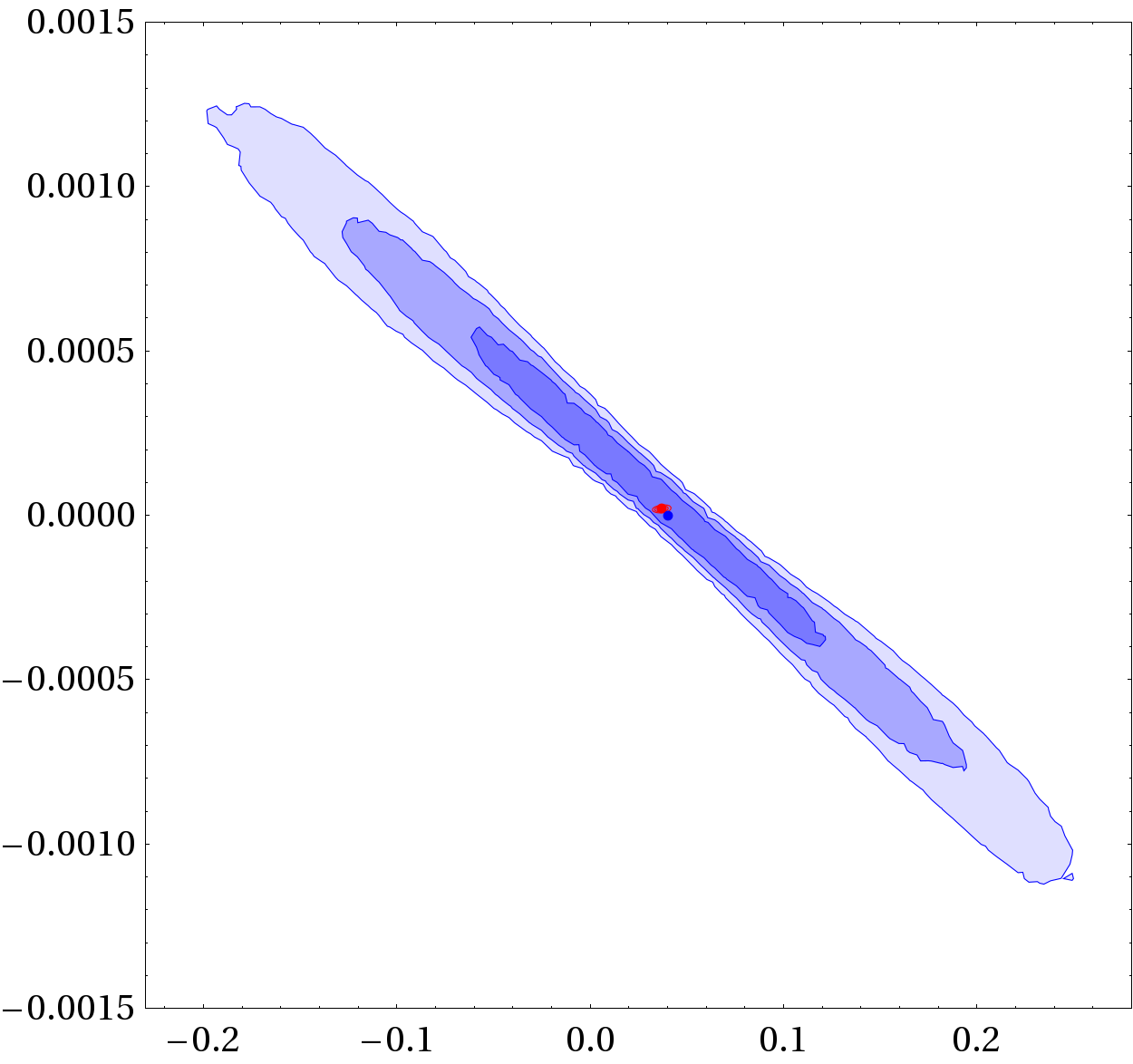}}\\
\subfigure[$\aslb$ vs. $|\V{ub}|$.\label{fig:NoUnNP:04a}]{\includegraphics[width=0.35\textwidth]{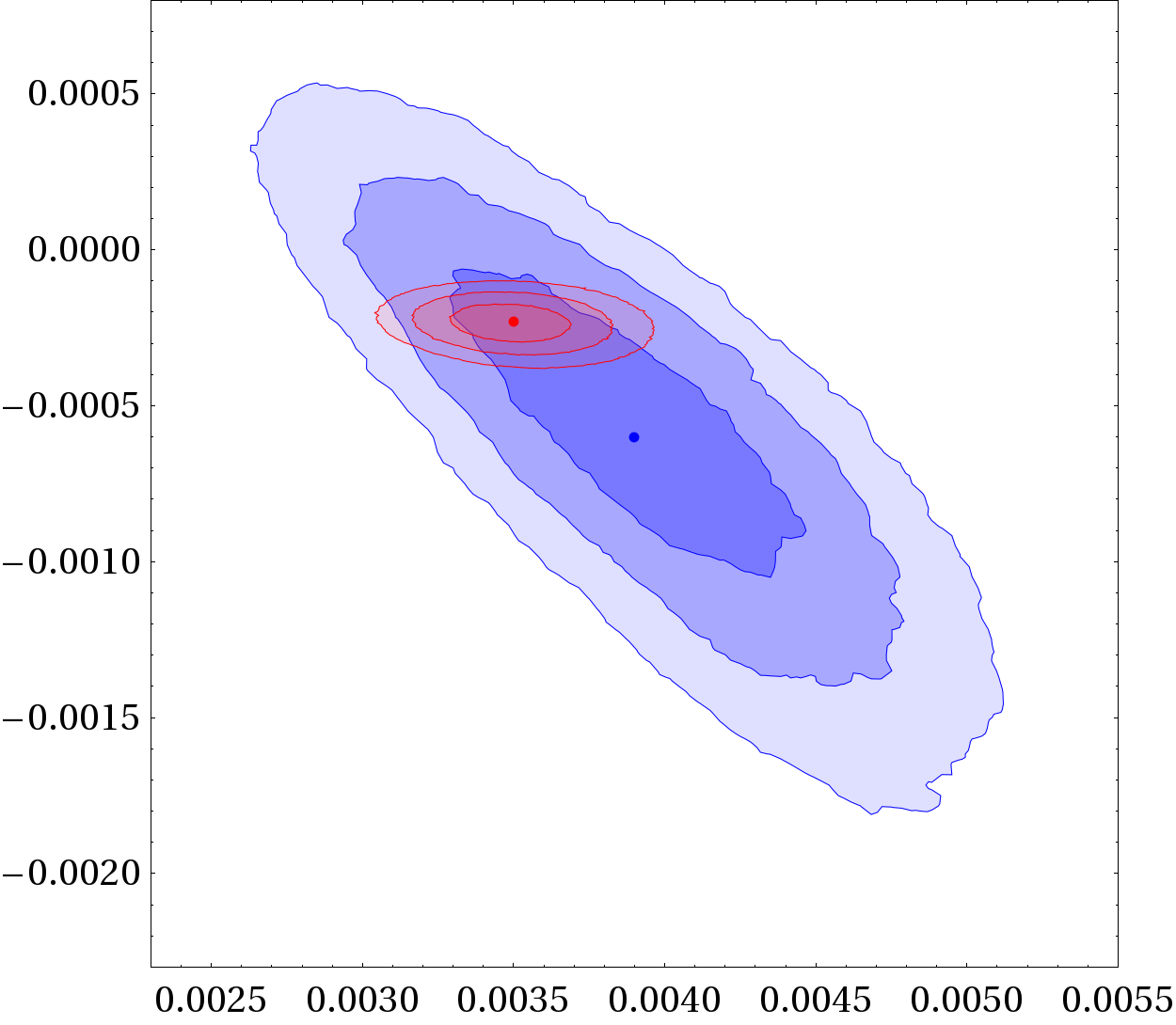}}\qquad
\subfigure[$\aslb$ vs. $\AJPP$.\label{fig:NoUnNP:04b}]{\includegraphics[width=0.35\textwidth]{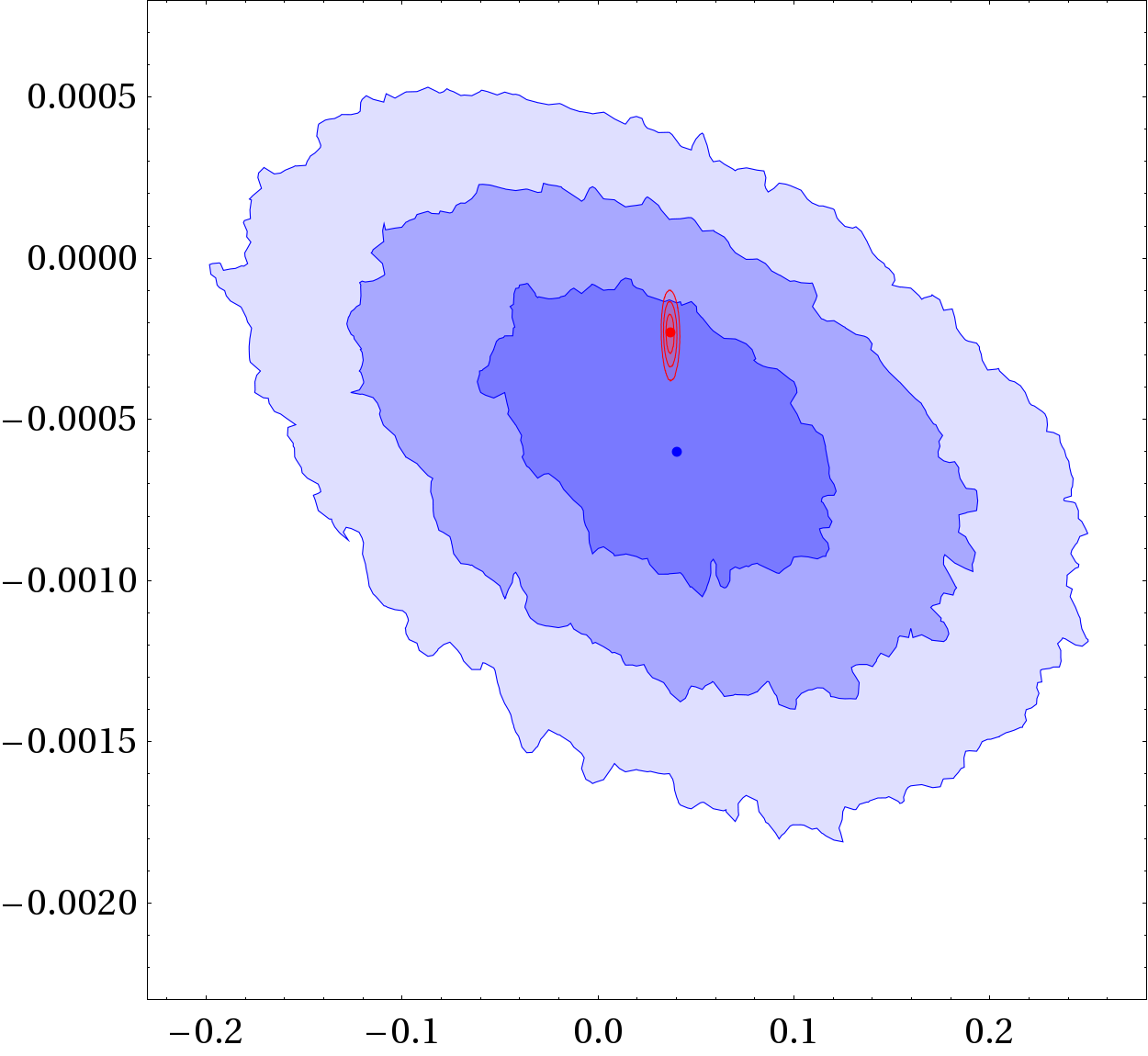}}\\
\caption{$\DC$ 68\%, 95\% and 99\% CL regions. {\color{blue} Blue} regions correspond to the $4\times 4$ unitary NP scenario, {\color{red} red} regions correspond to the SM case.\label{fig:NoUnNP:02}}
\end{figure*}
\noindent This confirms the basic picture that underlies deviations from SM expectations and establishes deviations from $3\times 3$ unitarity as a framework that accommodates them naturally: in the $bd$ sector, $\asld$ values at the $10^{-3}$ level can be reached when the tight connection between $|\V{ub}|$ and $\AJPKs$ present in the SM is relaxed; in the $bs$ sector, $\asls$ values at the $10^{-3}$ level can be reached when $\AJPP$ deviates from the SM expectation $\AJPP\simeq 0.04$. Both ingredients are present in this NP scenario with the CKM matrix part of a larger $4\times 4$ unitary matrix, as figures \ref{fig:NoUnNP:02a} and \ref{fig:NoUnNP:02b} illustrate. This behaviour is inherited by $\aslb$: deviations in $\aslb$ from SM expectations are correlated, as in the $3\times 3$ unitary case with NP in $\Mmixq$, with deviations in $|\V{ub}|$ and $\AJPP$, as figures \ref{fig:NoUnNP:04a} and \ref{fig:NoUnNP:04b} illustrate.
\clearpage
\noindent\underline{Deviations from ${3\times 3}$ unitarity}\\
In sections \ref{SEC:3x3NP} and \ref{SEC:no3x3NP} we have explored two NP avenues that induce deviations in $\asld$ and $\asls$. The experimental contraints entering both analyses are: (i) tree level measurements of the mixing matrix -- moduli $|\V{ij}|$ of the first two rows and the phase $\gamma$ --, (ii) measurements of \BBdmix\ and \BBsmix\ mixings -- $\DMBd$, $\DMBs$ and the effective phases $2\bar\beta$ (through $\AJPKs$) and $2\bar\beta_s$ (through $\AJPP$) --. An important question one can ask is the following: with those ingredients, to what extent could we distinguish the two NP scenarios? That is, could we uncover eventual deviations from $3\times 3$ unitarity if they are indeed originating some discrepancy with respect to SM expectations?\\
\noindent In figures \ref{fig:UTMixbd:01a}, \ref{fig:UTMixbd:02a} and \ref{fig:UTMixbd:03a}, the tree level measurements ``fix'' the $|\V{ud}\Vc{ub}|$ and $|\V{cd}\Vc{cb}|$ sides, together with their relative orientation given by $\gamma$. Then $\DMBd$ and $\AJPKs$ ``fix'' the mixing in figures \ref{fig:UTMixbd:01b}, \ref{fig:UTMixbd:02b} and \ref{fig:UTMixbd:03b}. If there is some NP hint it will manifest through an incompatibility among related quantities, for example among the value of $|\V{td}\Vc{tb}|$ (controlling $\DMBd$) and $\gamma$, or among the value of $|\V{cd}\Vc{cb}|$ and $\alpha=\pi-\gamma-\beta$, or among the value of $|\V{ud}\Vc{ub}|$ and $\beta$ (this incompatibility is none other than the ``$bd$ tension''). While this may seem straightforward, as soon as one concedes that only SM tree level dominated quantities are ``safe'' (not polluted by eventual NP contributions), none of these is useful: for the first, $\DMBd$ arises at one loop in the SM, invalidating the indirect obtention of $|\V{td}\Vc{tb}|$ from it\footnote{For completeness: there are no direct measurements of $|\V{td}|$, and not very constraining ones of $|\V{tb}|$ (see table \ref{AP:tab:data}).}; for the second and third, the phases that are in fact measured are not $\beta$ and $\alpha$ but the effective $\bar\beta$ and $\bar\alpha=\pi-\gamma-\bar\beta$, which may deviate from $\beta$, $\alpha$ through new contributions to $\Mmixd$.\\
One can establish a tension with respect to the SM expectations, but this mismatch involves both the structure of the mixing matrix (the unitarity triangle) \emph{and} the $\Mmixd$ prediction: as soon as NP introduces new parameters that break the SM connection between both, the minimal set of observables that we are considering cannot indicate whether we have deviations from $3\times 3$ unitarity or not \cite{Silva:1996ih}. The previous discussion concerns the $bd$ sector, but the situation in the $bs$ sector is not conceptually different. 
 This does not mean that unitarity deviations cannot be established, it only means that the rather restricted set of observables that we are considering for this general analysis is not sufficient for that task, and the following roads have to be explored. 
\begin{enumerate}
\item As soon as a specific model that incorporates mixings beyond the $3\times 3$ unitary case is considered, a specific pattern of deviations with respect to SM expectations in flavour changing processes like $B_d\to X_s\gamma$, $B_d\to X_s\ell^+\ell^-$, $B_{d,s}\to\mu^+\mu^-$ -- and others outside $B$ mesons systems like $K_L\to\mu^+\mu^-$, $K\to\pi\nu\bar\nu$ or \KKmix\ and \DDmix\ oscillations -- will emerge, and use made of a much larger set of experimental measurements.
\item On the other hand, deviations from $3\times3$ unitarity may be directly probed through
\begin{enumerate}
\item $|\V{tb}|\neq 1$ at the percent level -- which would be within the sensitivity of the LHC experiments \cite{Aad:2012xca,Chatrchyan:2012zca,Adelman:2013gis} --,
\item $|\V{ud}|^2+|\V{us}|^2+|\V{ub}|^2\neq 1$, and 
\item $|\V{cd}|^2+|\V{cs}|^2+|\V{cb}|^2\neq 1$.
\end{enumerate}
\end{enumerate}
The first possibility, followed for example in \cite{Botella:2012ju}, is completely model specific and thus of no use for the the present model independent approach. For the second possibility, we can directly explore to which extent all three signals of deviation from $3\times 3$ unitarity may arise. This is illustrated in figure \ref{fig:Undev:Rows}. In figure \ref{fig:Undev:Vtb} one can actually observe that $|\V{tb}|$ can depart from the $1-\mathcal O(10^{-4})$ ballpark that $3\times 3$ unitarity imposes, and do so at a level which the LHC experiments can probe. In figures \ref{fig:Undev:uRow} and \ref{fig:Undev:cRow} the deviation from $3\times 3$ unitarity in the first ($u$) and second ($c$) rows of the mixing matrix are displayed: in both cases deviations from $3\times 3$ unitarity at a level to be explored in the near future are allowed within our framework.
\clearpage
\section{Conclusions\label{SEC:Conc}}
Within the SM, the CP-violating asymmetries $\asld$ and $\asls$ in the neutral \BBdmix\ and \BBsmix\ mixings are expected to be naturally small.\\
The D0 collaboration has measured the like-sign dimuon asymmetry $\aslb$ -- which is a combination of the $\asld$ and $\asls$ asymmetries -- and obtained a large value, marginally compatible (at around the $3\sigma$ level) with the SM expectations.
Since this fact might hint to New Physics, we have considered two different NP scenarios. In the well known first scenario the CKM mixing matrix remains $3\times 3$ unitary and NP enters \BBdmix\ and \BBsmix\ mixings in a simple parametric manner. In the second scenario, which we analyse for the first time in detail, deviations from $3\times 3$ unitarity in the mixing matrix are allowed, and they are related to new contributions to $B$ meson mixings.
In both scenarios $\asld$ and $\asls$ can be sizably enhanced with respect to SM expectations. In the case of $\asld$, non-standard values are related to the tension between $|\V{ub}|$ and $\AJPKs$: NP alleviates that tension and, modifying $\Mmixd$ at the $20-30$\% level, can increase $\asld$ \emph{fivefold}. The case of $\asls$ is different: as NP crucially changes the relation between the phase $\beta_s$ and $\AJPP$, $\asls$ is allowed to reach values almost two orders of magnitude larger than the SM expectation. In $\asls$ too, deviations from SM expectations are related to other NP effects: $\AJPP$ in this case.
When both $\asld$ and $\asls$ are enhanced, $\aslb$ may reach values at the $10^{-3}$ to $2\cdot 10^{-3}$ level. Nevertheless, obtaining a prediction \emph{five times larger} than in the SM is not enough to reproduce the D0 measurement of $\aslb$.
Meanwhile, experimental results from the LHCb experiment are eagerly awaited to put some 
light on the issue. The SM predictions for $\asld$ and $\asls$ are really tight: a measurement that sees an increase in one or both will point, undoubtedly, to NP and new sources of CP violation.
\acknowledgments
This work was supported by Spanish MINECO under grant FPA2011-23596, by \emph{Generalitat Valenciana} under grant GVPROMETEOII 2014-049 and by \emph{Funda\c{c}\~ao para a Ci\^encia e a Tecnologia} (FCT, Portugal) through the projects CERN/FP/83503/2008,  EXPL/FIS-NUC/0460/2013 and CFTP-FCT Unit 777 (PEst-OE/FIS/UI0777/2011) which are partially funded through POCTI (FEDER).

\clearpage
\appendix
\section{Input\label{APP:Input}}
Table \ref{AP:tab:data} summarizes the experimental input \cite{Abazov:2011yk,Abazov:2013uma,Abazov:2012hha,PDG:2012,Amhis:2012bh,LHCb:2011ab,LHCb:2011aa,Aaij:2013oba,Aaij:2013mpa,Adachi:2012mm,Lees:2012ju,Nakano:2005jb,Aubert:2006nf,Aaij:2013gta} used for the different calculations; measurements are interpreted as gaussians with the quoted values for the central value and the uncertainty. The $\DC$ profiles and regions have been computed through adapted Markov Chain MonteCarlo techniques that allow for an efficient exploration of the different parameter spaces. For the additional theoretical input from lattice QCD, $f_{B_{s}}\sqrt{B_{B_s}}=266\pm 18$ MeV and $\xi\equiv\frac{f_{B_{s}}\sqrt{B_{B_s}}}{f_{B_{d}}\sqrt{B_{B_d}}}=1.268\pm 0.063$ have been used \cite{Aoki:2013ldr}; although the results presented correspond to modelling the theoretical uncertainties in a gaussian manner, it has been checked that modelling them with uniform uncertainties restricted to 1$\sigma$ or $2\sigma$ ranges produces no change.

\begin{table*}[h]
\begin{tabular}{|c|c|c|c|c|c|}
\hline
$|\V{ud}|$ & $0.97425\pm 0.00022$ & $|\V{us}|$ & $0.2252\pm 0.0009$ & $|\V{ub}|$ & $0.00375\pm 0.00040$\\ \hline	
$|\V{cd}|$ & $0.230\pm 0.011$     & $|\V{cs}|$ & $1.023\pm 0.036$ & $|\V{cb}|$ & $0.041\pm 0.001$   \\ \hline	
$R_t$ & $0.90\pm 0.05$\\ \hline	
$\gamma$   & $(68\pm 8)^\circ$ & $\AJPKs$ & $0.68\pm 0.02$ & $\AJPP$ & $ 0.01\pm 0.07$\\ \hline
$\sin(2\bar\alpha)$ & $ 0.00\pm 0.14$ & $\sin(2\bar\beta+\gamma)$ & $0.95\pm 0.40$\\ \hline
$\DMBd$ & $(0.507\pm 0.004)$ ps$^{-1}$ & $\DGd$ & $(-0.011\pm 0.014)$ ps$^{-1}$  & $\asld$ & $(3\pm 23)\times 10^{-4}$\\ \hline
$\DMBs$ & $(17.768\pm 0.024)$ ps$^{-1}$ & $\DGs$ & $(0.091\pm 0.008)$ ps$^{-1}$ & $\asls$ & $(-32\pm 52)\times 10^{-4}$\\ \hline
$\aslb$ \cite{Abazov:2013uma} & $(-4.96\pm 1.69)\times 10^{-3}$\\ \cline{1-2}
\end{tabular}
\caption{Experimental input (N.B. $R_t\equiv|\V{tb}|^2/(|\V{td}|^2+|\V{ts}|^2+|\V{tb}|^2)$.\label{AP:tab:data}}
\end{table*}

\section{Unitarity and mixings\label{APP:NonUnitarity}}
\begin{figure*}[h]
\begin{center}
\subfigure[$bd$ unitarity triangle in the SM.\label{fig:UTMixbd:01a}]{\includegraphics[width=0.275\textwidth]{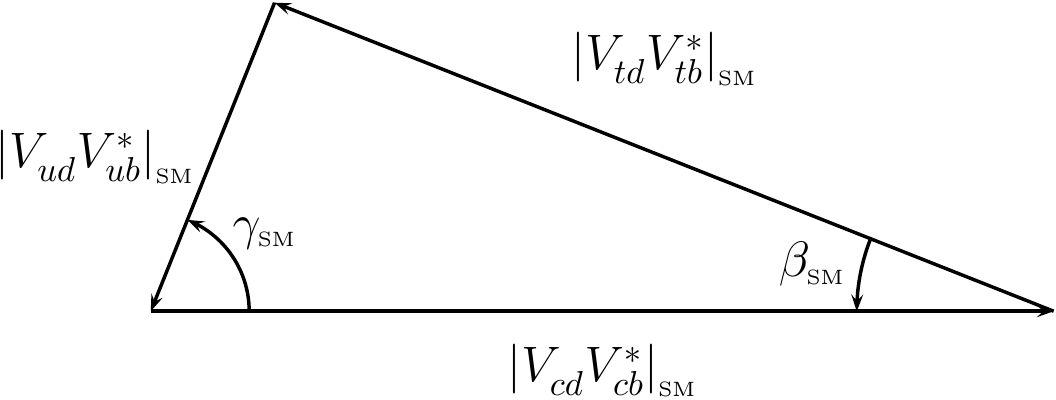}}\qquad
\subfigure[$\Mmixd$ in the SM.\label{fig:UTMixbd:01b}]{\includegraphics[width=0.15\textwidth]{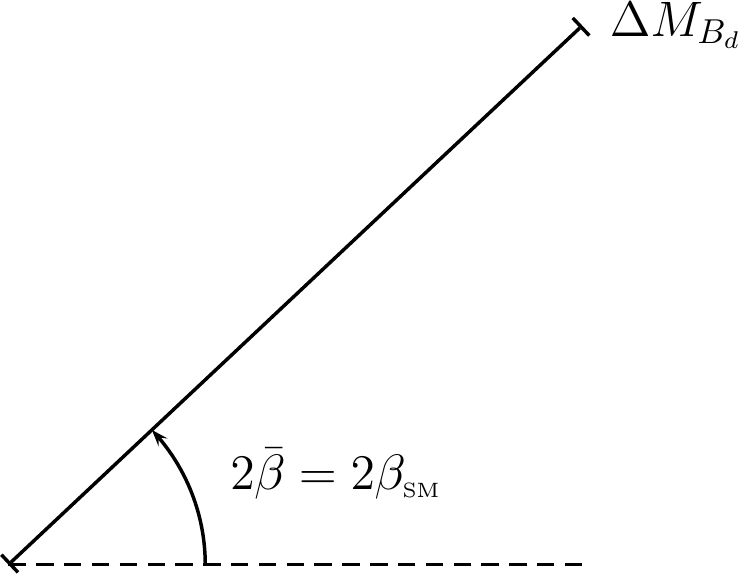}}\\
\subfigure[$bd$ unitarity triangle with NP in mixings.\label{fig:UTMixbd:02a}]{\includegraphics[width=0.275\textwidth]{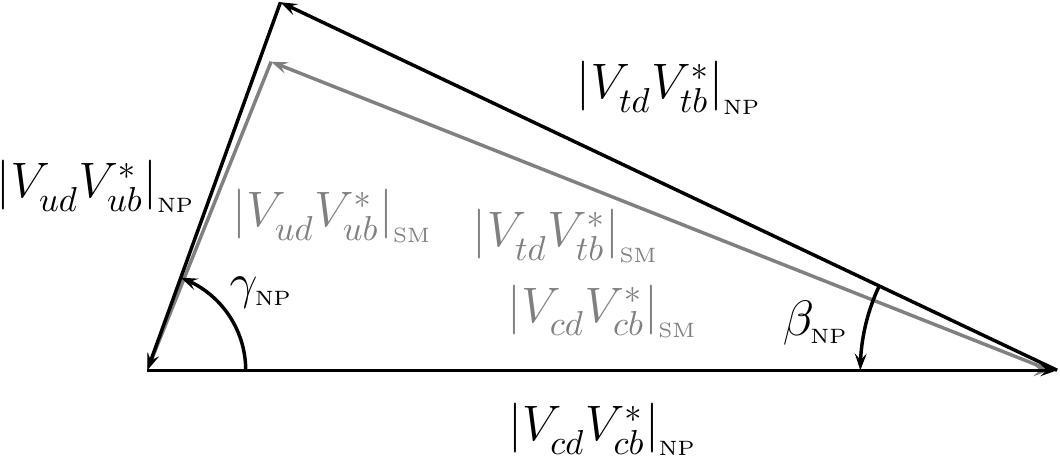}}\qquad
\subfigure[$\Mmixd$ with NP.\label{fig:UTMixbd:02b}]{\includegraphics[width=0.15\textwidth]{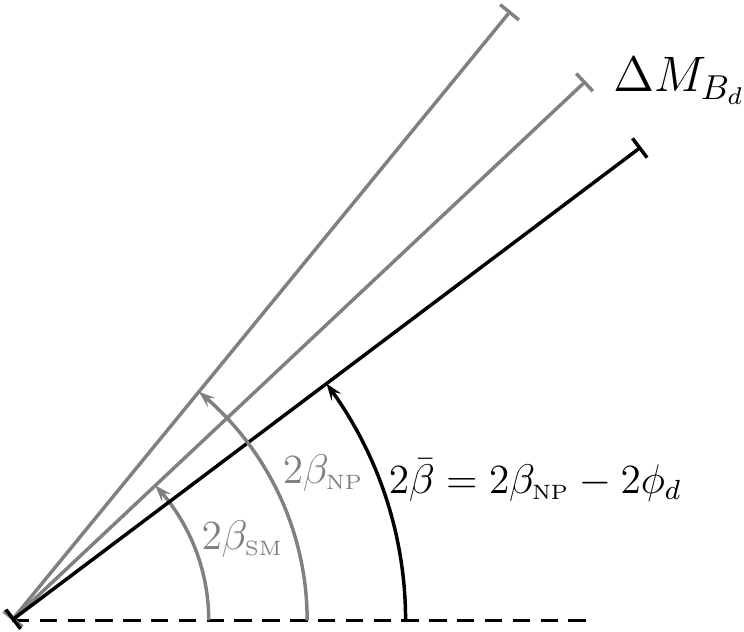}}\qquad
\subfigure[$bd$ unitarity quadrangle.\label{fig:UTMixbd:03a}]{\includegraphics[width=0.275\textwidth]{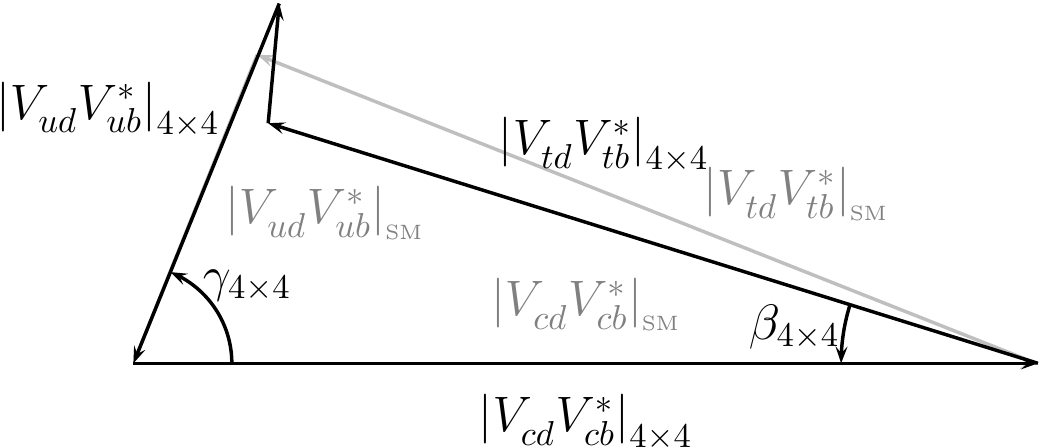}}\qquad
\subfigure[$\Mmixd$ beyond $3\times 3$ unitarity.\label{fig:UTMixbd:03b}]{\includegraphics[width=0.15\textwidth]{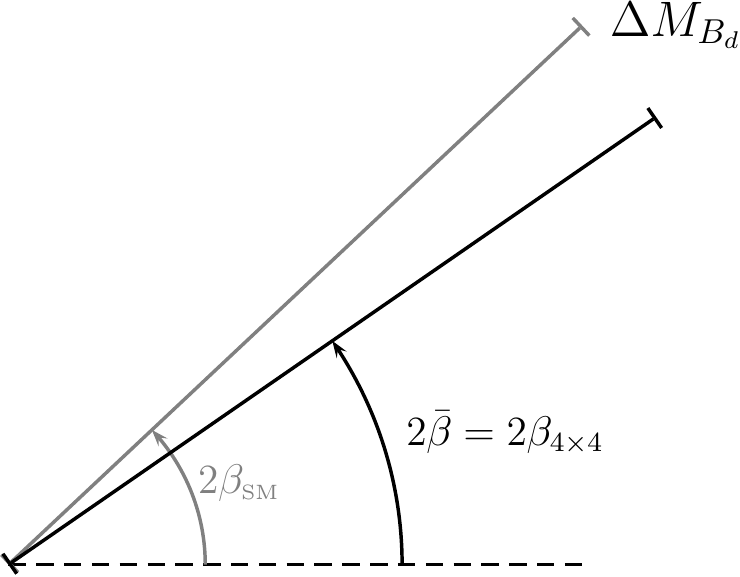}}
\caption{$bd$ unitarity and $\Mmixd$ in the SM and beyond.\label{fig:UTMixbd}}
\end{center}
\end{figure*}

\begin{figure*}[h]
\begin{center}
\subfigure[$bs$ unitarity triangle in the SM.\label{fig:UTMixbs:01a}]{\includegraphics[width=0.325\textwidth]{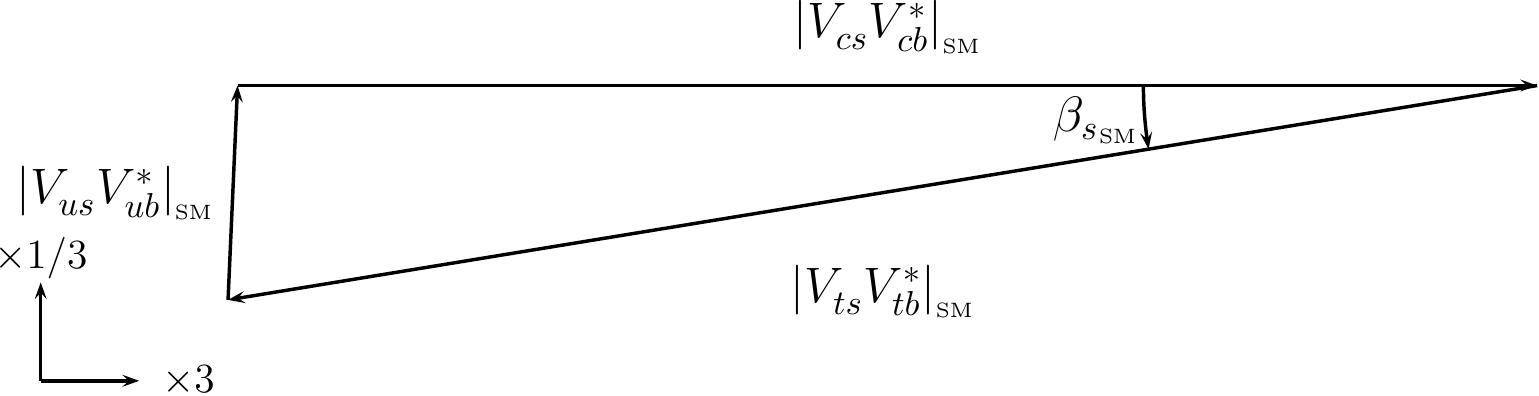}}\\
\subfigure[$bs$ unitarity triangle with NP in mixings.\label{fig:UTMixbs:02a}]{\includegraphics[width=0.325\textwidth]{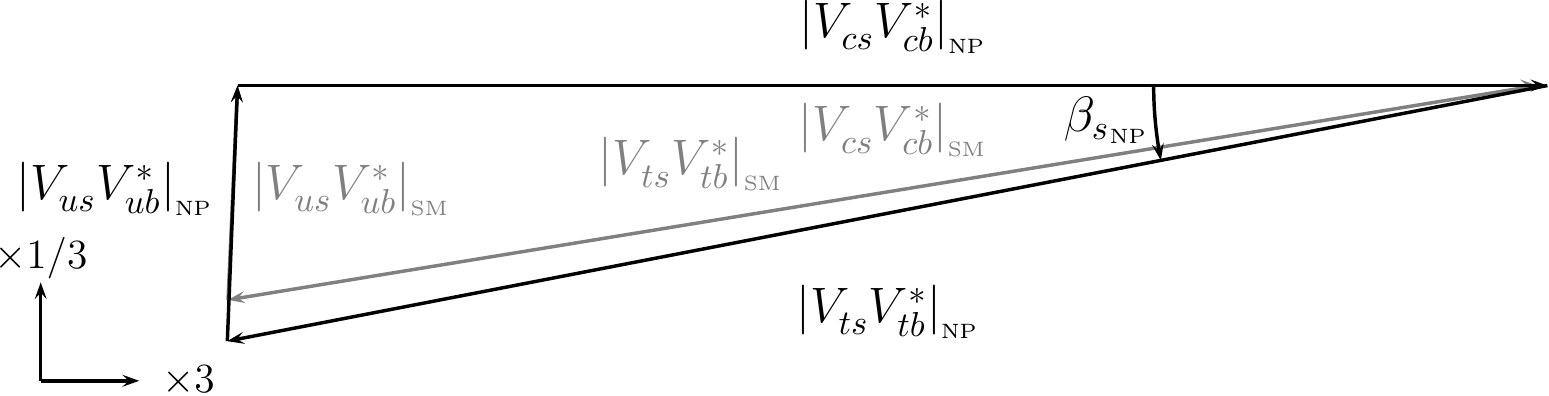}}\qquad
\subfigure[$bs$ unitarity quadrangle.\label{fig:UTMixbs:03a}]{\includegraphics[width=0.325\textwidth]{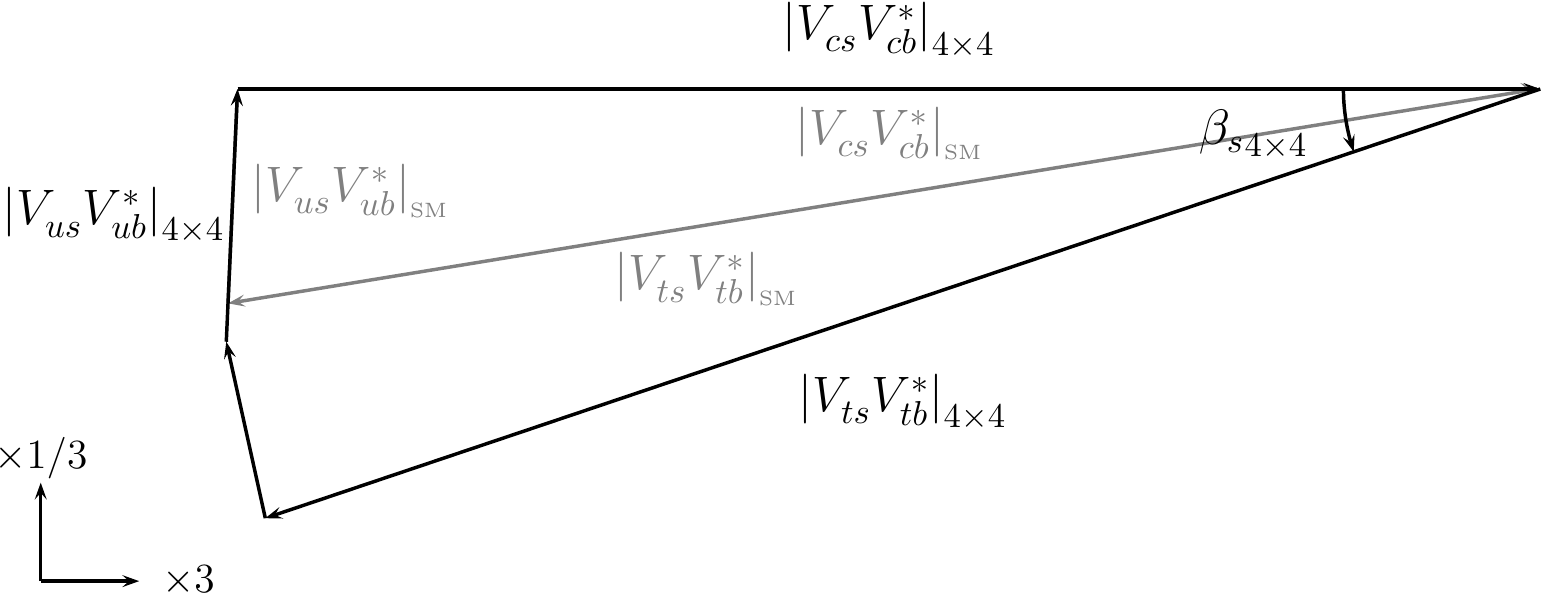}}\\
\caption{$bs$ unitarity and $\Mmixs$ in the SM and beyond.\label{fig:UTMixbs}}
\end{center}
\end{figure*}

\noindent Figures \ref{fig:UTMixbd:01a} and \ref{fig:UTMixbd:01b} show the unitarity triangle $bd$ (in the complex plane) and $\DMBd e^{i2\bar\beta}$ (that is $\Mmixd$) for the SM case. The ``$bd$ tension'' is, schematically, the coincidence of an experimental value of $|\V{ub}|$ which pushes towards larger values than the represented (illustrative) case, with an experimental value of $\AJPKs=\sin(2\bar\beta)$ which pulls in the opposite direction. In figures \ref{fig:UTMixbd:02a} and \ref{fig:UTMixbd:02b}, we illustrate the analysis of section \ref{SEC:3x3NP}: the presence of NP alleviates the tension allowing for larger $|\V{ub}|$ values, which trigger the sizable deviations in $\asld$, from SM expectations, which we are interested in. Figures \ref{fig:UTMixbd:03a} and \ref{fig:UTMixbd:03b} illustrate the non $3\times 3$ unitary scenario. In particular, figure \ref{fig:UTMixbd:03a} displays the unitarity quadrangle $bd$ corresponding to the enlarged $4\times 4$ unitary mixing matrix. One can easily see how the presence of the fourth side, i.e. deviation from $3\times 3$ unitarity, permits larger $|\V{ub}|$ values. Figure \ref{fig:UTMixbd:03b} shows how the corresponding mixing gives adequate values for $\DMBd$ and $\AJPKs$.
For the $bs$ case, fig. \ref{fig:UTMixbs} illustrates the situation (we omit $\Mmixs$ for conciseness): fig. \ref{fig:UTMixbs:01a} is just the SM squashed unitarity triangle $bs$; it does not change much (as analysed in section \ref{SEC:3x3NP}) upon inclusion of NP in $\Mmixq$, as fig. \ref{fig:UTMixbs:02a} shows: the relevant contribution in that case is directly provided by NP through $\phi_s$. Finally, fig. \ref{fig:UTMixbs:03a} shows how the departure from $3\times 3$ unitarity may induce significant departures in the value of the phase $\beta_s$ entering $\Mmixs$, as required to depart from SM values of $\asls$ (and $\AJPP$). 

\clearpage


%


\begin{thebibliography}{85}%
\makeatletter
\providecommand \@ifxundefined [1]{%
 \@ifx{#1\undefined}
}%
\providecommand \@ifnum [1]{%
 \ifnum #1\expandafter \@firstoftwo
 \else \expandafter \@secondoftwo
 \fi
}%
\providecommand \@ifx [1]{%
 \ifx #1\expandafter \@firstoftwo
 \else \expandafter \@secondoftwo
 \fi
}%
\providecommand \natexlab [1]{#1}%
\providecommand \enquote  [1]{``#1''}%
\providecommand \bibnamefont  [1]{#1}%
\providecommand \bibfnamefont [1]{#1}%
\providecommand \citenamefont [1]{#1}%
\providecommand \href@noop [0]{\@secondoftwo}%
\providecommand \href [0]{\begingroup \@sanitize@url \@href}%
\providecommand \@href[1]{\@@startlink{#1}\@@href}%
\providecommand \@@href[1]{\endgroup#1\@@endlink}%
\providecommand \@sanitize@url [0]{\catcode `\\12\catcode `\$12\catcode
  `\&12\catcode `\#12\catcode `\^12\catcode `\_12\catcode `\%12\relax}%
\providecommand \@@startlink[1]{}%
\providecommand \@@endlink[0]{}%
\providecommand \url  [0]{\begingroup\@sanitize@url \@url }%
\providecommand \@url [1]{\endgroup\@href {#1}{\urlprefix }}%
\providecommand \urlprefix  [0]{URL }%
\providecommand \Eprint [0]{\href }%
\providecommand \doibase [0]{http://dx.doi.org/}%
\providecommand \selectlanguage [0]{\@gobble}%
\providecommand \bibinfo  [0]{\@secondoftwo}%
\providecommand \bibfield  [0]{\@secondoftwo}%
\providecommand \translation [1]{[#1]}%
\providecommand \BibitemOpen [0]{}%
\providecommand \bibitemStop [0]{}%
\providecommand \bibitemNoStop [0]{.\EOS\space}%
\providecommand \EOS [0]{\spacefactor3000\relax}%
\providecommand \BibitemShut  [1]{\csname bibitem#1\endcsname}%
\let\auto@bib@innerbib\@empty
\bibitem [{\citenamefont {Abazov}\ and\ \citenamefont {others
  (D0~Collaboration)}(2011)}]{Abazov:2011yk}%
  \BibitemOpen
  \bibfield  {author} {\bibinfo {author} {\bibfnamefont {V.~M.}\ \bibnamefont
  {Abazov}}\ and\ \bibinfo {author} {\bibnamefont {others (D0~Collaboration)}}
  (\bibinfo {collaboration} {D0 Collaboration}),\ }\href {\doibase
  10.1103/PhysRevD.84.052007} {\bibfield  {journal} {\bibinfo  {journal}
  {Phys.Rev.}\ }\textbf {\bibinfo {volume} {D84}},\ \bibinfo {pages} {052007}
  (\bibinfo {year} {2011})},\ \Eprint {http://arxiv.org/abs/1106.6308}
  {arXiv:1106.6308 [hep-ex]} \BibitemShut {NoStop}%
\bibitem [{\citenamefont {Abazov}\ \emph {et~al.}(2014)\citenamefont {Abazov}
  \emph {et~al.}}]{Abazov:2013uma}%
  \BibitemOpen
  \bibfield  {author} {\bibinfo {author} {\bibfnamefont {V.~M.}\ \bibnamefont
  {Abazov}} \emph {et~al.} (\bibinfo {collaboration} {D0 Collaboration}),\
  }\href {\doibase 10.1103/PhysRevD.89.012002} {\bibfield  {journal} {\bibinfo
  {journal} {Phys.Rev.}\ }\textbf {\bibinfo {volume} {D89}},\ \bibinfo {pages}
  {012002} (\bibinfo {year} {2014})},\ \Eprint {http://arxiv.org/abs/1310.0447}
  {arXiv:1310.0447 [hep-ex]} \BibitemShut {NoStop}%
\bibitem [{Note1()}]{Note1}%
  \BibitemOpen
  \bibinfo {note} {We directly refer in the following to \protect \emph {muons}
  since they are the cleanest case from the experimental point of
  view.}\BibitemShut {Stop}%
\bibitem [{Note2()}]{Note2}%
  \BibitemOpen
  \bibinfo {note} {Although central in any experimental analysis, we omit any
  discussion on issues such as efficiencies or backgrounds.}\BibitemShut
  {Stop}%
\bibitem [{\citenamefont {Lees}\ \emph
  {et~al.}(2013{\natexlab{a}})\citenamefont {Lees} \emph
  {et~al.}}]{Lees:2013sua}%
  \BibitemOpen
  \bibfield  {author} {\bibinfo {author} {\bibfnamefont {J.}~\bibnamefont
  {Lees}} \emph {et~al.} (\bibinfo {collaboration} {BaBar Collaboration}),\
  }\href {\doibase 10.1103/PhysRevLett.111.101802} {\bibfield  {journal}
  {\bibinfo  {journal} {Phys.Rev.Lett.}\ }\textbf {\bibinfo {volume} {111}},\
  \bibinfo {pages} {101802} (\bibinfo {year} {2013}{\natexlab{a}})},\ \Eprint
  {http://arxiv.org/abs/1305.1575} {arXiv:1305.1575 [hep-ex]} \BibitemShut
  {NoStop}%
\bibitem [{\citenamefont {Aaij}\ \emph {et~al.}(2014)\citenamefont {Aaij} \emph
  {et~al.}}]{Aaij:2013gta}%
  \BibitemOpen
  \bibfield  {author} {\bibinfo {author} {\bibfnamefont {R.}~\bibnamefont
  {Aaij}} \emph {et~al.} (\bibinfo {collaboration} {LHCb collaboration}),\
  }\href {\doibase 10.1016/j.physletb.2013.12.030} {\bibfield  {journal}
  {\bibinfo  {journal} {Phys.Lett.}\ }\textbf {\bibinfo {volume} {B728}},\
  \bibinfo {pages} {607} (\bibinfo {year} {2014})},\ \Eprint
  {http://arxiv.org/abs/1308.1048} {arXiv:1308.1048 [hep-ex]} \BibitemShut
  {NoStop}%
\bibitem [{Note3()}]{Note3}%
  \BibitemOpen
  \bibinfo {note} {Nevertheless, as we will show, since significant
  cancellations are at work in the SM case, large NP contributions are not
  necessary to obtain significant enhancements in $A^d_{SL}$.}\BibitemShut
  {Stop}%
\bibitem [{\citenamefont {Ko}\ and\ \citenamefont {Park}(2010)}]{Ko:2010mn}%
  \BibitemOpen
  \bibfield  {author} {\bibinfo {author} {\bibfnamefont {P.}~\bibnamefont
  {Ko}}\ and\ \bibinfo {author} {\bibfnamefont {J.-h.}\ \bibnamefont {Park}},\
  }\href {\doibase 10.1103/PhysRevD.82.117701} {\bibfield  {journal} {\bibinfo
  {journal} {Phys.Rev.}\ }\textbf {\bibinfo {volume} {D82}},\ \bibinfo {pages}
  {117701} (\bibinfo {year} {2010})},\ \Eprint {http://arxiv.org/abs/1006.5821}
  {arXiv:1006.5821 [hep-ph]} \BibitemShut {NoStop}%
\bibitem [{\citenamefont {Parry}(2011)}]{Parry:2010ce}%
  \BibitemOpen
  \bibfield  {author} {\bibinfo {author} {\bibfnamefont {J.}~\bibnamefont
  {Parry}},\ }\href {\doibase 10.1016/j.physletb.2010.10.011} {\bibfield
  {journal} {\bibinfo  {journal} {Phys.Lett.}\ }\textbf {\bibinfo {volume}
  {B694}},\ \bibinfo {pages} {363} (\bibinfo {year} {2011})},\ \Eprint
  {http://arxiv.org/abs/1006.5331} {arXiv:1006.5331 [hep-ph]} \BibitemShut
  {NoStop}%
\bibitem [{\citenamefont {Ishimori}\ \emph {et~al.}(2011)\citenamefont
  {Ishimori}, \citenamefont {Kajiyama}, \citenamefont {Shimizu},\ and\
  \citenamefont {Tanimoto}}]{Ishimori:2011nv}%
  \BibitemOpen
  \bibfield  {author} {\bibinfo {author} {\bibfnamefont {H.}~\bibnamefont
  {Ishimori}}, \bibinfo {author} {\bibfnamefont {Y.}~\bibnamefont {Kajiyama}},
  \bibinfo {author} {\bibfnamefont {Y.}~\bibnamefont {Shimizu}}, \ and\
  \bibinfo {author} {\bibfnamefont {M.}~\bibnamefont {Tanimoto}},\ }\href
  {\doibase 10.1143/PTP.126.703} {\bibfield  {journal} {\bibinfo  {journal}
  {Prog.Theor.Phys.}\ }\textbf {\bibinfo {volume} {126}},\ \bibinfo {pages}
  {703} (\bibinfo {year} {2011})},\ \Eprint {http://arxiv.org/abs/1103.5705}
  {arXiv:1103.5705 [hep-ph]} \BibitemShut {NoStop}%
\bibitem [{\citenamefont {Datta}\ \emph {et~al.}(2011)\citenamefont {Datta},
  \citenamefont {Duraisamy},\ and\ \citenamefont {Khalil}}]{Datta:2010yq}%
  \BibitemOpen
  \bibfield  {author} {\bibinfo {author} {\bibfnamefont {A.}~\bibnamefont
  {Datta}}, \bibinfo {author} {\bibfnamefont {M.}~\bibnamefont {Duraisamy}}, \
  and\ \bibinfo {author} {\bibfnamefont {S.}~\bibnamefont {Khalil}},\ }\href
  {\doibase 10.1103/PhysRevD.83.094501} {\bibfield  {journal} {\bibinfo
  {journal} {Phys.Rev.}\ }\textbf {\bibinfo {volume} {D83}},\ \bibinfo {pages}
  {094501} (\bibinfo {year} {2011})},\ \Eprint {http://arxiv.org/abs/1011.5979}
  {arXiv:1011.5979 [hep-ph]} \BibitemShut {NoStop}%
\bibitem [{\citenamefont {Goertz}\ and\ \citenamefont
  {Pfoh}(2011)}]{Goertz:2011nx}%
  \BibitemOpen
  \bibfield  {author} {\bibinfo {author} {\bibfnamefont {F.}~\bibnamefont
  {Goertz}}\ and\ \bibinfo {author} {\bibfnamefont {T.}~\bibnamefont {Pfoh}},\
  }\href {\doibase 10.1103/PhysRevD.84.095016} {\bibfield  {journal} {\bibinfo
  {journal} {Phys.Rev.}\ }\textbf {\bibinfo {volume} {D84}},\ \bibinfo {pages}
  {095016} (\bibinfo {year} {2011})},\ \Eprint {http://arxiv.org/abs/1105.1507}
  {arXiv:1105.1507 [hep-ph]} \BibitemShut {NoStop}%
\bibitem [{\citenamefont {Deshpande}\ \emph {et~al.}(2010)\citenamefont
  {Deshpande}, \citenamefont {He},\ and\ \citenamefont
  {Valencia}}]{Deshpande:2010hy}%
  \BibitemOpen
  \bibfield  {author} {\bibinfo {author} {\bibfnamefont {N.}~\bibnamefont
  {Deshpande}}, \bibinfo {author} {\bibfnamefont {X.-G.}\ \bibnamefont {He}}, \
  and\ \bibinfo {author} {\bibfnamefont {G.}~\bibnamefont {Valencia}},\ }\href
  {\doibase 10.1103/PhysRevD.82.056013} {\bibfield  {journal} {\bibinfo
  {journal} {Phys.Rev.}\ }\textbf {\bibinfo {volume} {D82}},\ \bibinfo {pages}
  {056013} (\bibinfo {year} {2010})},\ \Eprint {http://arxiv.org/abs/1006.1682}
  {arXiv:1006.1682 [hep-ph]} \BibitemShut {NoStop}%
\bibitem [{\citenamefont {Alok}\ \emph {et~al.}(2011)\citenamefont {Alok},
  \citenamefont {Baek},\ and\ \citenamefont {London}}]{Alok:2010ij}%
  \BibitemOpen
  \bibfield  {author} {\bibinfo {author} {\bibfnamefont {A.~K.}\ \bibnamefont
  {Alok}}, \bibinfo {author} {\bibfnamefont {S.}~\bibnamefont {Baek}}, \ and\
  \bibinfo {author} {\bibfnamefont {D.}~\bibnamefont {London}},\ }\href
  {\doibase 10.1007/JHEP07(2011)111} {\bibfield  {journal} {\bibinfo  {journal}
  {JHEP}\ }\textbf {\bibinfo {volume} {1107}},\ \bibinfo {pages} {111}
  (\bibinfo {year} {2011})},\ \Eprint {http://arxiv.org/abs/1010.1333}
  {arXiv:1010.1333 [hep-ph]} \BibitemShut {NoStop}%
\bibitem [{\citenamefont {Kim}\ \emph {et~al.}(2011)\citenamefont {Kim},
  \citenamefont {Seo},\ and\ \citenamefont {Shin}}]{Kim:2010gx}%
  \BibitemOpen
  \bibfield  {author} {\bibinfo {author} {\bibfnamefont {J.~E.}\ \bibnamefont
  {Kim}}, \bibinfo {author} {\bibfnamefont {M.-S.}\ \bibnamefont {Seo}}, \ and\
  \bibinfo {author} {\bibfnamefont {S.}~\bibnamefont {Shin}},\ }\href {\doibase
  10.1103/PhysRevD.83.036003} {\bibfield  {journal} {\bibinfo  {journal}
  {Phys.Rev.}\ }\textbf {\bibinfo {volume} {D83}},\ \bibinfo {pages} {036003}
  (\bibinfo {year} {2011})},\ \Eprint {http://arxiv.org/abs/1010.5123}
  {arXiv:1010.5123 [hep-ph]} \BibitemShut {NoStop}%
\bibitem [{\citenamefont {Kim}\ \emph {et~al.}(2013)\citenamefont {Kim},
  \citenamefont {Kim},\ and\ \citenamefont {Shin}}]{Kim:2012rpa}%
  \BibitemOpen
  \bibfield  {author} {\bibinfo {author} {\bibfnamefont {H.~D.}\ \bibnamefont
  {Kim}}, \bibinfo {author} {\bibfnamefont {S.-G.}\ \bibnamefont {Kim}}, \ and\
  \bibinfo {author} {\bibfnamefont {S.}~\bibnamefont {Shin}},\ }\href {\doibase
  10.1103/PhysRevD.88.015005} {\bibfield  {journal} {\bibinfo  {journal}
  {Phys.Rev.}\ }\textbf {\bibinfo {volume} {D88}},\ \bibinfo {pages} {015005}
  (\bibinfo {year} {2013})},\ \Eprint {http://arxiv.org/abs/1205.6481}
  {arXiv:1205.6481 [hep-ph]} \BibitemShut {NoStop}%
\bibitem [{\citenamefont {Lee}\ and\ \citenamefont {Nam}(2012)}]{Lee:2011kn}%
  \BibitemOpen
  \bibfield  {author} {\bibinfo {author} {\bibfnamefont {K.~Y.}\ \bibnamefont
  {Lee}}\ and\ \bibinfo {author} {\bibfnamefont {S.-h.}\ \bibnamefont {Nam}},\
  }\href {\doibase 10.1103/PhysRevD.85.035001} {\bibfield  {journal} {\bibinfo
  {journal} {Phys.Rev.}\ }\textbf {\bibinfo {volume} {D85}},\ \bibinfo {pages}
  {035001} (\bibinfo {year} {2012})},\ \Eprint {http://arxiv.org/abs/1111.4666}
  {arXiv:1111.4666 [hep-ph]} \BibitemShut {NoStop}%
\bibitem [{\citenamefont {Jung}\ \emph {et~al.}(2010)\citenamefont {Jung},
  \citenamefont {Pich},\ and\ \citenamefont {Tuzon}}]{Jung:2010ik}%
  \BibitemOpen
  \bibfield  {author} {\bibinfo {author} {\bibfnamefont {M.}~\bibnamefont
  {Jung}}, \bibinfo {author} {\bibfnamefont {A.}~\bibnamefont {Pich}}, \ and\
  \bibinfo {author} {\bibfnamefont {P.}~\bibnamefont {Tuzon}},\ }\href
  {\doibase 10.1007/JHEP11(2010)003} {\bibfield  {journal} {\bibinfo  {journal}
  {JHEP}\ }\textbf {\bibinfo {volume} {1011}},\ \bibinfo {pages} {003}
  (\bibinfo {year} {2010})},\ \Eprint {http://arxiv.org/abs/1006.0470}
  {arXiv:1006.0470 [hep-ph]} \BibitemShut {NoStop}%
\bibitem [{\citenamefont {Dobrescu}\ \emph {et~al.}(2010)\citenamefont
  {Dobrescu}, \citenamefont {Fox},\ and\ \citenamefont
  {Martin}}]{Dobrescu:2010rh}%
  \BibitemOpen
  \bibfield  {author} {\bibinfo {author} {\bibfnamefont {B.~A.}\ \bibnamefont
  {Dobrescu}}, \bibinfo {author} {\bibfnamefont {P.~J.}\ \bibnamefont {Fox}}, \
  and\ \bibinfo {author} {\bibfnamefont {A.}~\bibnamefont {Martin}},\ }\href
  {\doibase 10.1103/PhysRevLett.105.041801} {\bibfield  {journal} {\bibinfo
  {journal} {Phys.Rev.Lett.}\ }\textbf {\bibinfo {volume} {105}},\ \bibinfo
  {pages} {041801} (\bibinfo {year} {2010})},\ \Eprint
  {http://arxiv.org/abs/1005.4238} {arXiv:1005.4238 [hep-ph]} \BibitemShut
  {NoStop}%
\bibitem [{\citenamefont {Trott}\ and\ \citenamefont
  {Wise}(2010)}]{Trott:2010iz}%
  \BibitemOpen
  \bibfield  {author} {\bibinfo {author} {\bibfnamefont {M.}~\bibnamefont
  {Trott}}\ and\ \bibinfo {author} {\bibfnamefont {M.~B.}\ \bibnamefont
  {Wise}},\ }\href {\doibase 10.1007/JHEP11(2010)157} {\bibfield  {journal}
  {\bibinfo  {journal} {JHEP}\ }\textbf {\bibinfo {volume} {1011}},\ \bibinfo
  {pages} {157} (\bibinfo {year} {2010})},\ \Eprint
  {http://arxiv.org/abs/1009.2813} {arXiv:1009.2813 [hep-ph]} \BibitemShut
  {NoStop}%
\bibitem [{\citenamefont {Bai}\ and\ \citenamefont
  {Nelson}(2010)}]{Bai:2010kf}%
  \BibitemOpen
  \bibfield  {author} {\bibinfo {author} {\bibfnamefont {Y.}~\bibnamefont
  {Bai}}\ and\ \bibinfo {author} {\bibfnamefont {A.~E.}\ \bibnamefont
  {Nelson}},\ }\href {\doibase 10.1103/PhysRevD.82.114027} {\bibfield
  {journal} {\bibinfo  {journal} {Phys.Rev.}\ }\textbf {\bibinfo {volume}
  {D82}},\ \bibinfo {pages} {114027} (\bibinfo {year} {2010})},\ \Eprint
  {http://arxiv.org/abs/1007.0596} {arXiv:1007.0596 [hep-ph]} \BibitemShut
  {NoStop}%
\bibitem [{\citenamefont {Chen}\ and\ \citenamefont
  {Faisel}(2011)}]{Chen:2010wv}%
  \BibitemOpen
  \bibfield  {author} {\bibinfo {author} {\bibfnamefont {C.-H.}\ \bibnamefont
  {Chen}}\ and\ \bibinfo {author} {\bibfnamefont {G.}~\bibnamefont {Faisel}},\
  }\href {\doibase 10.1016/j.physletb.2011.01.010} {\bibfield  {journal}
  {\bibinfo  {journal} {Phys.Lett.}\ }\textbf {\bibinfo {volume} {B696}},\
  \bibinfo {pages} {487} (\bibinfo {year} {2011})},\ \Eprint
  {http://arxiv.org/abs/1005.4582} {arXiv:1005.4582 [hep-ph]} \BibitemShut
  {NoStop}%
\bibitem [{\citenamefont {Hou}\ and\ \citenamefont
  {Mahajan}(2007)}]{Hou:2007ps}%
  \BibitemOpen
  \bibfield  {author} {\bibinfo {author} {\bibfnamefont {W.-S.}\ \bibnamefont
  {Hou}}\ and\ \bibinfo {author} {\bibfnamefont {N.}~\bibnamefont {Mahajan}},\
  }\href {\doibase 10.1103/PhysRevD.75.077501} {\bibfield  {journal} {\bibinfo
  {journal} {Phys.Rev.}\ }\textbf {\bibinfo {volume} {D75}},\ \bibinfo {pages}
  {077501} (\bibinfo {year} {2007})},\ \Eprint
  {http://arxiv.org/abs/hep-ph/0702163} {arXiv:hep-ph/0702163 [HEP-PH]}
  \BibitemShut {NoStop}%
\bibitem [{\citenamefont {Soni}\ \emph {et~al.}(2010)\citenamefont {Soni},
  \citenamefont {Alok}, \citenamefont {Giri}, \citenamefont {Mohanta},\ and\
  \citenamefont {Nandi}}]{Soni:2010xh}%
  \BibitemOpen
  \bibfield  {author} {\bibinfo {author} {\bibfnamefont {A.}~\bibnamefont
  {Soni}}, \bibinfo {author} {\bibfnamefont {A.~K.}\ \bibnamefont {Alok}},
  \bibinfo {author} {\bibfnamefont {A.}~\bibnamefont {Giri}}, \bibinfo {author}
  {\bibfnamefont {R.}~\bibnamefont {Mohanta}}, \ and\ \bibinfo {author}
  {\bibfnamefont {S.}~\bibnamefont {Nandi}},\ }\href {\doibase
  10.1103/PhysRevD.82.033009} {\bibfield  {journal} {\bibinfo  {journal}
  {Phys.Rev.}\ }\textbf {\bibinfo {volume} {D82}},\ \bibinfo {pages} {033009}
  (\bibinfo {year} {2010})},\ \Eprint {http://arxiv.org/abs/1002.0595}
  {arXiv:1002.0595 [hep-ph]} \BibitemShut {NoStop}%
\bibitem [{\citenamefont {Chen}\ \emph {et~al.}(2010)\citenamefont {Chen},
  \citenamefont {Geng},\ and\ \citenamefont {Wang}}]{Chen:2010aq}%
  \BibitemOpen
  \bibfield  {author} {\bibinfo {author} {\bibfnamefont {C.-H.}\ \bibnamefont
  {Chen}}, \bibinfo {author} {\bibfnamefont {C.-Q.}\ \bibnamefont {Geng}}, \
  and\ \bibinfo {author} {\bibfnamefont {W.}~\bibnamefont {Wang}},\ }\href
  {\doibase 10.1007/JHEP11(2010)089} {\bibfield  {journal} {\bibinfo  {journal}
  {JHEP}\ }\textbf {\bibinfo {volume} {1011}},\ \bibinfo {pages} {089}
  (\bibinfo {year} {2010})},\ \Eprint {http://arxiv.org/abs/1006.5216}
  {arXiv:1006.5216 [hep-ph]} \BibitemShut {NoStop}%
\bibitem [{\citenamefont {Botella}\ \emph {et~al.}(2009)\citenamefont
  {Botella}, \citenamefont {Branco},\ and\ \citenamefont
  {Nebot}}]{Botella:2008qm}%
  \BibitemOpen
  \bibfield  {author} {\bibinfo {author} {\bibfnamefont {F.~J.}\ \bibnamefont
  {Botella}}, \bibinfo {author} {\bibfnamefont {G.~C.}\ \bibnamefont {Branco}},
  \ and\ \bibinfo {author} {\bibfnamefont {M.}~\bibnamefont {Nebot}},\ }\href
  {\doibase 10.1103/PhysRevD.79.096009} {\bibfield  {journal} {\bibinfo
  {journal} {Phys.Rev.}\ }\textbf {\bibinfo {volume} {D79}},\ \bibinfo {pages}
  {096009} (\bibinfo {year} {2009})},\ \Eprint {http://arxiv.org/abs/0805.3995}
  {arXiv:0805.3995 [hep-ph]} \BibitemShut {NoStop}%
\bibitem [{\citenamefont {Botella}\ \emph {et~al.}(2012)\citenamefont
  {Botella}, \citenamefont {Branco},\ and\ \citenamefont
  {Nebot}}]{Botella:2012ju}%
  \BibitemOpen
  \bibfield  {author} {\bibinfo {author} {\bibfnamefont {F.}~\bibnamefont
  {Botella}}, \bibinfo {author} {\bibfnamefont {G.}~\bibnamefont {Branco}}, \
  and\ \bibinfo {author} {\bibfnamefont {M.}~\bibnamefont {Nebot}},\ }\href
  {\doibase 10.1007/JHEP12(2012)040} {\bibfield  {journal} {\bibinfo  {journal}
  {JHEP}\ }\textbf {\bibinfo {volume} {1212}},\ \bibinfo {pages} {040}
  (\bibinfo {year} {2012})},\ \Eprint {http://arxiv.org/abs/1207.4440}
  {arXiv:1207.4440 [hep-ph]} \BibitemShut {NoStop}%
\bibitem [{\citenamefont {Alok}\ and\ \citenamefont
  {Gangal}(2012)}]{Alok:2012xm}%
  \BibitemOpen
  \bibfield  {author} {\bibinfo {author} {\bibfnamefont {A.~K.}\ \bibnamefont
  {Alok}}\ and\ \bibinfo {author} {\bibfnamefont {S.}~\bibnamefont {Gangal}},\
  }\href {\doibase 10.1103/PhysRevD.86.114009} {\bibfield  {journal} {\bibinfo
  {journal} {Phys.Rev.}\ }\textbf {\bibinfo {volume} {D86}},\ \bibinfo {pages}
  {114009} (\bibinfo {year} {2012})},\ \Eprint {http://arxiv.org/abs/1209.1987}
  {arXiv:1209.1987 [hep-ph]} \BibitemShut {NoStop}%
\bibitem [{\citenamefont {Alok}\ \emph {et~al.}(2014)\citenamefont {Alok},
  \citenamefont {Banerjee}, \citenamefont {Kumar},\ and\ \citenamefont
  {Sankar}}]{Alok:2014yua}%
  \BibitemOpen
  \bibfield  {author} {\bibinfo {author} {\bibfnamefont {A.~K.}\ \bibnamefont
  {Alok}}, \bibinfo {author} {\bibfnamefont {S.}~\bibnamefont {Banerjee}},
  \bibinfo {author} {\bibfnamefont {D.}~\bibnamefont {Kumar}}, \ and\ \bibinfo
  {author} {\bibfnamefont {S.~U.}\ \bibnamefont {Sankar}},\ }\href@noop {} {\
  (\bibinfo {year} {2014})},\ \Eprint {http://arxiv.org/abs/1402.1023}
  {arXiv:1402.1023 [hep-ph]} \BibitemShut {NoStop}%
\bibitem [{\citenamefont {Ligeti}\ \emph {et~al.}(2010)\citenamefont {Ligeti},
  \citenamefont {Papucci}, \citenamefont {Perez},\ and\ \citenamefont
  {Zupan}}]{Ligeti:2010ia}%
  \BibitemOpen
  \bibfield  {author} {\bibinfo {author} {\bibfnamefont {Z.}~\bibnamefont
  {Ligeti}}, \bibinfo {author} {\bibfnamefont {M.}~\bibnamefont {Papucci}},
  \bibinfo {author} {\bibfnamefont {G.}~\bibnamefont {Perez}}, \ and\ \bibinfo
  {author} {\bibfnamefont {J.}~\bibnamefont {Zupan}},\ }\href {\doibase
  10.1103/PhysRevLett.105.131601} {\bibfield  {journal} {\bibinfo  {journal}
  {Phys.Rev.Lett.}\ }\textbf {\bibinfo {volume} {105}},\ \bibinfo {pages}
  {131601} (\bibinfo {year} {2010})},\ \Eprint {http://arxiv.org/abs/1006.0432}
  {arXiv:1006.0432 [hep-ph]} \BibitemShut {NoStop}%
\bibitem [{\citenamefont {Bauer}\ and\ \citenamefont
  {Dunn}(2011)}]{Bauer:2010dga}%
  \BibitemOpen
  \bibfield  {author} {\bibinfo {author} {\bibfnamefont {C.~W.}\ \bibnamefont
  {Bauer}}\ and\ \bibinfo {author} {\bibfnamefont {N.~D.}\ \bibnamefont
  {Dunn}},\ }\href {\doibase 10.1016/j.physletb.2010.12.039} {\bibfield
  {journal} {\bibinfo  {journal} {Phys.Lett.}\ }\textbf {\bibinfo {volume}
  {B696}},\ \bibinfo {pages} {362} (\bibinfo {year} {2011})},\ \Eprint
  {http://arxiv.org/abs/1006.1629} {arXiv:1006.1629 [hep-ph]} \BibitemShut
  {NoStop}%
\bibitem [{\citenamefont {Bobeth}\ and\ \citenamefont
  {Haisch}(2013)}]{Bobeth:2011st}%
  \BibitemOpen
  \bibfield  {author} {\bibinfo {author} {\bibfnamefont {C.}~\bibnamefont
  {Bobeth}}\ and\ \bibinfo {author} {\bibfnamefont {U.}~\bibnamefont
  {Haisch}},\ }\href@noop {} {\bibfield  {journal} {\bibinfo  {journal} {Acta
  Phys.Polon.}\ }\textbf {\bibinfo {volume} {B44}},\ \bibinfo {pages} {127}
  (\bibinfo {year} {2013})},\ \Eprint {http://arxiv.org/abs/1109.1826}
  {arXiv:1109.1826 [hep-ph]} \BibitemShut {NoStop}%
\bibitem [{\citenamefont {Bobeth}\ \emph {et~al.}(2014)\citenamefont {Bobeth},
  \citenamefont {Haisch}, \citenamefont {Lenz}, \citenamefont {Pecjak},\ and\
  \citenamefont {Tetlalmatzi-Xolocotzi}}]{Bobeth:2014rda}%
  \BibitemOpen
  \bibfield  {author} {\bibinfo {author} {\bibfnamefont {C.}~\bibnamefont
  {Bobeth}}, \bibinfo {author} {\bibfnamefont {U.}~\bibnamefont {Haisch}},
  \bibinfo {author} {\bibfnamefont {A.}~\bibnamefont {Lenz}}, \bibinfo {author}
  {\bibfnamefont {B.}~\bibnamefont {Pecjak}}, \ and\ \bibinfo {author}
  {\bibfnamefont {G.}~\bibnamefont {Tetlalmatzi-Xolocotzi}},\ }\href {\doibase
  10.1007/JHEP06(2014)040} {\  (\bibinfo {year} {2014}),\
  10.1007/JHEP06(2014)040},\ \Eprint {http://arxiv.org/abs/1404.2531}
  {arXiv:1404.2531 [hep-ph]} \BibitemShut {NoStop}%
\bibitem [{\citenamefont {Descotes-Genon}\ and\ \citenamefont
  {Kamenik}(2013)}]{DescotesGenon:2012kr}%
  \BibitemOpen
  \bibfield  {author} {\bibinfo {author} {\bibfnamefont {S.}~\bibnamefont
  {Descotes-Genon}}\ and\ \bibinfo {author} {\bibfnamefont {J.~F.}\
  \bibnamefont {Kamenik}},\ }\href {\doibase 10.1103/PhysRevD.87.074036}
  {\bibfield  {journal} {\bibinfo  {journal} {Phys.Rev.}\ }\textbf {\bibinfo
  {volume} {D87}},\ \bibinfo {pages} {074036} (\bibinfo {year} {2013})},\
  \Eprint {http://arxiv.org/abs/1207.4483} {arXiv:1207.4483 [hep-ph]}
  \BibitemShut {NoStop}%
\bibitem [{\citenamefont {Branco}\ \emph {et~al.}(1999)\citenamefont {Branco},
  \citenamefont {Lavoura},\ and\ \citenamefont {Silva}}]{Branco:1999fs}%
  \BibitemOpen
  \bibfield  {author} {\bibinfo {author} {\bibfnamefont {G.~C.}\ \bibnamefont
  {Branco}}, \bibinfo {author} {\bibfnamefont {L.}~\bibnamefont {Lavoura}}, \
  and\ \bibinfo {author} {\bibfnamefont {J.~P.}\ \bibnamefont {Silva}},\
  }\href@noop {} {\emph {\bibinfo {title} {{CP Violation}}}},\ Vol.\ \bibinfo
  {volume} {103}\ (\bibinfo {year} {1999})\ pp.\ \bibinfo {pages}
  {1--536}\BibitemShut {NoStop}%
\bibitem [{Note4()}]{Note4}%
  \BibitemOpen
  \bibinfo {note} {Equation (\ref {eq:M12q:01}) includes perturbative QCD
  corrections $\eta _B$, and non-perturbative information, i.e. the decay
  constant $f_{B_q}$ and the bag parameter $B_q$. Subleading contributions from
  virtual $u$ or $c$ quarks are neglected.}\BibitemShut {Stop}%
\bibitem [{\citenamefont {Beneke}\ \emph {et~al.}(1999)\citenamefont {Beneke},
  \citenamefont {Buchalla}, \citenamefont {Greub}, \citenamefont {Lenz},\ and\
  \citenamefont {Nierste}}]{Beneke:1998sy}%
  \BibitemOpen
  \bibfield  {author} {\bibinfo {author} {\bibfnamefont {M.}~\bibnamefont
  {Beneke}}, \bibinfo {author} {\bibfnamefont {G.}~\bibnamefont {Buchalla}},
  \bibinfo {author} {\bibfnamefont {C.}~\bibnamefont {Greub}}, \bibinfo
  {author} {\bibfnamefont {A.}~\bibnamefont {Lenz}}, \ and\ \bibinfo {author}
  {\bibfnamefont {U.}~\bibnamefont {Nierste}},\ }\href@noop {} {\bibfield
  {journal} {\bibinfo  {journal} {Phys.Lett.}\ }\textbf {\bibinfo {volume}
  {B459}},\ \bibinfo {pages} {631} (\bibinfo {year} {1999})},\ \Eprint
  {http://arxiv.org/abs/hep-ph/9808385} {arXiv:hep-ph/9808385 [hep-ph]}
  \BibitemShut {NoStop}%
\bibitem [{\citenamefont {Beneke}\ \emph {et~al.}(2003)\citenamefont {Beneke},
  \citenamefont {Buchalla}, \citenamefont {Lenz},\ and\ \citenamefont
  {Nierste}}]{Beneke:2003az}%
  \BibitemOpen
  \bibfield  {author} {\bibinfo {author} {\bibfnamefont {M.}~\bibnamefont
  {Beneke}}, \bibinfo {author} {\bibfnamefont {G.}~\bibnamefont {Buchalla}},
  \bibinfo {author} {\bibfnamefont {A.}~\bibnamefont {Lenz}}, \ and\ \bibinfo
  {author} {\bibfnamefont {U.}~\bibnamefont {Nierste}},\ }\href@noop {}
  {\bibfield  {journal} {\bibinfo  {journal} {Phys.Lett.}\ }\textbf {\bibinfo
  {volume} {B576}},\ \bibinfo {pages} {173} (\bibinfo {year} {2003})},\ \Eprint
  {http://arxiv.org/abs/hep-ph/0307344} {arXiv:hep-ph/0307344 [hep-ph]}
  \BibitemShut {NoStop}%
\bibitem [{\citenamefont {Ciuchini}\ \emph {et~al.}(2003)\citenamefont
  {Ciuchini}, \citenamefont {Franco}, \citenamefont {Lubicz}, \citenamefont
  {Mescia},\ and\ \citenamefont {Tarantino}}]{Ciuchini:2003ww}%
  \BibitemOpen
  \bibfield  {author} {\bibinfo {author} {\bibfnamefont {M.}~\bibnamefont
  {Ciuchini}}, \bibinfo {author} {\bibfnamefont {E.}~\bibnamefont {Franco}},
  \bibinfo {author} {\bibfnamefont {V.}~\bibnamefont {Lubicz}}, \bibinfo
  {author} {\bibfnamefont {F.}~\bibnamefont {Mescia}}, \ and\ \bibinfo {author}
  {\bibfnamefont {C.}~\bibnamefont {Tarantino}},\ }\href@noop {} {\bibfield
  {journal} {\bibinfo  {journal} {JHEP}\ }\textbf {\bibinfo {volume} {0308}},\
  \bibinfo {pages} {031} (\bibinfo {year} {2003})},\ \Eprint
  {http://arxiv.org/abs/hep-ph/0308029} {arXiv:hep-ph/0308029 [hep-ph]}
  \BibitemShut {NoStop}%
\bibitem [{\citenamefont {Lenz}(2012)}]{Lenz:2012mb}%
  \BibitemOpen
  \bibfield  {author} {\bibinfo {author} {\bibfnamefont {A.}~\bibnamefont
  {Lenz}},\ }\href@noop {} {\  (\bibinfo {year} {2012})},\ \Eprint
  {http://arxiv.org/abs/1205.1444} {arXiv:1205.1444 [hep-ph]} \BibitemShut
  {NoStop}%
\bibitem [{\citenamefont {Hagelin}(1981)}]{Hagelin:1981zk}%
  \BibitemOpen
  \bibfield  {author} {\bibinfo {author} {\bibfnamefont {J.}~\bibnamefont
  {Hagelin}},\ }\href {\doibase 10.1016/0550-3213(81)90521-6} {\bibfield
  {journal} {\bibinfo  {journal} {Nucl.Phys.}\ }\textbf {\bibinfo {volume}
  {B193}},\ \bibinfo {pages} {123} (\bibinfo {year} {1981})}\BibitemShut
  {NoStop}%
\bibitem [{\citenamefont {Lenz}\ and\ \citenamefont
  {Nierste}(2007)}]{Lenz:2006hd}%
  \BibitemOpen
  \bibfield  {author} {\bibinfo {author} {\bibfnamefont {A.}~\bibnamefont
  {Lenz}}\ and\ \bibinfo {author} {\bibfnamefont {U.}~\bibnamefont {Nierste}},\
  }\href {\doibase 10.1088/1126-6708/2007/06/072} {\bibfield  {journal}
  {\bibinfo  {journal} {JHEP}\ }\textbf {\bibinfo {volume} {0706}},\ \bibinfo
  {pages} {072} (\bibinfo {year} {2007})},\ \Eprint
  {http://arxiv.org/abs/hep-ph/0612167} {arXiv:hep-ph/0612167 [hep-ph]}
  \BibitemShut {NoStop}%
\bibitem [{\citenamefont {Botella}\ \emph {et~al.}(2007)\citenamefont
  {Botella}, \citenamefont {Branco},\ and\ \citenamefont
  {Nebot}}]{Botella:2006va}%
  \BibitemOpen
  \bibfield  {author} {\bibinfo {author} {\bibfnamefont {F.~J.}\ \bibnamefont
  {Botella}}, \bibinfo {author} {\bibfnamefont {G.~C.}\ \bibnamefont {Branco}},
  \ and\ \bibinfo {author} {\bibfnamefont {M.}~\bibnamefont {Nebot}},\ }\href
  {\doibase 10.1016/j.nuclphysb.2006.12.022} {\bibfield  {journal} {\bibinfo
  {journal} {Nucl.Phys.}\ }\textbf {\bibinfo {volume} {B768}},\ \bibinfo
  {pages} {1} (\bibinfo {year} {2007})},\ \Eprint
  {http://arxiv.org/abs/hep-ph/0608100} {arXiv:hep-ph/0608100 [hep-ph]}
  \BibitemShut {NoStop}%
\bibitem [{Note5()}]{Note5}%
  \BibitemOpen
  \bibinfo {note} {In both $B^0_d$--$\protect \mathaccentV {bar}016B^0_d$\ and
  $B^0_s$--$\protect \mathaccentV {bar}016B^0_s$\ systems, $\Delta
  M_{B_q}=2|M_{12}^{({q})}|$ since $|\Gamma _{12}^{({q})}|\ll |M_{12}^{({q})}|$
  \cite {Branco:1999fs}.}\BibitemShut {Stop}%
\bibitem [{Note6()}]{Note6}%
  \BibitemOpen
  \bibinfo {note} {Notice that eq.(\ref {eq:G12d:01}) is written, as it should,
  in terms of quantities invariant under rephasings of the CKM elements and of
  the $B_d^0$ and $\protect \mathaccentV {bar}016B_d^0$ states, even if, for
  the sake of brevity, intermediate expressions such as eq.(\ref {eq:M12d:01})
  are not.}\BibitemShut {Stop}%
\bibitem [{Note7()}]{Note7}%
  \BibitemOpen
  \bibinfo {note} {Besides $2\protect \mathaccentV {bar}016\beta $ from the
  golden channel $B_d\to J/\Psi K_s$, $\gamma $ is accessed through \protect
  \emph {tree} level decays such as $B_d\to DK$, while the combination
  $2(\protect \mathaccentV {bar}016\beta +\gamma )$ is obtained in decay
  channels $B_d\to \pi \pi , \rho \pi ,\rho \rho $.}\BibitemShut {Stop}%
\bibitem [{\citenamefont {Bird}()}]{Bird:Discrete2012}%
  \BibitemOpen
  \bibfield  {author} {\bibinfo {author} {\bibfnamefont {T.}~\bibnamefont
  {Bird}},\ }\href@noop {} {\enquote {\bibinfo {title} {{Studies of CP
  violation using semileptonic B decays}},}\ }\bibinfo {howpublished} {Discrete
  12, Lisbon}\BibitemShut {NoStop}%
\bibitem [{\citenamefont {Laplace}\ \emph {et~al.}(2002)\citenamefont
  {Laplace}, \citenamefont {Ligeti}, \citenamefont {Nir},\ and\ \citenamefont
  {Perez}}]{Laplace:2002ik}%
  \BibitemOpen
  \bibfield  {author} {\bibinfo {author} {\bibfnamefont {S.}~\bibnamefont
  {Laplace}}, \bibinfo {author} {\bibfnamefont {Z.}~\bibnamefont {Ligeti}},
  \bibinfo {author} {\bibfnamefont {Y.}~\bibnamefont {Nir}}, \ and\ \bibinfo
  {author} {\bibfnamefont {G.}~\bibnamefont {Perez}},\ }\href {\doibase
  10.1103/PhysRevD.65.094040} {\bibfield  {journal} {\bibinfo  {journal}
  {Phys.Rev.}\ }\textbf {\bibinfo {volume} {D65}},\ \bibinfo {pages} {094040}
  (\bibinfo {year} {2002})},\ \Eprint {http://arxiv.org/abs/hep-ph/0202010}
  {arXiv:hep-ph/0202010 [hep-ph]} \BibitemShut {NoStop}%
\bibitem [{\citenamefont {Ligeti}\ \emph {et~al.}(2006)\citenamefont {Ligeti},
  \citenamefont {Papucci},\ and\ \citenamefont {Perez}}]{Ligeti:2006pm}%
  \BibitemOpen
  \bibfield  {author} {\bibinfo {author} {\bibfnamefont {Z.}~\bibnamefont
  {Ligeti}}, \bibinfo {author} {\bibfnamefont {M.}~\bibnamefont {Papucci}}, \
  and\ \bibinfo {author} {\bibfnamefont {G.}~\bibnamefont {Perez}},\ }\href
  {\doibase 10.1103/PhysRevLett.97.101801} {\bibfield  {journal} {\bibinfo
  {journal} {Phys.Rev.Lett.}\ }\textbf {\bibinfo {volume} {97}},\ \bibinfo
  {pages} {101801} (\bibinfo {year} {2006})},\ \Eprint
  {http://arxiv.org/abs/hep-ph/0604112} {arXiv:hep-ph/0604112 [hep-ph]}
  \BibitemShut {NoStop}%
\bibitem [{\citenamefont {Ball}\ and\ \citenamefont
  {Fleischer}(2006)}]{Ball:2006xx}%
  \BibitemOpen
  \bibfield  {author} {\bibinfo {author} {\bibfnamefont {P.}~\bibnamefont
  {Ball}}\ and\ \bibinfo {author} {\bibfnamefont {R.}~\bibnamefont
  {Fleischer}},\ }\href {\doibase 10.1140/epjc/s10052-006-0034-4} {\bibfield
  {journal} {\bibinfo  {journal} {Eur.Phys.J.}\ }\textbf {\bibinfo {volume}
  {C48}},\ \bibinfo {pages} {413} (\bibinfo {year} {2006})},\ \Eprint
  {http://arxiv.org/abs/hep-ph/0604249} {arXiv:hep-ph/0604249 [hep-ph]}
  \BibitemShut {NoStop}%
\bibitem [{\citenamefont {Grossman}\ \emph {et~al.}(2006)\citenamefont
  {Grossman}, \citenamefont {Nir},\ and\ \citenamefont
  {Raz}}]{Grossman:2006ce}%
  \BibitemOpen
  \bibfield  {author} {\bibinfo {author} {\bibfnamefont {Y.}~\bibnamefont
  {Grossman}}, \bibinfo {author} {\bibfnamefont {Y.}~\bibnamefont {Nir}}, \
  and\ \bibinfo {author} {\bibfnamefont {G.}~\bibnamefont {Raz}},\ }\href
  {\doibase 10.1103/PhysRevLett.97.151801} {\bibfield  {journal} {\bibinfo
  {journal} {Phys.Rev.Lett.}\ }\textbf {\bibinfo {volume} {97}},\ \bibinfo
  {pages} {151801} (\bibinfo {year} {2006})},\ \Eprint
  {http://arxiv.org/abs/hep-ph/0605028} {arXiv:hep-ph/0605028 [hep-ph]}
  \BibitemShut {NoStop}%
\bibitem [{\citenamefont {Bona}\ and\ \citenamefont {others
  (UTfit~Collaboration)}(2006)}]{Bona:2006sa}%
  \BibitemOpen
  \bibfield  {author} {\bibinfo {author} {\bibfnamefont {M.}~\bibnamefont
  {Bona}}\ and\ \bibinfo {author} {\bibnamefont {others (UTfit~Collaboration)}}
  (\bibinfo {collaboration} {UTfit Collaboration}),\ }\href {\doibase
  10.1103/PhysRevLett.97.151803} {\bibfield  {journal} {\bibinfo  {journal}
  {Phys.Rev.Lett.}\ }\textbf {\bibinfo {volume} {97}},\ \bibinfo {pages}
  {151803} (\bibinfo {year} {2006})},\ \Eprint
  {http://arxiv.org/abs/hep-ph/0605213} {arXiv:hep-ph/0605213 [hep-ph]}
  \BibitemShut {NoStop}%
\bibitem [{\citenamefont {Botella}\ \emph {et~al.}(2005)\citenamefont
  {Botella}, \citenamefont {Branco}, \citenamefont {Nebot},\ and\ \citenamefont
  {Rebelo}}]{Botella:2005fc}%
  \BibitemOpen
  \bibfield  {author} {\bibinfo {author} {\bibfnamefont {F.}~\bibnamefont
  {Botella}}, \bibinfo {author} {\bibfnamefont {G.}~\bibnamefont {Branco}},
  \bibinfo {author} {\bibfnamefont {M.}~\bibnamefont {Nebot}}, \ and\ \bibinfo
  {author} {\bibfnamefont {M.}~\bibnamefont {Rebelo}},\ }\href {\doibase
  10.1016/j.nuclphysb.2005.07.006} {\bibfield  {journal} {\bibinfo  {journal}
  {Nucl.Phys.}\ }\textbf {\bibinfo {volume} {B725}},\ \bibinfo {pages} {155}
  (\bibinfo {year} {2005})},\ \Eprint {http://arxiv.org/abs/hep-ph/0502133}
  {arXiv:hep-ph/0502133 [hep-ph]} \BibitemShut {NoStop}%
\bibitem [{\citenamefont {Bona}\ \emph {et~al.}(2008)\citenamefont {Bona} \emph
  {et~al.}}]{Bona:2007vi}%
  \BibitemOpen
  \bibfield  {author} {\bibinfo {author} {\bibfnamefont {M.}~\bibnamefont
  {Bona}} \emph {et~al.} (\bibinfo {collaboration} {UTfit Collaboration}),\
  }\href {\doibase 10.1088/1126-6708/2008/03/049} {\bibfield  {journal}
  {\bibinfo  {journal} {JHEP}\ }\textbf {\bibinfo {volume} {0803}},\ \bibinfo
  {pages} {049} (\bibinfo {year} {2008})},\ \Eprint
  {http://arxiv.org/abs/0707.0636} {arXiv:0707.0636 [hep-ph]} \BibitemShut
  {NoStop}%
\bibitem [{\citenamefont {Lenz}\ \emph {et~al.}(2012)\citenamefont {Lenz},
  \citenamefont {Nierste}, \citenamefont {Charles}, \citenamefont
  {Descotes-Genon}, \citenamefont {Lacker} \emph {et~al.}}]{Lenz:2012az}%
  \BibitemOpen
  \bibfield  {author} {\bibinfo {author} {\bibfnamefont {A.}~\bibnamefont
  {Lenz}}, \bibinfo {author} {\bibfnamefont {U.}~\bibnamefont {Nierste}},
  \bibinfo {author} {\bibfnamefont {J.}~\bibnamefont {Charles}}, \bibinfo
  {author} {\bibfnamefont {S.}~\bibnamefont {Descotes-Genon}}, \bibinfo
  {author} {\bibfnamefont {H.}~\bibnamefont {Lacker}},  \emph {et~al.},\ }\href
  {\doibase 10.1103/PhysRevD.86.033008} {\bibfield  {journal} {\bibinfo
  {journal} {Phys.Rev.}\ }\textbf {\bibinfo {volume} {D86}},\ \bibinfo {pages}
  {033008} (\bibinfo {year} {2012})},\ \Eprint {http://arxiv.org/abs/1203.0238}
  {arXiv:1203.0238 [hep-ph]} \BibitemShut {NoStop}%
\bibitem [{Note8()}]{Note8}%
  \BibitemOpen
  \bibinfo {note} {Another popular alternative uses the NP parameters $h_q$ and
  $\sigma _q$ with $r_q^2e^{-i2\phi _q}\equiv 1+h_qe^{i2\sigma _q}$, where this
  separation of NP is less straightforward. Our results, in any case, do not
  depend on adopting one parametrization or the other.}\BibitemShut {Stop}%
\bibitem [{Note9()}]{Note9}%
  \BibitemOpen
  \bibinfo {note} {In fact, \protect \emph {tightly} constrained, except for
  the argument of $M_{12}^{({s})}$, accessed through $B_s\to J/\Psi \Phi $,
  where, despite the excellent performance of LHCb, the smallness of the SM
  expectation still allows for deviations.}\BibitemShut {Stop}%
\bibitem [{\citenamefont {Bona}\ and\ \citenamefont {others
  (UTfit~Collaboration)}(2010)}]{Bona:2009cj}%
  \BibitemOpen
  \bibfield  {author} {\bibinfo {author} {\bibfnamefont {M.}~\bibnamefont
  {Bona}}\ and\ \bibinfo {author} {\bibnamefont {others (UTfit~Collaboration)}}
  (\bibinfo {collaboration} {UTfit Collaboration}),\ }\href {\doibase
  10.1016/j.physletb.2010.02.063} {\bibfield  {journal} {\bibinfo  {journal}
  {Phys.Lett.}\ }\textbf {\bibinfo {volume} {B687}},\ \bibinfo {pages} {61}
  (\bibinfo {year} {2010})},\ \Eprint {http://arxiv.org/abs/0908.3470}
  {arXiv:0908.3470 [hep-ph]} \BibitemShut {NoStop}%
\bibitem [{\citenamefont {Lunghi}\ and\ \citenamefont
  {Soni}(2011)}]{Lunghi:2010gv}%
  \BibitemOpen
  \bibfield  {author} {\bibinfo {author} {\bibfnamefont {E.}~\bibnamefont
  {Lunghi}}\ and\ \bibinfo {author} {\bibfnamefont {A.}~\bibnamefont {Soni}},\
  }\href {\doibase 10.1016/j.physletb.2011.02.016} {\bibfield  {journal}
  {\bibinfo  {journal} {Phys.Lett.}\ }\textbf {\bibinfo {volume} {B697}},\
  \bibinfo {pages} {323} (\bibinfo {year} {2011})},\ \Eprint
  {http://arxiv.org/abs/1010.6069} {arXiv:1010.6069 [hep-ph]} \BibitemShut
  {NoStop}%
\bibitem [{Note10()}]{Note10}%
  \BibitemOpen
  \bibinfo {note} {For these and successive two dimensional $\Delta \chi ^2$
  profiles, we display, for clarity, 68\%, 95\% and 99\% CL
  regions.}\BibitemShut {Stop}%
\bibitem [{\citenamefont {Barenboim}\ and\ \citenamefont
  {Botella}(1998)}]{Barenboim:1997pf}%
  \BibitemOpen
  \bibfield  {author} {\bibinfo {author} {\bibfnamefont {G.}~\bibnamefont
  {Barenboim}}\ and\ \bibinfo {author} {\bibfnamefont {F.}~\bibnamefont
  {Botella}},\ }\href {\doibase 10.1016/S0370-2693(98)00695-9} {\bibfield
  {journal} {\bibinfo  {journal} {Phys.Lett.}\ }\textbf {\bibinfo {volume}
  {B433}},\ \bibinfo {pages} {385} (\bibinfo {year} {1998})},\ \Eprint
  {http://arxiv.org/abs/hep-ph/9708209} {arXiv:hep-ph/9708209 [hep-ph]}
  \BibitemShut {NoStop}%
\bibitem [{\citenamefont {Barenboim}\ \emph {et~al.}(1998)\citenamefont
  {Barenboim}, \citenamefont {Botella}, \citenamefont {Branco},\ and\
  \citenamefont {Vives}}]{Barenboim:1997qx}%
  \BibitemOpen
  \bibfield  {author} {\bibinfo {author} {\bibfnamefont {G.}~\bibnamefont
  {Barenboim}}, \bibinfo {author} {\bibfnamefont {F.}~\bibnamefont {Botella}},
  \bibinfo {author} {\bibfnamefont {G.}~\bibnamefont {Branco}}, \ and\ \bibinfo
  {author} {\bibfnamefont {O.}~\bibnamefont {Vives}},\ }\href {\doibase
  10.1016/S0370-2693(97)01515-3} {\bibfield  {journal} {\bibinfo  {journal}
  {Phys.Lett.}\ }\textbf {\bibinfo {volume} {B422}},\ \bibinfo {pages} {277}
  (\bibinfo {year} {1998})},\ \Eprint {http://arxiv.org/abs/hep-ph/9709369}
  {arXiv:hep-ph/9709369 [hep-ph]} \BibitemShut {NoStop}%
\bibitem [{\citenamefont {Barenboim}\ \emph
  {et~al.}(2001{\natexlab{a}})\citenamefont {Barenboim}, \citenamefont
  {Botella},\ and\ \citenamefont {Vives}}]{Barenboim:2000zz}%
  \BibitemOpen
  \bibfield  {author} {\bibinfo {author} {\bibfnamefont {G.}~\bibnamefont
  {Barenboim}}, \bibinfo {author} {\bibfnamefont {F.}~\bibnamefont {Botella}},
  \ and\ \bibinfo {author} {\bibfnamefont {O.}~\bibnamefont {Vives}},\ }\href
  {\doibase 10.1103/PhysRevD.64.015007} {\bibfield  {journal} {\bibinfo
  {journal} {Phys.Rev.}\ }\textbf {\bibinfo {volume} {D64}},\ \bibinfo {pages}
  {015007} (\bibinfo {year} {2001}{\natexlab{a}})},\ \Eprint
  {http://arxiv.org/abs/hep-ph/0012197} {arXiv:hep-ph/0012197 [hep-ph]}
  \BibitemShut {NoStop}%
\bibitem [{\citenamefont {Barenboim}\ \emph
  {et~al.}(2001{\natexlab{b}})\citenamefont {Barenboim}, \citenamefont
  {Botella},\ and\ \citenamefont {Vives}}]{Barenboim:2001fd}%
  \BibitemOpen
  \bibfield  {author} {\bibinfo {author} {\bibfnamefont {G.}~\bibnamefont
  {Barenboim}}, \bibinfo {author} {\bibfnamefont {F.}~\bibnamefont {Botella}},
  \ and\ \bibinfo {author} {\bibfnamefont {O.}~\bibnamefont {Vives}},\
  }\href@noop {} {\bibfield  {journal} {\bibinfo  {journal} {Nucl.Phys.}\
  }\textbf {\bibinfo {volume} {B613}},\ \bibinfo {pages} {285} (\bibinfo {year}
  {2001}{\natexlab{b}})},\ \Eprint {http://arxiv.org/abs/hep-ph/0105306}
  {arXiv:hep-ph/0105306 [hep-ph]} \BibitemShut {NoStop}%
\bibitem [{\citenamefont {Eyal}\ and\ \citenamefont {Nir}(1999)}]{Eyal:1999ii}%
  \BibitemOpen
  \bibfield  {author} {\bibinfo {author} {\bibfnamefont {G.}~\bibnamefont
  {Eyal}}\ and\ \bibinfo {author} {\bibfnamefont {Y.}~\bibnamefont {Nir}},\
  }\href@noop {} {\bibfield  {journal} {\bibinfo  {journal} {JHEP}\ }\textbf
  {\bibinfo {volume} {9909}},\ \bibinfo {pages} {013} (\bibinfo {year}
  {1999})},\ \Eprint {http://arxiv.org/abs/hep-ph/9908296}
  {arXiv:hep-ph/9908296 [hep-ph]} \BibitemShut {NoStop}%
\bibitem [{\citenamefont {Frampton}\ \emph {et~al.}(2000)\citenamefont
  {Frampton}, \citenamefont {Hung},\ and\ \citenamefont
  {Sher}}]{Frampton:1999xi}%
  \BibitemOpen
  \bibfield  {author} {\bibinfo {author} {\bibfnamefont {P.~H.}\ \bibnamefont
  {Frampton}}, \bibinfo {author} {\bibfnamefont {P.}~\bibnamefont {Hung}}, \
  and\ \bibinfo {author} {\bibfnamefont {M.}~\bibnamefont {Sher}},\ }\href
  {\doibase 10.1016/S0370-1573(99)00095-2} {\bibfield  {journal} {\bibinfo
  {journal} {Phys.Rept.}\ }\textbf {\bibinfo {volume} {330}},\ \bibinfo {pages}
  {263} (\bibinfo {year} {2000})},\ \Eprint
  {http://arxiv.org/abs/hep-ph/9903387} {arXiv:hep-ph/9903387 [hep-ph]}
  \BibitemShut {NoStop}%
\bibitem [{\citenamefont {Branco}\ \emph {et~al.}(1993)\citenamefont {Branco},
  \citenamefont {Parada}, \citenamefont {Morozumi},\ and\ \citenamefont
  {Rebelo}}]{Branco:1992uy}%
  \BibitemOpen
  \bibfield  {author} {\bibinfo {author} {\bibfnamefont {G.}~\bibnamefont
  {Branco}}, \bibinfo {author} {\bibfnamefont {P.}~\bibnamefont {Parada}},
  \bibinfo {author} {\bibfnamefont {T.}~\bibnamefont {Morozumi}}, \ and\
  \bibinfo {author} {\bibfnamefont {M.}~\bibnamefont {Rebelo}},\ }\href
  {\doibase 10.1016/0370-2693(93)90098-3} {\bibfield  {journal} {\bibinfo
  {journal} {Phys.Lett.}\ }\textbf {\bibinfo {volume} {B306}},\ \bibinfo
  {pages} {398} (\bibinfo {year} {1993})}\BibitemShut {NoStop}%
\bibitem [{Note11()}]{Note11}%
  \BibitemOpen
  \bibinfo {note} {Those are sufficiently generous ranges: for example, for an
  additional up quark $T$, we will have $C_{1}\to S_0(x_t,x_T)$ and $C_{2}\to
  S_0(x_T)$, ($x_T=m_T^2/M_W^2$); with a mass $m_T$ ranging up to $5$ TeV,
  $C_{1}\leq 9.3$ and $C_{2}\leq 980$.}\BibitemShut {Stop}%
\bibitem [{Note12()}]{Note12}%
  \BibitemOpen
  \bibinfo {note} {For completeness: there are no direct measurements of
  $|V_{td}^{\protect \tmspace +\thinmuskip {.1667em}}|$, and not very
  constraining ones of $|V_{tb}^{\protect \tmspace +\thinmuskip {.1667em}}|$
  (see table \ref {AP:tab:data}).}\BibitemShut {Stop}%
\bibitem [{\citenamefont {Silva}\ and\ \citenamefont
  {Wolfenstein}(1997)}]{Silva:1996ih}%
  \BibitemOpen
  \bibfield  {author} {\bibinfo {author} {\bibfnamefont {J.~P.}\ \bibnamefont
  {Silva}}\ and\ \bibinfo {author} {\bibfnamefont {L.}~\bibnamefont
  {Wolfenstein}},\ }\href {\doibase 10.1103/PhysRevD.55.5331} {\bibfield
  {journal} {\bibinfo  {journal} {Phys.Rev.}\ }\textbf {\bibinfo {volume}
  {D55}},\ \bibinfo {pages} {5331} (\bibinfo {year} {1997})},\ \Eprint
  {http://arxiv.org/abs/hep-ph/9610208} {arXiv:hep-ph/9610208 [hep-ph]}
  \BibitemShut {NoStop}%
\bibitem [{\citenamefont {Aad}\ \emph {et~al.}(2012)\citenamefont {Aad} \emph
  {et~al.}}]{Aad:2012xca}%
  \BibitemOpen
  \bibfield  {author} {\bibinfo {author} {\bibfnamefont {G.}~\bibnamefont
  {Aad}} \emph {et~al.} (\bibinfo {collaboration} {ATLAS Collaboration}),\
  }\href {\doibase 10.1016/j.physletb.2012.08.011} {\bibfield  {journal}
  {\bibinfo  {journal} {Phys.Lett.}\ }\textbf {\bibinfo {volume} {B716}},\
  \bibinfo {pages} {142} (\bibinfo {year} {2012})},\ \Eprint
  {http://arxiv.org/abs/1205.5764} {arXiv:1205.5764 [hep-ex]} \BibitemShut
  {NoStop}%
\bibitem [{\citenamefont {Chatrchyan}\ \emph {et~al.}(2013)\citenamefont
  {Chatrchyan} \emph {et~al.}}]{Chatrchyan:2012zca}%
  \BibitemOpen
  \bibfield  {author} {\bibinfo {author} {\bibfnamefont {S.}~\bibnamefont
  {Chatrchyan}} \emph {et~al.} (\bibinfo {collaboration} {CMS Collaboration}),\
  }\href {\doibase 10.1103/PhysRevLett.110.022003} {\bibfield  {journal}
  {\bibinfo  {journal} {Phys.Rev.Lett.}\ }\textbf {\bibinfo {volume} {110}},\
  \bibinfo {pages} {022003} (\bibinfo {year} {2013})},\ \Eprint
  {http://arxiv.org/abs/1209.3489} {arXiv:1209.3489 [hep-ex]} \BibitemShut
  {NoStop}%
\bibitem [{\citenamefont {Adelman}\ \emph {et~al.}(2013)\citenamefont
  {Adelman}, \citenamefont {Alvarez~Gonzalez}, \citenamefont {Bai},
  \citenamefont {Baumgart}, \citenamefont {Ellis} \emph
  {et~al.}}]{Adelman:2013gis}%
  \BibitemOpen
  \bibfield  {author} {\bibinfo {author} {\bibfnamefont {J.}~\bibnamefont
  {Adelman}}, \bibinfo {author} {\bibfnamefont {B.}~\bibnamefont
  {Alvarez~Gonzalez}}, \bibinfo {author} {\bibfnamefont {Y.}~\bibnamefont
  {Bai}}, \bibinfo {author} {\bibfnamefont {M.}~\bibnamefont {Baumgart}},
  \bibinfo {author} {\bibfnamefont {R.~K.}\ \bibnamefont {Ellis}},  \emph
  {et~al.},\ }\href@noop {} {\  (\bibinfo {year} {2013})},\ \Eprint
  {http://arxiv.org/abs/1309.1947} {arXiv:1309.1947 [hep-ex]} \BibitemShut
  {NoStop}%
\bibitem [{\citenamefont {Abazov}\ \emph {et~al.}(2012)\citenamefont {Abazov}
  \emph {et~al.}}]{Abazov:2012hha}%
  \BibitemOpen
  \bibfield  {author} {\bibinfo {author} {\bibfnamefont {V.~M.}\ \bibnamefont
  {Abazov}} \emph {et~al.} (\bibinfo {collaboration} {D0 Collaboration}),\
  }\href {\doibase 10.1103/PhysRevD.86.072009} {\bibfield  {journal} {\bibinfo
  {journal} {Phys.Rev.}\ }\textbf {\bibinfo {volume} {D86}},\ \bibinfo {pages}
  {072009} (\bibinfo {year} {2012})},\ \Eprint {http://arxiv.org/abs/1208.5813}
  {arXiv:1208.5813 [hep-ex]} \BibitemShut {NoStop}%
\bibitem [{\citenamefont {Beringer}\ and\ \citenamefont {others (Particle
  Data~Group)}(2012)}]{PDG:2012}%
  \BibitemOpen
  \bibfield  {author} {\bibinfo {author} {\bibfnamefont {J.}~\bibnamefont
  {Beringer}}\ and\ \bibinfo {author} {\bibnamefont {others (Particle
  Data~Group)}} (\bibinfo {collaboration} {Particle Data Group}),\ }\href
  {\doibase 10.1103/PhysRevD.86.010001} {\bibfield  {journal} {\bibinfo
  {journal} {Phys.Rev.}\ }\textbf {\bibinfo {volume} {D86}},\ \bibinfo {pages}
  {010001} (\bibinfo {year} {2012})}\BibitemShut {NoStop}%
\bibitem [{\citenamefont {Amhis}\ and\ \citenamefont {others (Heavy Flavor
  Averaging~Group)}(2012)}]{Amhis:2012bh}%
  \BibitemOpen
  \bibfield  {author} {\bibinfo {author} {\bibfnamefont {Y.}~\bibnamefont
  {Amhis}}\ and\ \bibinfo {author} {\bibnamefont {others (Heavy Flavor
  Averaging~Group)}} (\bibinfo {collaboration} {Heavy Flavor Averaging
  Group}),\ }\href@noop {} {\  (\bibinfo {year} {2012})},\ \Eprint
  {http://arxiv.org/abs/1207.1158} {arXiv:1207.1158 [hep-ex]} \BibitemShut
  {NoStop}%
\bibitem [{\citenamefont {Aaij}\ and\ \citenamefont {others
  (LHCb~Collaboration)}(2012{\natexlab{a}})}]{LHCb:2011ab}%
  \BibitemOpen
  \bibfield  {author} {\bibinfo {author} {\bibfnamefont {R.}~\bibnamefont
  {Aaij}}\ and\ \bibinfo {author} {\bibnamefont {others (LHCb~Collaboration)}}
  (\bibinfo {collaboration} {LHCb Collaboration}),\ }\href {\doibase
  10.1016/j.physletb.2012.01.017} {\bibfield  {journal} {\bibinfo  {journal}
  {Phys.Lett.}\ }\textbf {\bibinfo {volume} {B707}},\ \bibinfo {pages} {497}
  (\bibinfo {year} {2012}{\natexlab{a}})},\ \Eprint
  {http://arxiv.org/abs/1112.3056} {arXiv:1112.3056 [hep-ex]} \BibitemShut
  {NoStop}%
\bibitem [{\citenamefont {Aaij}\ and\ \citenamefont {others
  (LHCb~Collaboration)}(2012{\natexlab{b}})}]{LHCb:2011aa}%
  \BibitemOpen
  \bibfield  {author} {\bibinfo {author} {\bibfnamefont {R.}~\bibnamefont
  {Aaij}}\ and\ \bibinfo {author} {\bibnamefont {others (LHCb~Collaboration)}}
  (\bibinfo {collaboration} {LHCb Collaboration}),\ }\href {\doibase
  10.1103/PhysRevLett.108.101803} {\bibfield  {journal} {\bibinfo  {journal}
  {Phys.Rev.Lett.}\ }\textbf {\bibinfo {volume} {108}},\ \bibinfo {pages}
  {101803} (\bibinfo {year} {2012}{\natexlab{b}})},\ \Eprint
  {http://arxiv.org/abs/1112.3183} {arXiv:1112.3183 [hep-ex]} \BibitemShut
  {NoStop}%
\bibitem [{\citenamefont {Aaij}\ \emph
  {et~al.}(2013{\natexlab{a}})\citenamefont {Aaij} \emph
  {et~al.}}]{Aaij:2013oba}%
  \BibitemOpen
  \bibfield  {author} {\bibinfo {author} {\bibfnamefont {R.}~\bibnamefont
  {Aaij}} \emph {et~al.} (\bibinfo {collaboration} {LHCb collaboration}),\
  }\href {\doibase 10.1103/PhysRevD.87.112010} {\bibfield  {journal} {\bibinfo
  {journal} {Phys.Rev.}\ }\textbf {\bibinfo {volume} {D87}},\ \bibinfo {pages}
  {112010} (\bibinfo {year} {2013}{\natexlab{a}})},\ \Eprint
  {http://arxiv.org/abs/1304.2600} {arXiv:1304.2600 [hep-ex]} \BibitemShut
  {NoStop}%
\bibitem [{\citenamefont {Aaij}\ \emph
  {et~al.}(2013{\natexlab{b}})\citenamefont {Aaij} \emph
  {et~al.}}]{Aaij:2013mpa}%
  \BibitemOpen
  \bibfield  {author} {\bibinfo {author} {\bibfnamefont {R.}~\bibnamefont
  {Aaij}} \emph {et~al.} (\bibinfo {collaboration} {LHCb collaboration}),\
  }\href {\doibase 10.1088/1367-2630/15/5/053021} {\bibfield  {journal}
  {\bibinfo  {journal} {New J.Phys.}\ }\textbf {\bibinfo {volume} {15}},\
  \bibinfo {pages} {053021} (\bibinfo {year} {2013}{\natexlab{b}})},\ \Eprint
  {http://arxiv.org/abs/1304.4741} {arXiv:1304.4741 [hep-ex]} \BibitemShut
  {NoStop}%
\bibitem [{\citenamefont {Adachi}\ \emph {et~al.}(2013)\citenamefont {Adachi}
  \emph {et~al.}}]{Adachi:2012mm}%
  \BibitemOpen
  \bibfield  {author} {\bibinfo {author} {\bibfnamefont {I.}~\bibnamefont
  {Adachi}} \emph {et~al.} (\bibinfo {collaboration} {Belle Collaboration}),\
  }\href {\doibase 10.1103/PhysRevLett.110.131801} {\bibfield  {journal}
  {\bibinfo  {journal} {Phys.Rev.Lett.}\ }\textbf {\bibinfo {volume} {110}},\
  \bibinfo {pages} {131801} (\bibinfo {year} {2013})},\ \Eprint
  {http://arxiv.org/abs/1208.4678} {arXiv:1208.4678 [hep-ex]} \BibitemShut
  {NoStop}%
\bibitem [{\citenamefont {Lees}\ \emph
  {et~al.}(2013{\natexlab{b}})\citenamefont {Lees} \emph
  {et~al.}}]{Lees:2012ju}%
  \BibitemOpen
  \bibfield  {author} {\bibinfo {author} {\bibfnamefont {J.}~\bibnamefont
  {Lees}} \emph {et~al.} (\bibinfo {collaboration} {BaBar Collaboration}),\
  }\href {\doibase 10.1103/PhysRevD.88.031102} {\bibfield  {journal} {\bibinfo
  {journal} {Phys.Rev.}\ }\textbf {\bibinfo {volume} {D88}},\ \bibinfo {pages}
  {031102} (\bibinfo {year} {2013}{\natexlab{b}})},\ \Eprint
  {http://arxiv.org/abs/1207.0698} {arXiv:1207.0698 [hep-ex]} \BibitemShut
  {NoStop}%
\bibitem [{\citenamefont {Nakano}\ \emph {et~al.}(2006)\citenamefont {Nakano}
  \emph {et~al.}}]{Nakano:2005jb}%
  \BibitemOpen
  \bibfield  {author} {\bibinfo {author} {\bibfnamefont {E.}~\bibnamefont
  {Nakano}} \emph {et~al.} (\bibinfo {collaboration} {Belle Collaboration}),\
  }\href {\doibase 10.1103/PhysRevD.73.112002} {\bibfield  {journal} {\bibinfo
  {journal} {Phys.Rev.}\ }\textbf {\bibinfo {volume} {D73}},\ \bibinfo {pages}
  {112002} (\bibinfo {year} {2006})},\ \Eprint
  {http://arxiv.org/abs/hep-ex/0505017} {arXiv:hep-ex/0505017 [hep-ex]}
  \BibitemShut {NoStop}%
\bibitem [{\citenamefont {Aubert}\ \emph {et~al.}(2006)\citenamefont {Aubert}
  \emph {et~al.}}]{Aubert:2006nf}%
  \BibitemOpen
  \bibfield  {author} {\bibinfo {author} {\bibfnamefont {B.}~\bibnamefont
  {Aubert}} \emph {et~al.} (\bibinfo {collaboration} {BaBar Collaboration}),\
  }\href {\doibase 10.1103/PhysRevLett.96.251802} {\bibfield  {journal}
  {\bibinfo  {journal} {Phys.Rev.Lett.}\ }\textbf {\bibinfo {volume} {96}},\
  \bibinfo {pages} {251802} (\bibinfo {year} {2006})},\ \Eprint
  {http://arxiv.org/abs/hep-ex/0603053} {arXiv:hep-ex/0603053 [hep-ex]}
  \BibitemShut {NoStop}%
\bibitem [{\citenamefont {Aoki}\ \emph {et~al.}(2013)\citenamefont {Aoki},
  \citenamefont {Aoki}, \citenamefont {Bernard}, \citenamefont {Blum},
  \citenamefont {Colangelo} \emph {et~al.}}]{Aoki:2013ldr}%
  \BibitemOpen
  \bibfield  {author} {\bibinfo {author} {\bibfnamefont {S.}~\bibnamefont
  {Aoki}}, \bibinfo {author} {\bibfnamefont {Y.}~\bibnamefont {Aoki}}, \bibinfo
  {author} {\bibfnamefont {C.}~\bibnamefont {Bernard}}, \bibinfo {author}
  {\bibfnamefont {T.}~\bibnamefont {Blum}}, \bibinfo {author} {\bibfnamefont
  {G.}~\bibnamefont {Colangelo}},  \emph {et~al.},\ }\href@noop {} {\
  (\bibinfo {year} {2013})},\ \Eprint {http://arxiv.org/abs/1310.8555}
  {arXiv:1310.8555 [hep-lat]} \BibitemShut {NoStop}%
\end{thebibliography}

\end{document}